\documentclass[paperpaper,twocolumn,english,superscriptaddress,aps,prb,floatfix,amsfonts,amssymb]{revtex4}
\usepackage{float}
\usepackage{epsfig}
\usepackage{graphicx}
\usepackage{dcolumn}
\usepackage{bm}
\usepackage{appendix}
\usepackage{subfigure}
\usepackage{color}
\usepackage{natbib}

\begin{document}


\title{Bethe strings in the spin dynamical structure factor of the Mott-Hubbard phase in one-dimensional fermionic Hubbard model}
\author{Jos\'e M. P. Carmelo}
\affiliation{Center of Physics of University of Minho and University of Porto, P-4169-007 Oporto, Portugal}
\affiliation{Department of Physics, University of Minho, Campus Gualtar, P-4710-057 Braga, Portugal}
\affiliation{Boston University, Department of Physics, 590 Commonwealth Avenue, Boston, 
Massachusetts 02215, USA}
\author{Tilen \v{C}ade\v{z}}
\affiliation{Center for Theoretical Physics of Complex Systems, Institute for Basic Science (IBS), Daejeon 34126, Republic of Korea}

\date{6 August 2020; revised 9 October 2020; accepted 24 December 2020; published 15 January 2021}


\begin{abstract}
The spectra and role in the spin dynamical properties of bound states of elementary magnetic excitations named 
Bethe strings that occur in some integrable spin and electronic one-dimensional models 
have recently been identified and realized in several materials by experiments. Corresponding theoretical studies have usually 
relied on the one-dimensional spin-$1/2$ Heisenberg antiferromagnet in a magnetic field. At the isotropic point, 
it describes the large onsite repulsion $U$ limit of the spin degrees of freedom of the one-dimensional fermionic Hubbard 
model with one electron per site in a magnetic field $h$. In this paper we consider the thermodynamic limit and 
study the effects of lowering the latter quantum problem ratio $u=U/4t$, where $t$ is the first-neighbor transfer integral,
on the line-shape singularities in $(k,\omega)$-plane regions at and just above the lower thresholds of the transverse and 
longitudinal spin dynamical structure factors. The most significant spectral weight contribution from Bethe strings leads to a 
gapped continuum in the spectrum of the spin dynamical structure factor $S^{+-} (k,\omega)$. 
Our study focuses on the line shape singularities at and just above the gapped lower threshold 
of that continuum, which have been identified in experiments. Our results are consistent with the contribution of
Bethe strings to $S^{zz} (k,\omega)$ being small at low spin densities and becoming negligible upon increasing
that density. Our results provide physically important information about how electron itinerancy affects the spin dynamics.
 \end{abstract}


\maketitle

\section{Introduction}
\label{SECI}

Recently, there has been a renewed interest in the experimental identification and realization of bound states of elementary magnetic excitations named Bethe strings in materials whose magnetic properties are described by the one-dimensional (1D)
spin-$1/2$ Heisenberg antiferromagnet in magnetic fields \cite{Wang_19,Bera_20,Wang_18,Kohno_09,Stone_03}.
This applies to that model isotropic point in the case of experimental studies of CuCl$_2$$\cdot$2N(C$_5$D$_5$) 
and Cu(C$_4$H$_4$N$_2$)(NO$_3$)$_2$ \cite{Kohno_09,Stone_03,Heilmann_78}.

The isotropic spin-$1/2$ Heisenberg $XXX$ chain describes the spin degrees of freedom of the 1D fermionic Hubbard model's 
Mott-Hubbard insulator phase in the limit of large onsite repulsion $U$. That phase is reached at a density of one electron
per site. Interesting related physical questions are whether lowering the ratio $u=U/4t$ leads to a description of the spin dynamical properties suitable to spin-chain compounds and how electron itinerancy affects the spin dynamics.
Here $t$ is the model first-neighbor transfer integral.

In the case of the 1D fermionic Hubbard model, there are in its exact solution \cite{Lieb,Lieb-03,Martins}
two types of Bethe strings described by complex nonreal Bethe-ansatz rapidities. They refer to the model spin 
and charge degrees of freedom, respectively, \cite{Takahashi,Carmelo_18,Carmelo_18A}. Here
we call them charge and spin $n$-strings. The nature of their configurations becomes clearer
in terms of the {\it rotated electrons} that are generated from the electrons by a unitary transformation.
It is such that $\sigma=\uparrow,\downarrow$ rotated-electron single-site occupancy,
rotated-electron double-site occupancy, and rotated-electron no site occupancy are
good quantum numbers for the whole $u>0$ range. (For electrons they are good quantum
numbers only for large $u$.) The corresponding electron - rotated-electron 
unitary operator is uniquely defined in Ref. \onlinecite{Carmelo_17} by its
set of $4^L\times 4^L = 4^{2L}$ matrix elements between {\it all}  $4^L$ energy
eigenstates that span the model's Hilbert space. Here $L$ is the number of sites
and lattice length in units of lattice spacing one.

The spin $n$-strings are for $n>1$ bound states of a number $n$ of spin-singlet 
pairs of rotated electrons with opposite spin projection that singly occupy sites.
The charge $n$-strings are for $n>1$ bound states of 
$n$ charge $\eta$-spin singlet pairs of rotated-electron doubly and unoccupied sites \cite{Carmelo_18,Carmelo_18A}. 
However, energy eigenstates described by
only real Bethe-ansatz rapidities do not contain $n>1$ charge and spin $n$-strings and are
populated by unbound spin-singlet pairs and unbound charge $\eta$-spin singlet pairs \cite{Carmelo_18,Carmelo_18A}. 
Ground states are not populated by the latter type of pairs.

Previous studies focused on contributions to the spin dynamical structure factors of the
1D fermionic Hubbard model with one electron per site from excited energy eigenstates
described by real Bethe-anstaz rapidities at zero magnetic field
\cite{Benthien_07,Bhaseen_05,Essler_99} and in a finite magnetic field \cite{Carmelo_16}.
There were also studies of structure factors of the 1D Hubbard model in a magnetic field
in the limit of low excitation energy $\omega$ \cite{Carmelo_93A}.

Our study addresses the 1D Hubbard model with one electron per site in the spin subspace spanned by energy
eigenstates without charge $\eta$-spin singlet pairs. Some of these energy eigenstates are described by
complex nonreal spin Bethe-ansatz rapidities and thus are populated by spin $n$-strings. 

The general goal of this paper is the study of the contribution from spin $n$-string states to the spin dynamical structure
factors of the 1D Hubbard model with one electron per site in a magnetic field $h$. 
Our study relies on the dynamical theory introduced for the 1D Hubbard model in Ref. \onlinecite{Carmelo_05}.
It has been adapted to the 1D Hubbard model with one electron per site in a spin subspace spanned
by energy eigenstates described by real Bethe-ansatz rapidities in Ref. \onlinecite{Carmelo_16}. The studies of this paper 
use the latter dynamical theory in an extended spin subspace spanned
by two classes of energy eigenstates, populated and not populated by spin $n$-strings, respectively.

In the case of integrable models, the general dynamical theory of Refs.
\onlinecite{Carmelo_16,Carmelo_05,Carmelo_08} reaches the same finite-energy
dynamical correlation functions expressions as the mobile quantum impurity model scheme 
of Refs. \onlinecite{Imambekov_09,Imambekov_12}. Such expressions 
apply at and in the $(k,\omega)$-plane vicinity of the corresponding
spectra's lower thresholds's. That for the former dynamical theory
and the mobile quantum impurity model scheme 
such dynamical correlation functions expressions 
are for arbitrary finite values of the excitation energy 
indeed the same and account for the same 
microscopic processes is an issue discussed and confirmed in Appendix A of
Ref. \onlinecite{Carmelo_18} and in Ref. \onlinecite{Carmelo_16A} for
a representative integrable model and several dynamical correlation functions.

The dynamical theory of Refs. \onlinecite{Carmelo_16,Carmelo_05,Carmelo_08} is a generalization to
the whole $u=U/4t>0$ range of the approach used in the $u\rightarrow\infty$ limit in Refs.
\onlinecite{Karlo,Karlo_97}. Momentum dependent exponents in the expressions
of spectral functions have also been obtained in Refs. \onlinecite{Sorella_96,Sorella_98}.

Beyond the studies of Ref. \onlinecite{Carmelo_16}, here the 
application of the dynamical theory is extended to the contribution to the
spin dynamical structure factors from excited energy eigenstates populated by spin $n$-strings. 

The theory refers to the thermodynamic limit, in which the 
expression of the square of the matrix elements of the dynamical structure factors between the
ground state and the excited states behind most spectral weight
has the general form given in Eq. (\ref{ME}). It does not provide the precise values of
the $u$ and $m$ dependent constant $0<B_s\leq 1$ and $u$ dependent constants $0<f_l<1$ where $l=0,2,4$
in that expression. In spite of this limitation, our results provide important physical
information on the dynamical structure factors under study.

In the case of the related isotropic spin $1/2$ Heisenberg chain in a magnetic field, 
it is known\cite{Kohno_09} that the only contribution from excited energy eigenstates populated by spin $n$-strings that
leads to a $(k,\omega)$-plane gapped continuum with a significant amount
of spectral wight refers to $S^{+-} (k,\omega)$. 

Based on a relation between the
level of negativity of the momentum dependent exponents that control
the spin dynamical structure factors $(k,\omega)$-plane singularities and
the amount of spectral weight existing near them, we confirm that that result
applies to the whole $u>0$ range of the 1D Hubbard model with one
electron per site in a magnetic field. However, the contribution of
spin $n$-strings states to $S^{zz} (k,\omega)$ is found to be small at low spin densities and 
to become negligible upon increasing it beyond a spin density $\tilde{m}$ that 
decreases upon decreasing $u$, reading $\tilde{m}=0$ for $u\rightarrow\infty$ and $\tilde{m}\approx 0.317$ for $u\gg 1$.
Finally, the contribution of these states to $S^{-+} (k,\omega)$ is found to be negligible 
at finite magnetic fields. 

The main aim of this paper is the study of the line shape singularities of 
$S^{+-} (k,\omega)$, $S^{xx} (k,\omega)$, and $S^{zz} (k,\omega)$
at and just above the $(k,\omega)$-plane gapped lower threshold 
of the spectra associated with spin $n$-string states. The corresponding
singularity peaks have been identified in neutron scattering experiments
\cite{Kohno_09,Stone_03,Heilmann_78}. 

As a side result, we address the more general problem of the line-shape of the transverse and 
longitudinal spin dynamical structure factors at finite magnetic field $h$ 
in the $(k,\omega)$-plane vicinity of singularities
at and above the lower thresholds of the spectra of the excited energy
eigenstates of the 1D Hubbard model with one electron per site that produce a significant amount of
spectral weight. This includes both excited states with and without spin $n$-strings.
The contribution from the latter states leads to the largest amount of
spin dynamical structure factors's spectral weight \cite{Carmelo_16}.

Our secondary goal is to provide an overall physical picture that includes the
relative $(k,\omega)$-plane location of all spectra with a significant amount
of spectral weight and accounting for the contributions of different
types of states to {\it both} the gapped and gapless lower threshold singularities that 
emerge in the spin dynamical structure factors.

The paper is organized as follows. The model and the spin dynamical structure factors are
the issues addressed in Sec. \ref{SECII}. In Sec. \ref{SECIII} the $(k,\omega)$-plane spectra of the excited states that
lead to most dynamical structure factors's spectral weight are studied, with
emphasis on those of the spin $n$-string states.
The line shape at and above the gapped lower thresholds of the $n$-string states's
dynamical structure factors spectra is the main subject of Sec.
\ref{SECIV}. As a side result, in that section the problem is revisited at and above 
the lower thresholds of the $(k,\omega)$-plane continua associated with excited
states described by real Bethe-anstaz rapidities. In Sec. \ref{SECV} the limiting behaviors 
of the spin dynamical structure factors are addressed. Finally, the discussion and concluding 
remarks are presented in Sec. \ref{SECVI}.

A set of useful results needed for our studies are presented in five Appendices.
This includes the selection rules and sum rule provided in Appendix \ref{A}.
In Appendix \ref{B} the gapless transverse and longitudinal continuum spectra
are revisited. The energy gaps between the gapped lower thresholds of the spin $n$-string
states's spectra and the lower $(k,\omega)$-plane continua is the issue addressed in
Appendix \ref{C}. In Appendix \ref{D} the number and current number deviations and 
the spectral functionals that control the momentum dependent exponents in
the spin dynamical structure factors's expressions are given.
Some useful quantities also needed for our studies are defined 
and provided in Appendix \ref{E}. 

\section{The model and the spin dynamical structure factors}
\label{SECII}

In this paper we use in general units of lattice constant and Planck constant one.
Our study refers to spin subspaces spanned by energy
eigenstates for which the number of lattice sites $N_a$ equals that of
electrons $N=N_{\uparrow}+N_{\downarrow}$, of which $N_{\uparrow}$ and $N_{\downarrow}$
have up and down spin projection, respectively.

The Hubbard model with one electron per site at vanishing chemical potential 
in a magnetic field $h$ under periodic boundary conditions on a 1D lattice 
of length $L\rightarrow\infty$ is given by,
\begin{equation}
{\hat{H}} = t\,\hat{T}+U\,\hat{V}_D + 2\mu_B h\,{\hat{S}}^{z} \, .
\label{H}
\end{equation}
Here $\mu_B$ is the Bohr magneton and for simplicity in $g\mu_B$ we have taken $g=2$. The operators read,
\begin{eqnarray}
\hat{T} & = & -\sum_{\sigma=\uparrow,\downarrow }\sum_{j=1}^{N}\left(c_{j,\sigma}^{\dag}\,
c_{j+1,\sigma} + c_{j+1,\sigma}^{\dag}\,c_{j,\sigma}\right)\hspace{0.20cm}{\rm and}
\nonumber \\
\hat{V}_D & = & \sum_{j=1}^{N}\hat{\rho}_{j,\uparrow}\hat{\rho}_{j,\downarrow} 
\hspace{0.20cm}{\rm where}\hspace{0.20cm}
\hat{\rho}_{j,\sigma} = c_{j,\sigma}^{\dag}\,c_{j,\sigma} -1/2 \, ,
\label{HH}
\end{eqnarray}
where $\hat{T}$ is the kinetic-energy operator in units of $t$, $\hat{V}_D$ is the electron 
(or spin $1/2$ atom) on-site repulsion operator in units of $U$, 
the operator $c_{j,\sigma}^{\dagger}$ (and $c_{j,\sigma}$)
creates (and annihilates) a spin-projection $\sigma = \uparrow,\downarrow$ electron at lattice site
$j=1,...,N$, and the electron number operators read
${\hat{N}}=\sum_{\sigma=\uparrow ,\downarrow }\,\hat{N}_{\sigma}$ and
${\hat{N}}_{\sigma}=\sum_{j=1}^{N}\hat{n}_{j,\sigma}= \sum_{j=1}^{N}c_{j,\sigma}^{\dag}\,c_{j,\sigma}$.
Moreover, ${\hat{S}}^{z}=\sum_{j=1}^{N}\hat{S}^{z}_j$ is the diagonal generator of the global spin $SU(2)$ symmetry algebra.
We denote the energy eigenstate's spin projection by 
$S^{z}=-(N_{\uparrow} -N_{\downarrow})/2\in [-S,S]$ where $S\in [0,N/2]$ denotes their spin.

Our results refer to magnetic fields $0<h<h_c$ and corresponding spin densities
$0<m<1$. Here $m = (N_{\uparrow} -N_{\downarrow})/N_a$
and $h_c$ is the critical magnetic field above which there is
fully polarized ferromagnetism. The corresponding spin-density curve
that relates $h$ and $m$ is given by,
\begin{eqnarray}
h (m) & = & - {\varepsilon_{s}^0 (k_{F\downarrow})\over 2\mu_B}\vert_{m = 1 - 2k_{F\downarrow}/\pi} \in [0,h_c]
\hspace{0.20cm}{\rm where}
\nonumber \\
2\mu_B\,h_c & = &2\mu_B\,h (m)\vert_{m=1}= \sqrt{(4t)^2+U^2} - U \, ,
\label{hc}
\end{eqnarray}
$\varepsilon_{s}^0 (q)$ is the $s$ band energy dispersion, Eq. (\ref{equA11}),
whose zero-energy level is shifted relative to that in Eq. (\ref{equA4}), such that
$\varepsilon_{s} (k_{F\downarrow})=0$, and the magnetic energy scale $2\mu_B\,h_c$
is associated with the quantum phase transition from the Mott-Hubbard insulator phase
to fully polarized ferromagnetism. It defines the corresponding critical magnetic field,
$h_c = (\sqrt{(4t)^2+U^2} - U)/2\mu_B$.

The spin dynamical structure factors studied in this paper in the $(k,\omega)$-plane vicinity of well defined singularities
are quantities of both theoretical interest and of interest for comparison with experimentally measurable quantities.
They can be written as,
\begin{eqnarray}
S^{aa} (k,\omega) & = & \sum_{j=1}^N e^{-ik j}\int_{-\infty}^{\infty}dt\,e^{-i\omega t}\langle GS\vert\hat{S}^{a}_j (t)\hat{S}^{a}_j (0)\vert GS\rangle 
\nonumber \\
& = & \sum_{\nu}\vert\langle \nu\vert\hat{S}^{a}_k\vert GS\rangle\vert^2
\delta (\omega - \omega^{aa}_{\nu} (k)) \, .
\label{SDSF}
\end{eqnarray}
Here $a =x,y,z$, the spectra read $\omega^{aa}_{\nu} (k) = (E_{\nu}^{aa} - E_{GS})$, $E_{\nu}^{aa}$ refers to
the energies of the excited energy eigenstates that contribute to the $aa= xx,yy,zz$
dynamical structure factors, $E_{GS}$ is the 
initial ground state energy, and $\hat{S}^{a}_k$ are for $a =x,y,z$ the Fourier transforms of the usual local $a =x,y,z$ 
spin operators $\hat{S}^{a}_j$, respectively.

Due to the rotational symmetry in spin space, off-diagonal components 
of the spin dynamical structure factor vanish, $S^{aa'} (k,\omega) = 0$ for
$a\neq a'$, and the two transverse components are identical, $S^{xx} (k,\omega)=S^{yy} (k,\omega)$.
At zero and finite magnetic field, one has that $S^{zz} (k,\omega)=S^{xx} (k,\omega)$
and $S^{zz} (k,\omega)\neq S^{xx} (k,\omega)$, respectively.

In the transverse case, we often address the problem in terms of
the dynamical structure factors $S^{+-} (k,\omega)$ and $S^{-+} (k,\omega)$ in
$S^{xx} (k,\omega) = {1\over 4}\left(S^{+-} (k,\omega)+S^{-+} (k,\omega)\right)$. 
We rely on the symmetry that exists for the problems under study
between the spin density intervals $m\in ]-1,0]$ and $m\in ]0,1[$, such that,
\begin{eqnarray}
S^{-+} (k,\omega)\vert_m & = &S^{+-} (k,\omega)\vert_{-m}
\hspace{0.20cm}{\rm and}
\nonumber \\
S^{+-} (k,\omega)\vert_m & = & S^{-+} (k,\omega)\vert_{-m}
\nonumber \\
& & {\rm for}\hspace{0.20cm}m \in ]0,1[ \,  .
\label{SPMMPmm}
\end{eqnarray}
Hence we only consider explicitly the spin density interval $m\in ]0,1[$.
Since $S^{aa} (k,\omega)=S^{aa} (-k,\omega)$ and the same applies
to $S^{+-} (k,\omega)$ and $S^{-+} (k,\omega)$, for simplicity
the results of this paper refer to $k>0$ momenta in the first Brillouin zone, $k\in [0,\pi]$. 

Some useful selection rules tell us which classes of energy eigenstates have nonzero matrix elements with the 
ground state \cite{Muller}. Such selection rules as well as some useful sum rules are given in Appendix \ref{A}.

The selection rules in Eq. (\ref{SRh0}) reveal that at $h=0$ and thus $m=0$ when
$S^{zz} (k,\omega) = S^{xx} (k,\omega)$, the longitudinal dynamical structure factor
is fully controlled by transitions from the ground state for which $S^z =S=0$ to excited states with
spin numbers $S^z=0$ and $S=1$. However, following such rules the transverse dynamical structure factors
are controlled by transitions from that ground state to excited states with
spin numbers $S^z=\pm 1$ and $S=1$. 

This is different from the case for magnetic fields $0<h<h_c$ considered in this paper.
According to the selection rules, Eq. (\ref{SRhfinite}), the longitudinal dynamical structure factor
$S^{zz} (k,\omega) \neq S^{xx} (k,\omega)$ is controlled by transitions from the ground state 
with spin numbers $S^z = -S$ to excited states with
the same spin numbers $S^z = -S$. According to the same selection rules, the dynamical structure factors
$S^{+-} (k,\omega)$ and $S^{-+} (k,\omega)$ are controlled by transitions from the ground state 
with spin numbers $S^z = -S$ to excited states with spin numbers $S^z = -S \pm 1$.
\begin{figure}
\begin{center}
\centerline{\includegraphics[width=9cm]{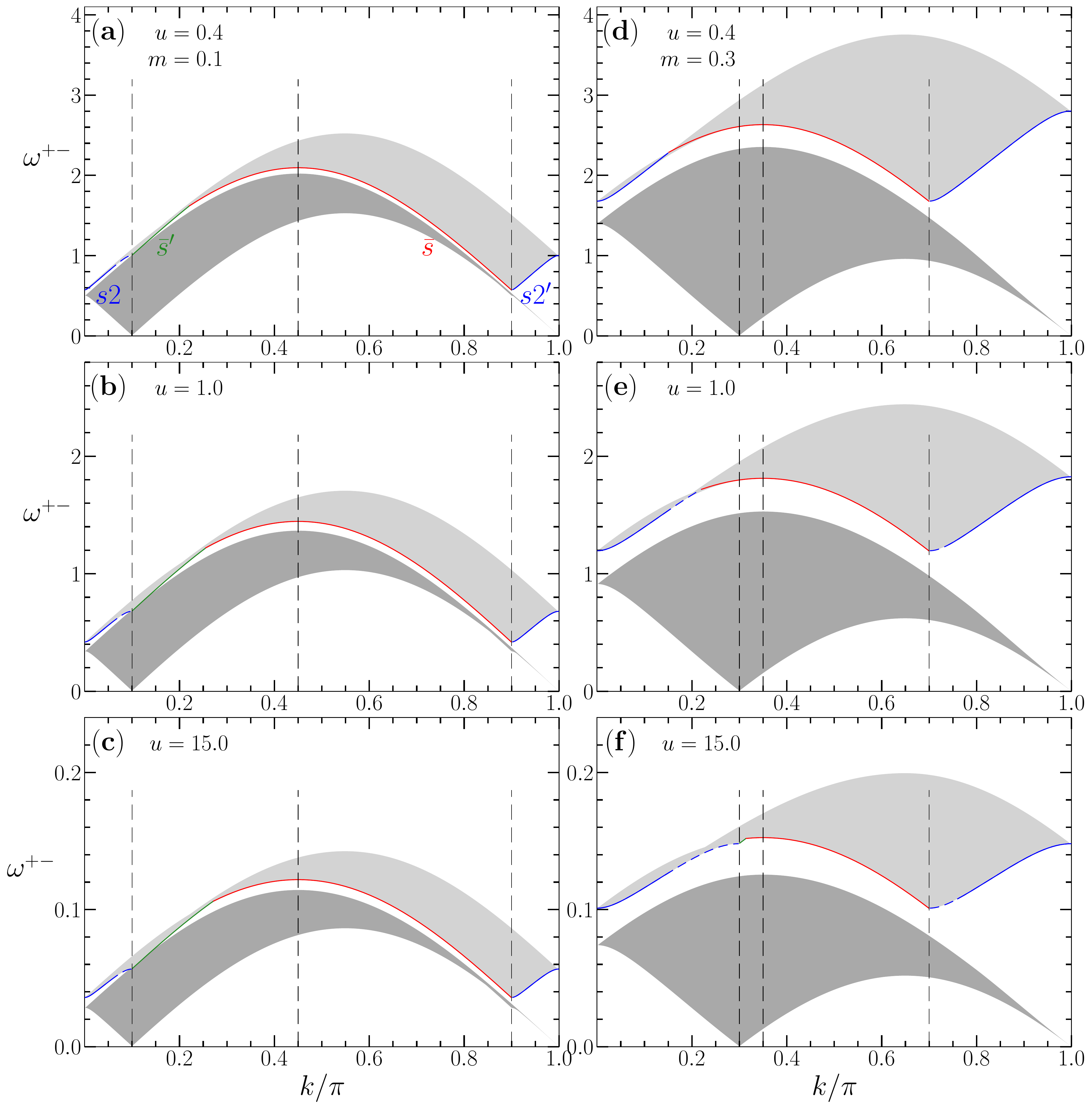}}
\caption{The two $(k,\omega)$-plane lower and upper continuum regions 
where for spin densities (a-c) $m=0.1$ and (d-f) $m =0.3$ 
and $u=0.4,1.0,15.0$ there is in the thermodynamic limit more spectral weight in $S^{+-} (k,\omega)$. 
The sketch of the $(k,\omega)$-plane distributions represented here and in Figs. \ref{figure2}-\ref{figure6} 
does not provide information on the relative amount of spectral weight contained within each spectrum's 
gray continuum. [The three reference vertical lines mark the momenta (a-c) $k=k_{F\uparrow} - k_{F\downarrow}=\pi/10$,
$k=k_{F\downarrow}=9\pi/20$, and $k=2k_{F\downarrow}=9\pi/10$
and (d-f) $k=k_{F\uparrow} - k_{F\downarrow}=3\pi/10$,
$k=k_{F\downarrow}=7\pi/20$, and $k=2k_{F\downarrow}=7\pi/10$,
where $k_{F\downarrow} = {\pi\over 2}(1-m)$ and $k_{F\uparrow} = {\pi\over 2}(1+m)$,
Eq. (\ref{kkk}).] The lower and upper 
continuum spectra are associated with excited energy eigenstates without and with spin $n$-strings, respectively.
In the thermodynamic limit, the $(k,\omega)$-plane region between the upper threshold of
the lower continuum and the gapped lower threshold of the upper $n$-string continuum has nearly no spectral weight.
In the case of the gapped lower threshold of the spin $n$-string continuum,
the analytical expressions given in this paper refer to near and just above that threshold 
whose subintervals correspond to branch lines parts represented in the figure
by solid and dashed lines. The latter refer to $k$ intervals where the momentum dependent
exponents plotted in Figs. \ref{figure7}-\ref{figure10} are negative and positive, respectively. In the former intervals, $S^{+-} (k,\omega)$ 
displays singularity peaks, seen also in experimental studies of CuCl$_2$$\cdot$2N(C$_5$D$_5$) 
and Cu(C$_4$H$_4$N$_2$)(NO$_3$)$_2$ \cite{Kohno_09,Stone_03,Heilmann_78}.}
\label{figure1}
\end{center}
\end{figure}

\section{Dynamical structure factors spectra}
\label{SECIII}

Our study of the spin dynamical structure factors relies on the representation of the energy eigenstates
suitable to the dynamical theory used in this paper\cite{Carmelo_16}. It involves ``quasiparticles''
that in this paper we call $sn$ particles. Here $n=1,...,\infty$ is the number of spin-singlet pairs that describes
their internal degrees of freedom. 

For $n>1$ a $sn$ particle contains $n$ bound spin-singlet pairs,
the integer $n$ being also the length of the corresponding spin $n$-string. For simplicity, we denote the $s1$
particles by $s$ particles. Their internal degrees of freedom correspond to a single singlet pair.
Energy eigenstates that are not populated and are populated by $sn$ particles with $n>1$
pairs are described by real and complex nonreal Bethe-anstaz rapidities, respectively.
\begin{figure}
\begin{center}
\centerline{\includegraphics[width=9cm]{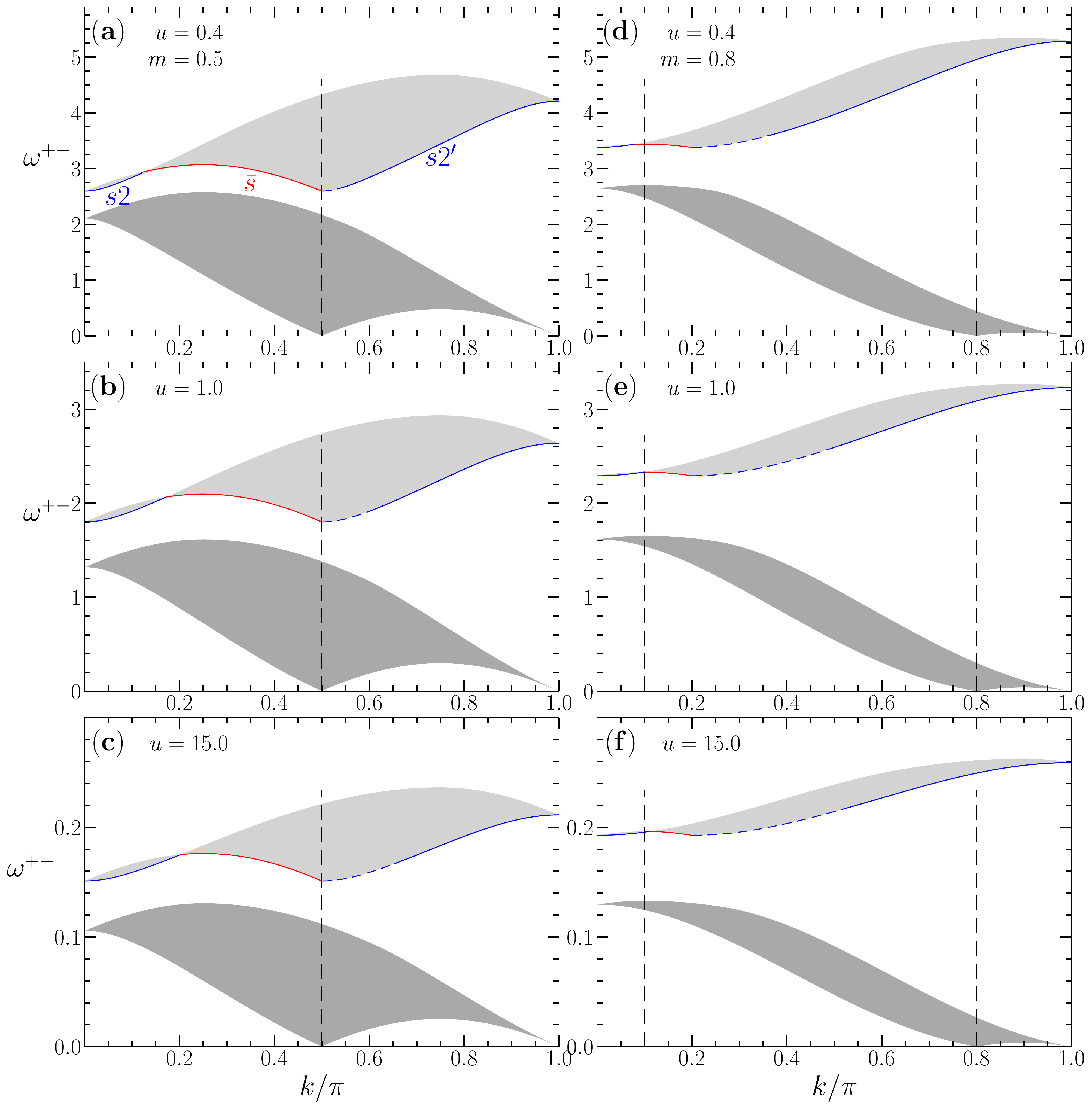}}
\caption{The same continuum spectra as in Fig. \ref{figure1} for 
spin densities (a-c) $m=0.5$ and (d-f) $m =0.8$ 
and $u=0.4,1.0,15.0$. [The three reference vertical lines mark the momenta (a-c) $k=k_{F\downarrow}=\pi/4$ and
$k=k_{F\uparrow} - k_{F\downarrow}=2k_{F\downarrow}=\pi/2$ and (d-f) $k=k_{F\downarrow}=\pi/10$, 
$k=2k_{F\downarrow}=\pi/5$, and $k=k_{F\uparrow} - k_{F\downarrow}=4\pi/5$,
where $k_{F\downarrow} = {\pi\over 2}(1-m)$ and $k_{F\uparrow} = {\pi\over 2}(1+m)$,
Eq. (\ref{kkk}).]}
\label{figure2}
\end{center}
\end{figure}

As mentioned in Sec. \ref{SECI} and confirmed in Appendix \ref{D}, 
there is a direct relation between the values of the momentum dependent exponents
that within the dynamical theory used here control the line shape in the $(k,\omega)$-plane vicinity
of the spin dynamical structure factors spectral features and the amount of spectral weight
located near them: Negative exponents imply the occurrence of singularities associated
with a significant amount of spectral weight in their $(k,\omega)$-plane vicinity.

The use of this criterion reveals that in the present thermodynamic limit and for magnetic fields $0<h<h_c$,
the only significant contribution to $S^{+-} (k,\omega)$ from energy eigenstates populated
by $sn$ particles refers to those populated by $N_{\downarrow}-2$ $s$ particles and one $s2$ particle.
Here $N_{\downarrow}= N_{\downarrow}^0 +1\in [2,N/2]$ is the excited energy eigenstate's number of down-spin electrons
in the case of initial ground states with $N_{\downarrow}^0 \in [1,N/2-1]$. 

There is as well a much weaker contribution at small spin densities from
states populated by $N_{\downarrow}-3$ $s$ particles and one $s3$ particle.
Here $N_{\downarrow}= N_{\downarrow}^0 +1\in [3,N/2]$ for the excited energy eigenstate
in the case of initial ground states with $N_{\downarrow}^0 \in [2,N/2-1]$. 
\begin{figure}
\begin{center}
\centerline{\includegraphics[width=9cm]{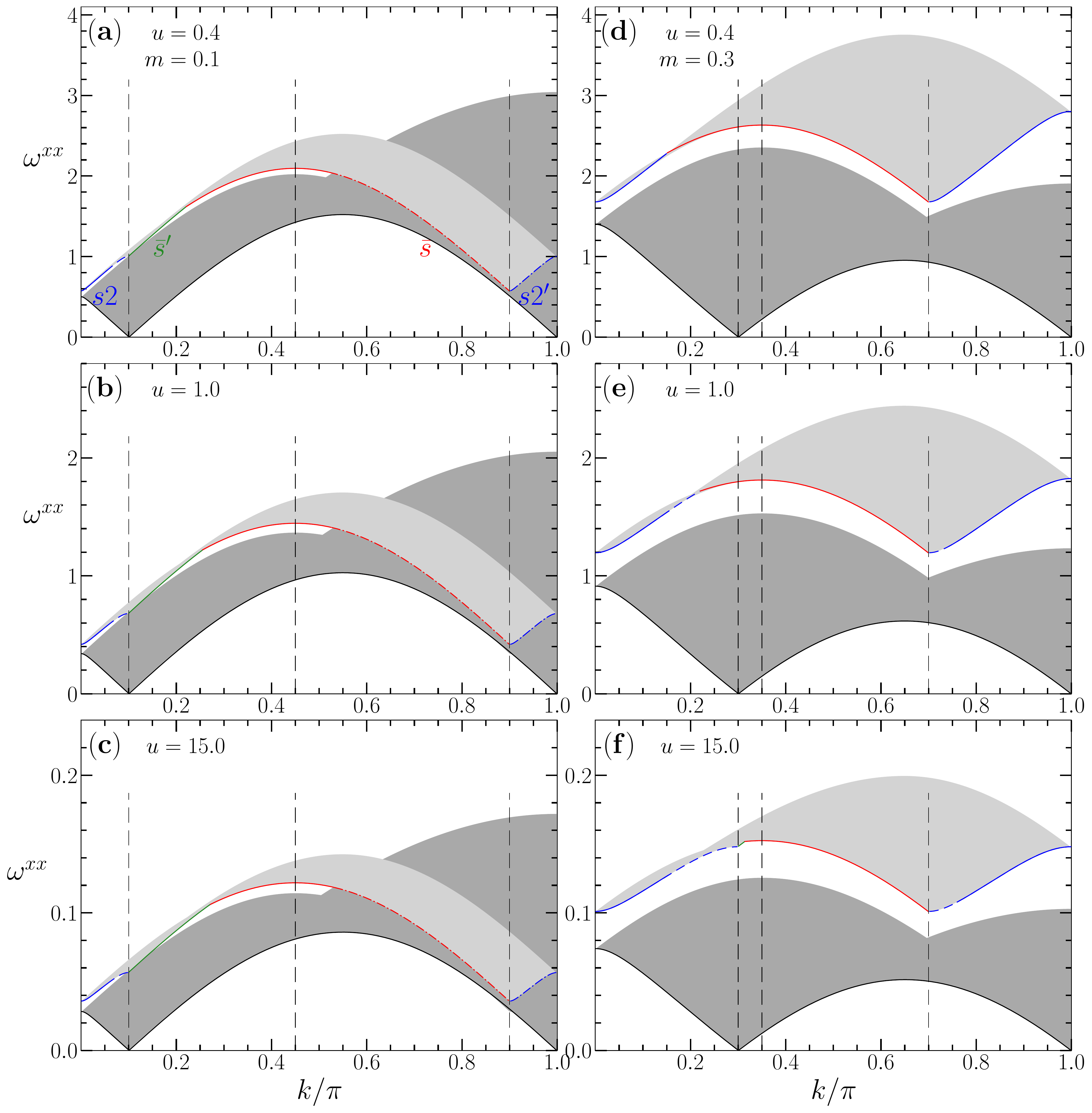}}
\caption{The two $(k,\omega)$-plane lower and upper continuum regions 
where for spin densities (a-c) $m=0.1$ and (d-f) $m =0.3$ 
and $u=0.4,1.0,15.0$ there is in the thermodynamic limit more spectral weight in $S^{xx} (k,\omega)$.
The notations are the same as in Fig. \ref{figure1}.
[The three reference vertical lines mark the momenta (a-c) $k=k_{F\uparrow} - k_{F\downarrow}=\pi/10$,
$k=k_{F\downarrow}=9\pi/20$, and $k=2k_{F\downarrow}=9\pi/10$
and (d-f) $k=k_{F\uparrow} - k_{F\downarrow}=3\pi/10$,
$k=k_{F\downarrow}=7\pi/20$, and $k=2k_{F\downarrow}=7\pi/10$, where 
$k_{F\downarrow} = {\pi\over 2}(1-m)$ and $k_{F\uparrow} = {\pi\over 2}(1+m)$, Eq. (\ref{kkk}).]
The additional part of the lower continuum relative to that of $S^{+-} (k,\omega)$ in Figs. \ref{figure1} and \ref{figure2}
stems from the contributions of $S^{-+} (k,\omega)$. As a result, for
some $k$ intervals the upper spin $n$-string continuum overlaps with
the lower continuum.}
\label{figure3}
\end{center}
\end{figure}

In the case of $S^{zz} (k,\omega)$, this refers only to energy eigenstates populated by 
$N_{\downarrow}-2$ $s$ particles and one $s2$ particle. Here $N_{\downarrow}= N_{\downarrow}^0\in [2,N/2]$ 
both for the excited energy eigenstate and initial ground states. The contribution from such states
to $S^{-+} (k,\omega)$ is found to be negligible, since all relevant exponents are 
both positive and large.

The contribution to $S^{+-} (k,\omega)$ from energy eigenstates populated
by $N_{\downarrow}-3$ $s$ particles and one $s3$ particle that occurs for small values
of the spin density is very weak and is negligible near the $(k,\omega)$-plane 
singularities to which the analytical expressions obtained in our study refer to. In addition, the latter 
very weak contributions occur in $(k,\omega)$-plane regions 
above the gapped lower threshold of the spectrum continuum associated with 
energy eigenstates populated by $N_{\downarrow}-2$ $s$ particles and one $s2$ particle. 
[The expression of that spectrum is given below in Eq. (\ref{dkEdPPM}).] 

Hence, the energy eigenstates described by complex nonreal Bethe ansatz rapidities considered
in our study are populated by $N_{\downarrow}-2$ $s$ particles and one $s2$ particle. Such states
contain thus a single spin $n$-string of length $n=2$. In addition, we account for the contribution
from energy eigenstates populated by $N_{\downarrow}$ $s$ particles that are
described by real Bethe ansatz rapidities. 
\begin{figure}
\begin{center}
\centerline{\includegraphics[width=9cm]{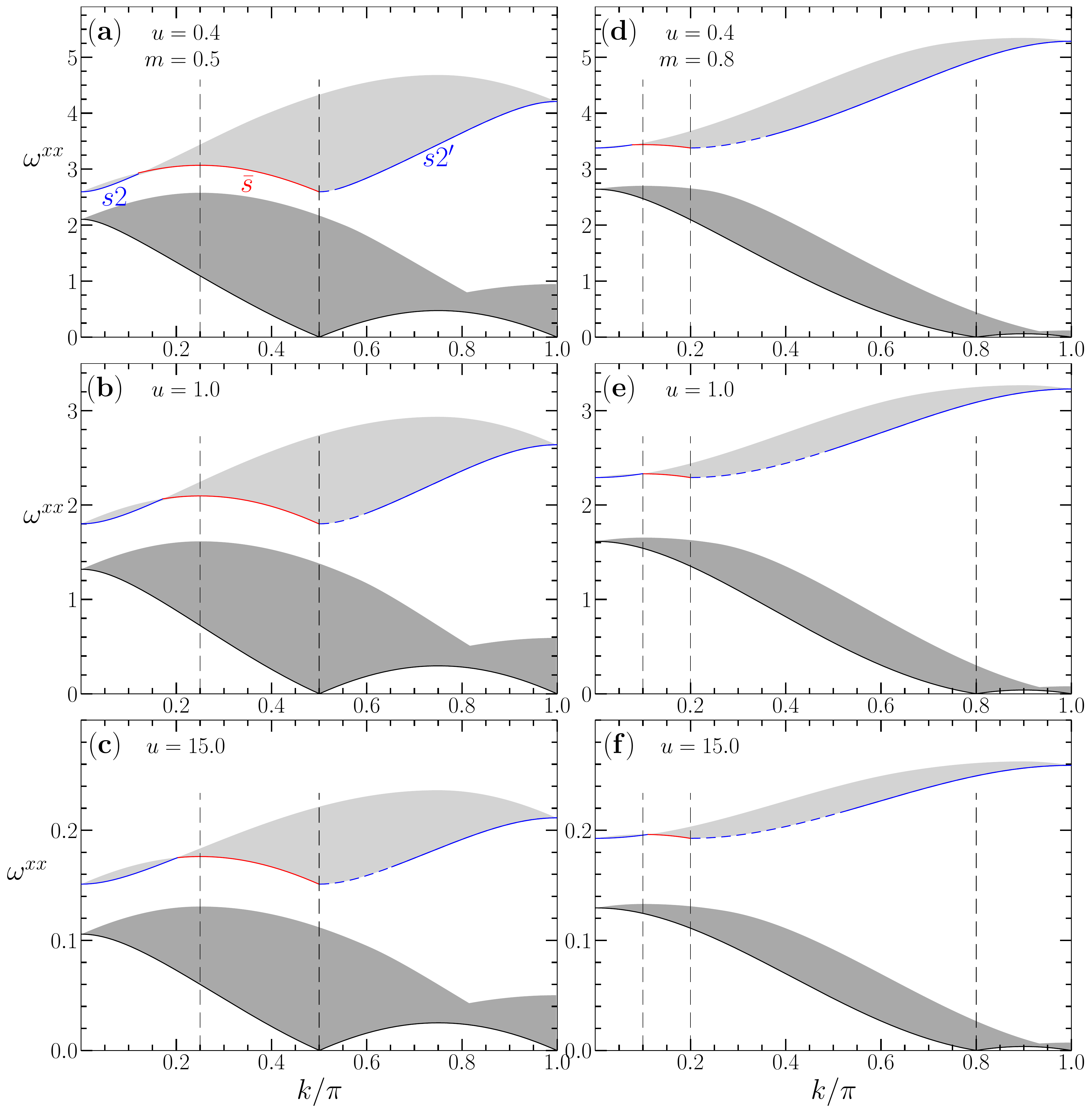}}
\caption{The same continuum spectra as in Fig. \ref{figure3} for spin densities (a-c) $m=0.5$ and (d-f) $m =0.8$ 
and $u=0.4,1.0,15.0$. For such spin densities, there is no overlap between the upper spin $n$-string continuum and
the lower continuum. [The three reference vertical lines mark the momenta (a-c) $k=k_{F\downarrow}=\pi/4$ and
$k=k_{F\uparrow} - k_{F\downarrow}=2k_{F\downarrow}=\pi/2$ and (d-f) $k=k_{F\downarrow}=\pi/10$, 
$k=2k_{F\downarrow}=\pi/5$, and $k=k_{F\uparrow} - k_{F\downarrow}=4\pi/5$,
where $k_{F\downarrow} = {\pi\over 2}(1-m)$ and $k_{F\uparrow} = {\pi\over 2}(1+m)$,
Eq. (\ref{kkk}).] }
\label{figure4}
\end{center}
\end{figure}

\begin{figure}
\begin{center}
\centerline{\includegraphics[width=9cm]{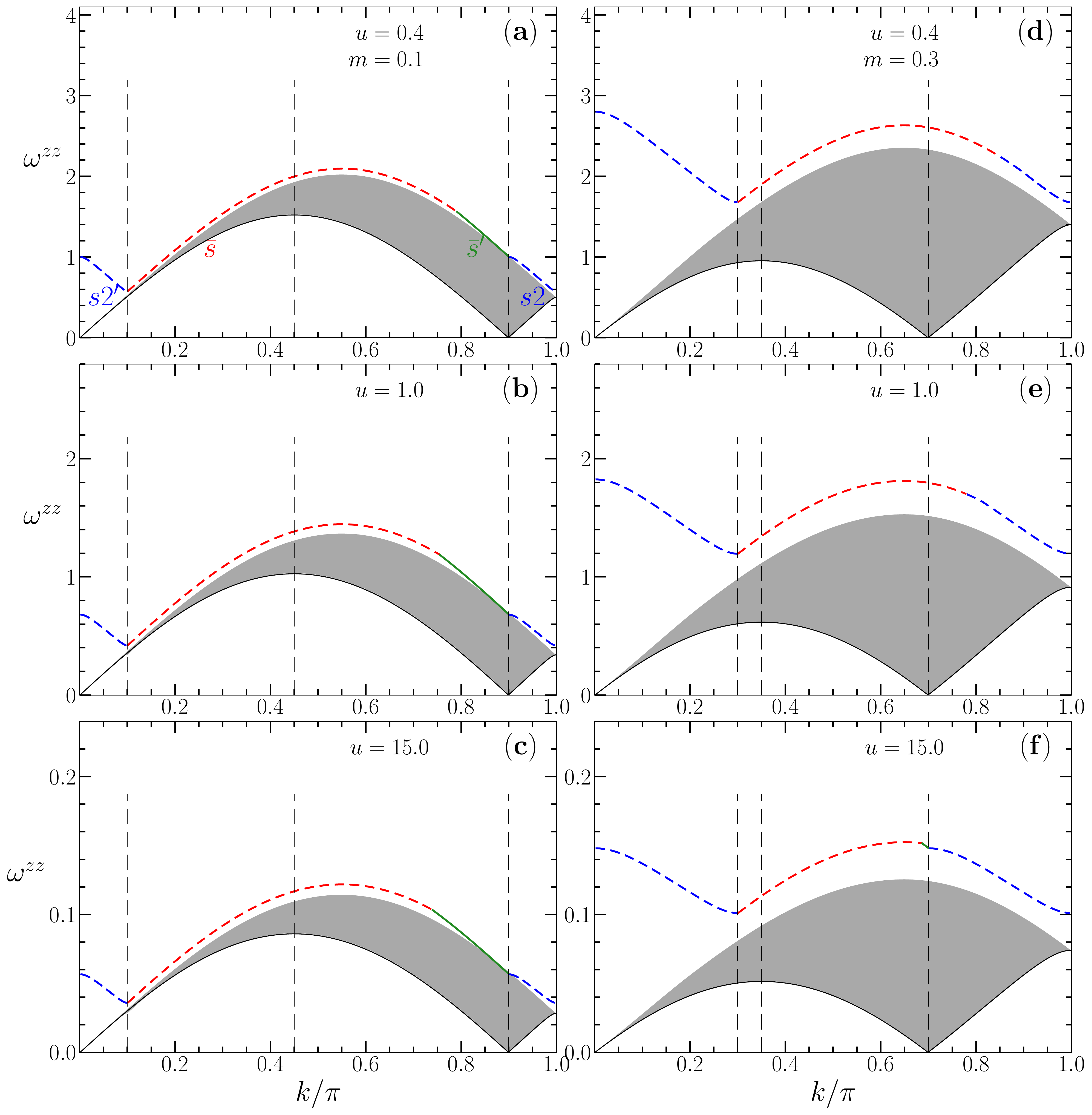}}
\caption{The $(k,\omega)$-plane continuum region
where for spin densities (a-c) $m=0.1$ and (d-f) $m =0.3$ 
and $u=0.4,1.0,15.0$ there is in the thermodynamic limit more spectral weight in $S^{zz} (k,\omega)$.
[The three reference vertical lines mark the momenta (a-c) $k=k_{F\uparrow} - k_{F\downarrow}=\pi/10$,
$k=k_{F\downarrow}=9\pi/20$, and $k=2k_{F\downarrow}=9\pi/10$
and (d-f) $k=k_{F\uparrow} - k_{F\downarrow}=3\pi/10$,
$k=k_{F\downarrow}=7\pi/20$, and $k=2k_{F\downarrow}=7\pi/10$
where $k_{F\downarrow} = {\pi\over 2}(1-m)$ and $k_{F\uparrow} = {\pi\over 2}(1+m)$,
Eq. (\ref{kkk}).] 
Contributions from excited states containing spin $n$-strings are much smaller
than for $S^{+-} (k,\omega)$ and $S^{xx} (k,\omega)$ and do not lead to an
upper continuum. The gapped lower threshold of such states is though
displayed. Only when for spin densities $0<m<\tilde{m}$ where $\tilde{m}=0$ for $u\rightarrow 0$ 
and $\tilde{m}\approx 0.317$ for $u\gg 1$ that threshold coincides with the $\bar{s}'$ branch
line, singularities occur near and just above it. That line is represented as
a solid (green) line. In the remaining parts of the gapped lower threshold, which
for spin densities $\tilde{m}<m<1$ means all of it, the momentum dependent
exponents are positive and there are no singularities. This
reveals there is a negligible amount of spectral weight near such lines.}
\label{figure5}
\end{center}
\end{figure}

\begin{figure}
\begin{center}
\centerline{\includegraphics[width=9cm]{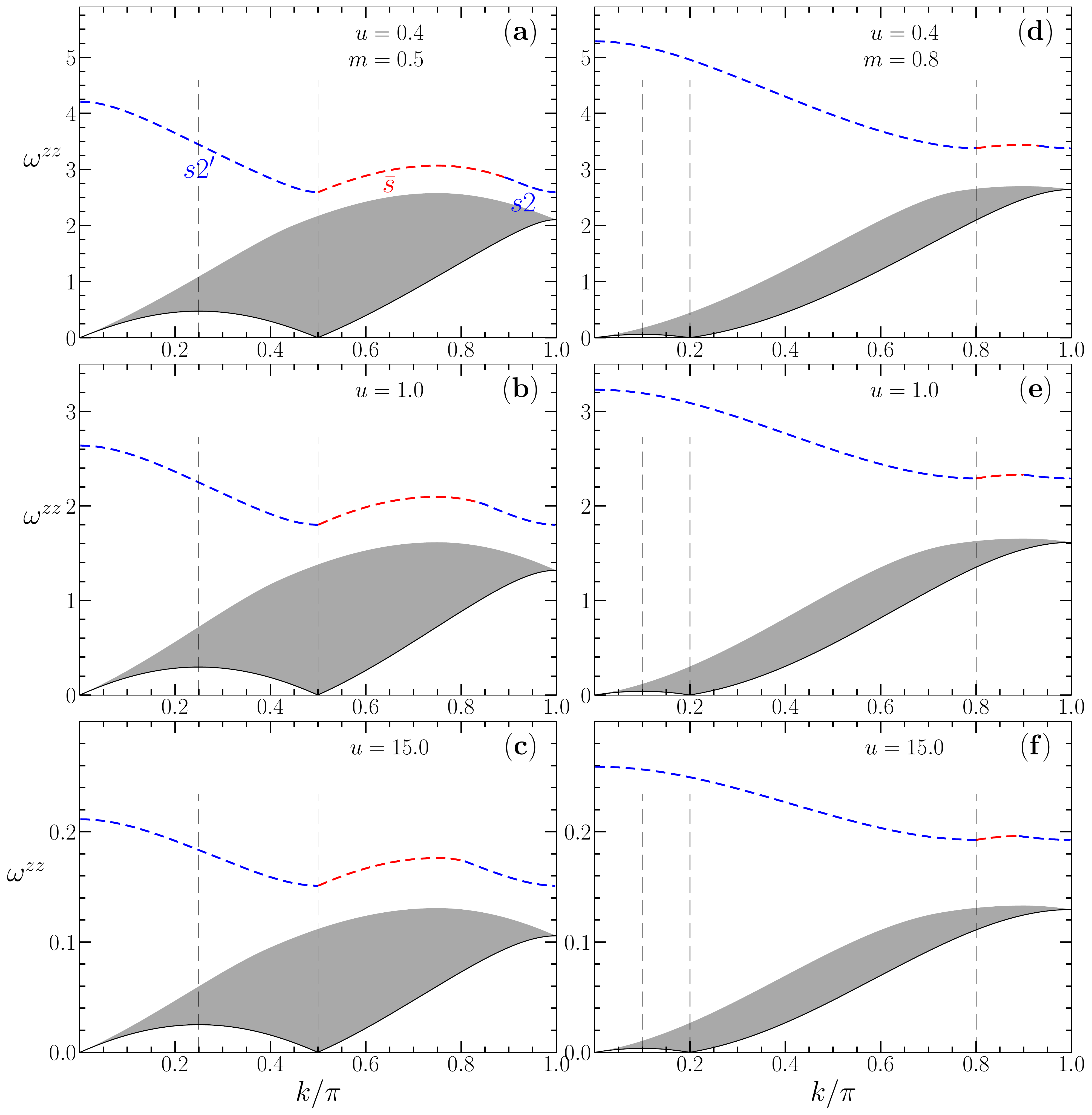}}
\caption{The same continuum spectra as in Fig. \ref{figure5} for spin densities (a-c) $m=0.5$ and (d-f) $m =0.8$ 
and $u=0.4,1.0,15.0$. For these spin densities, there are no singularities near the gapped lower threshold of the 
spin $n$-string excited states. For these spin densities the contribution of such states to $S^{zz} (k,\omega)$ are
actually negligible over the whole $(k,\omega)$ plane. [The three reference vertical lines mark the momenta 
(a-c) $k=k_{F\downarrow}=\pi/4$ and $k=k_{F\uparrow} - k_{F\downarrow}=2k_{F\downarrow}=\pi/2$ and 
(d-f) $k=k_{F\downarrow}=\pi/10$, $k=2k_{F\downarrow}=\pi/5$, and $k=k_{F\uparrow} - k_{F\downarrow}=4\pi/5$,
where $k_{F\downarrow} = {\pi\over 2}(1-m)$ and $k_{F\uparrow} = {\pi\over 2}(1+m)$,
Eq. (\ref{kkk}).]}
\label{figure6}
\end{center}
\end{figure}

The goal of this section is to introduce the spectra associated with $(k,\omega)$-plane regions
that contain most spectral weight of the spin dynamical structure factors. The
$(k,\omega)$-plane distribution of such spectra is represented for $S^{+-} (k,\omega)$, $S^{xx} (k,\omega)$,
and $S^{zz} (k,\omega)$ in Figs. \ref{figure1} and \ref{figure2}, \ref{figure3} and \ref{figure4},
and \ref{figure5} and \ref{figure6}, respectively. 
[In these figures, the spectra of the branch lines studied below are such that
the $s2$ and $s2'$ branch lines are represented by blue lines and the 
$\bar{s}$ and  $\bar{s}'$ branch lines by red and green lines, respectively;
The $U=0$ electronic Fermi points $k_{F\downarrow} = {\pi\over 2}(1-m)$ and $k_{F\uparrow} = {\pi\over 2}(1+m)$
define at $u>0$ the ground-state $s$ band Fermi points $\pm k_{F\downarrow}$ and
the $s$ band limiting momentum values $\pm k_{F\uparrow}$.]
The spectra displayed in Figs. \ref{figure1}, \ref{figure3}, and \ref{figure5}
refer to spin densities (a-c) $m=0.1$ and (d-f) $m =0.3$ and $u=0.4,1.0,15.0$. 
In Figs. \ref{figure2}, \ref{figure4}, and \ref{figure6} they correspond to spin densities (a-c) 
$m=0.5$ and (d-f) $m =0.8$ and the same set $u=0.4,1.0,15.0$ of $u$ values.

In the cases of $S^{+-} (k,\omega)$ and $S^{xx} (k,\omega)$, the figures show both a
lower continuum $(k,\omega)$-plane region whose spectral weight is associated with
excited states without spin $n$-strings and an upper continuum whose spectral
weight stems from excited states populated by spin $n$-strings. 
In the case of $S^{zz} (k,\omega)$, the contribution to the spectral weight from excited states 
containing spin $n$-strings is much weaker
than for $S^{+-} (k,\omega)$ and $S^{xx} (k,\omega)$ and does not lead to an
upper continuum. The gapped lower threshold of such states's spectrum is 
represented in Figs. \ref{figure5} and \ref{figure6} by a $(k,\omega)$-plane line.

Since at finite magnetic fields the contribution to the spectral weight from excited states 
containing spin $n$-strings is negligible in the case of $S^{-+} (k,\omega)$ and their lower continuum
spectrum was previously studied \cite{Carmelo_16}, its $(k,\omega)$-plane spectrum
distribution is not shown here. Note though that in Figs. \ref{figure3} and \ref{figure4} 
for $S^{xx} (k,\omega)$, the additional part of the lower 
continuum relative to that of $S^{+-} (k,\omega)$ represented in Figs. \ref{figure1} and \ref{figure2}
stems from contributions of $S^{-+} (k,\omega)$. As a result, 
for small spin densities and some $k$ intervals the upper spin $n$-string continuum of $S^{xx} (k,\omega)$ overlaps with
its lower continuum. 

In the case of both $S^{+-} (k,\omega)$ and $S^{zz} (k,\omega)$, 
there is in the present thermodynamic limit for spin densities $0<m<1$ and thus finite
magnetic fields $0<h<h_c$ very little spectral weight between the upper threshold 
of the lower continuum associated with spin $n$-string-less excited states and the gapped lower threshold 
of the spin $n$-string states's spectra in Figs. \ref{figure1}-\ref{figure2} and \ref{figure5} and \ref{figure6},
respectively. The same applies to $S^{xx} (k,\omega)$ in the $k$ intervals of Figs. \ref{figure3} and \ref{figure4}
for which there is a gap between the upper continuum associated with spin $n$-string states 
and the lower continuum. 

Indeed, in the thermodynamic limit nearly all the small amount of spectral weight associated with the 
spin $n$-string-less excited energy eigenstates named in the literature
four-spinon states, is contained inside the lower continuum in such figures. This also applies to large finite
systems. In the large $u$ limit, in which the spin degrees of freedom of the present model with one electron per site
are described by the isotropic spin-$1/2$ Heisenberg chain, this is so for the latter model
both at the isotropic point $\Delta =1$ (see Fig. 4 of Ref. \onlinecite{Caux_06}) and for
anisotropy $\Delta < 1$ (see Fig. 1 of Ref. \onlinecite{Caux_05}). 

Concerning this key issue for our study that the amount of spectral weight in the $(k,\omega)$-plane gap regions
shown in Figs. \ref{figure1}-\ref{figure6} is negligible, let us consider the more involved case of $S^{+-} (k,\omega)$. 
Similar conclusions apply to the simpler
problems of the other spin dynamical structure factors. The behavior of spin operators matrix elements between energy eigenstates 
in the selection rules valid for $u>0$ and magnetic fields $0<h<h_c$, Eq. (\ref{SRhfinite}), has important
physical consequences. It implies that the spectral weight stemming from 
excited energy eigenstates described by only real Bethe-ansatz rapidities
existing in finite systems in a $(k,\omega)$-plane region corresponding to the momentum interval
$k\in [2k_{F\downarrow},\pi]$ and excitation energy values $\omega$ above the 
upper threshold of the lower continuum in Figs. \ref{figure1} and \ref{figure2}, whose spectrum's expression
is given in Eq. (\ref{dkEdPxxPM}), becomes negligible in the present thermodynamic limit
for a macroscopic system. 

Our thermodynamic limit's study is complementary to and consistent with results obtained by completely different methods 
for finite-size systems and small yet finite $t^2/U$ \cite{Kohno_09,Muller}. 
The spectral weight located in that $(k,\omega)$-plane region
is found to decrease upon increasing the system size \cite{Kohno_09}. This is confirmed by comparing the spectra 
represented in the first row frames of Figs. 3 (a) and 3 (b) of Ref. \onlinecite{Kohno_09} for two finite-size systems 
with $N=320$ and $N=2240$ spins, respectively, in the case under consideration of the spin dynamical structure 
factor $S^{+-} (k,\omega)$.

More generally, the selection rules in Eqs. (\ref{SRh0}) and (\ref{SRhfinite}) 
valid for $u>0$ are behind in the thermodynamic limit nearly all spectral weight generated by transitions
to excited energy eigenstates described only by real Bethe-ansatz rapidities being
contained in the $(k,\omega)$-plane lower continuum shown in Figs. \ref{figure1} and \ref{figure2}, whose 
spectrum is given in Eq. (\ref{dkEdPxxPM}).

Let us consider the $(k,\omega)$-plane spectral weight distributions shown in Fig. 18 of Ref. 
\onlinecite{Muller} for $S^{+-} (k,\omega)$, which apply to the half-filled 1D Hubbard
model for small yet finite $t^2/U$. 
As reported in that reference, due to the interplay of the selections rules given 
in Eqs. (\ref{SRh0}) and (\ref{SRhfinite}) for $h=0$ and $0<h<h_c$, respectively,
the spectral weight existing between the continuous lower boundary $\epsilon_{4L}$ and the upper boundary 
$\epsilon_{4U}$ at $h=0$ becomes negligible for finite magnetic fields $0<h<h_c$. 
In addition, the spectral weight existing between the continuous lower boundary $\epsilon_{5L}$ and the upper boundary 
$\epsilon_{5U}$ for small finite-size systems, becomes negligible in
the thermodynamic limit for a macroscopic system. This is indeed due to the selection rules, 
Eq. (\ref{SRhfinite}), as discussed
in that reference, which for the 1D Hubbard model with one fermion per site
are valid for $u>0$. As also reported in Ref. \onlinecite{Muller}, only the spectral weight 
below the continuous lower boundary $\epsilon_{5L} (q)$, 
located in the $(k,\omega)$-plane between the lower boundary $\epsilon_{6L}$ and the upper boundary 
$\epsilon_{6U}$ has a significant amount of spectral weight. 

This refers to the $(k,\omega)$-plane region where, according to the analysis of Ref. \onlinecite{Muller}, for 
magnetic fields $0<h<h_c$ a macroscopic system has nearly the whole spectral weight stemming
from transitions to excited energy eigenstates described by only real Bethe-ansatz rapidities. Consistently
with the spectral weight in the present gap region being negligible, the $(k,\omega)$-plane
between the continuous lower boundary $\epsilon_{6L}$ and the upper boundary 
$\epsilon_{6U}$ in Fig. 18 of that reference corresponds precisely to the
lower continuum shown in Figs. \ref{figure1} and \ref{figure2}, whose spectrum is provided 
in Eq. (\ref{dkEdPxxPM}). 

Besides the $s$ and $s2$ particles, there is in the present spin subspace 
a $c$ particle branch of Bethe ansatz quantum
numbers associated with the charge degrees of freedom \cite{Carmelo_16,Carmelo_05}.
However, it refers to a corresponding full $c$ momentum band that does not contribute
to the spin dynamical properties. 
Its only contribution to the spin problem studied in this paper stems from microscopic momentum shifts 
$-{\pi\over L}$ or ${\pi\over L}$ of all the corresponding $c$ band $N$ discrete momentum values
$q_j = {2\pi\over L}\,I^{c}_j$. Here $I_j^{c} = 0,\pm 1,\pm 2,...$ for $N_s + N_{s2}$ even and
$I_j^{c} = \pm 1/2,\pm 3/2,\pm 5/2,...$ for $N_s + N_{s2}$ odd are the Bethe-ansatz $c$ band 
quantum numbers in Eq. (\ref{Ic-an}). Those lead to 
macroscopic momentum variations $-\pi$ or $\pi$, respectively, upon changes in the value of the numbers
of $s$ and $s2$ particles, according to the boundary conditions given in Eq. (\ref{Ic-an}).

The line shape near the gapped lower threshold of the $S^{+-} (k,\omega)$'s continuum spectrum 
represented in Figs. \ref{figure1} and \ref{figure2} is controlled by the above class
of excited states that are generated by the occupancy configurations of both $N_s = N_{\downarrow} - 2$ 
$s$ particles over $N_{\uparrow}$ discrete momentum values $q_j = {2\pi\over L}\,I_j^{s}$
and one $s2$ particle over $N_{\uparrow}-N_{\uparrow}+1$ discrete momentum values 
$q_j = {2\pi\over L}\,I_j^{s2}$. Here (i) $I_j^{s} = 0,\pm 1,\pm 2,...$ for $N_{\uparrow}$ odd and
$I_j^{s} =\pm 1/2,\pm 3/2,\pm 5/2,...$ for $N_{\uparrow}$ even and (ii)
$I_j^{s2} = 0,\pm 1,\pm 2,...$ for $N_{s2}=1$ are the Bethe-ansatz $s$ and $s2$ band 
quantum numbers, respectively, in Eq. (\ref{Ic-an}). 
However, the line shape in the vicinity of the lower threshold of the $S^{+-} (k,\omega)$'s lower continuum spectrum 
in the same figures is controlled by excited energy eigenstates described by real Bethe ansatz rapidities. 
Those are described by occupancy configurations of $N_s = N_{\downarrow}$ $s$ particles over 
$N_{\uparrow}$ discrete momentum values $q_j = {2\pi\over L}\,I_j^{s}$.

The Bethe-ansatz equations and quantum numbers whose occupancy configurations generate 
the energy eigenstates that span the spin subspaces 
used in our studies are given in Eqs. (\ref{Taps}) and (\ref{Taps2}) in
functional form, in terms of $s$ and $s2$ bands momentum distributions. Those
describe the momentum occupancy configurations that generate such states.

As further discussed in Appendix \ref{D}, ground states are for spin densities $0<m<1$
only populated by the $N_c=N$ nondynamical $c$ particles and $N_s = N_{\downarrow}$
$s$ particles that symmetrically or quasi-symmetrically occupy the $s$ band, which 
also contains $N_s^h = N_{\uparrow}-N_{\downarrow}$ holes.

The gapped upper spectrum in Figs. \ref{figure1} and \ref{figure2}
associated with the $(k,\omega)$-plane continuum of $S^{+-} (k,\omega)$ that stems
from transitions from the ground state to excited energy eigenstates populated by 
$N_s = N_{\downarrow} - 2$ $s$ particles and one $s2$ particle is given by,
\begin{eqnarray}
\omega^{+-}_{\Delta} (k) & = & - \varepsilon_{s} (q_1) + \varepsilon_{s2} (q_2) 
\nonumber \\
& & {\rm where}\hspace{0.20cm} k = \iota k_{F\downarrow} - q_1 + q_2 \hspace{0.20cm}
{\rm and}\hspace{0.20cm}\iota = \pm 1
\nonumber \\
& & {\rm for}\hspace{0.20cm} q_1 \in [-k_{F\downarrow},k_{F\downarrow}]\hspace{0.20cm}{\rm and}
\nonumber \\
& & q_2 \in [0,(k_{F\uparrow}-k_{F\downarrow})]\hspace{0.20cm}{\rm for}\hspace{0.20cm}\iota = 1 
\nonumber \\
& & q_2 \in [-(k_{F\uparrow}-k_{F\downarrow}),0]\hspace{0.20cm}{\rm for}\hspace{0.20cm}\iota = - 1 \, .
\label{dkEdPPM}
\end{eqnarray}
This spectrum has two branches corresponding to $\iota =\pm 1$ such that,
\begin{eqnarray}
k & = & k_{F\downarrow} - q_1 + q_2 \in [0,\pi]
\nonumber \\
k & = & - k_{F\downarrow} - q_1 + q_2 \in [-\pi,0] \, .
\label{dkEdPPM2}
\end{eqnarray}

In Eq. (\ref{dkEdPPM}) and other expressions of spin dynamical structure factors's
spectra given below and in Appendices \ref{B} and \ref{C}, $\varepsilon_{s} (q)$ and $\varepsilon_{s2} (q)$ 
are the $s$ and $s2$ band energy dispersions, respectively, defined by Eqs. (\ref{equA4}), (\ref{vares2}), and
(\ref{equA5})-(\ref{qtwoprimelimits}).
Limiting behaviors of such dispersions and corresponding $s$ and $s2$ group velocities that provide useful information 
on the corresponding spin dynamical structure factors's spectra momentum, spin density, and interaction dependences 
are provided in Eqs. (\ref{varesm0})-(\ref{vvm1}).
 
We denote by $\Delta^{ab} (k)$ where $ab=+-,xx,zz$ the spectra of the spin $n$-string excited states's gapped lower 
thresholds of $S^{ab} (k,\omega)$. They play an important role in our study, since for some $k$ intervals there are 
singularities at and just above them. 

For $S^{+-} (k,\omega)$, $S^{xx} (k,\omega)$, and $S^{zz} (k,\omega)$ such gapped thresholds have 
a different form for two spin density intervals $m\in ]0,\tilde{m}]$ and $m\in [\tilde{m},1[$, 
respectively. Here $\tilde{m}$ is a $u$ dependent spin density at which the following equality holds,
\begin{equation}
W_{s2}\vert_{m=\tilde{m}} = -  \varepsilon_{s} (2k_{F\downarrow}-k_{F\uparrow})\vert_{m=\tilde{m}} \, .
\label{mtilde}
\end{equation}
From the use of the $\varepsilon_{s2} (0)$'s expression given in Eq. (\ref{vares2limits}), 
the $s2$ energy bandwidth $W_{s2}$ appearing here can be expressed as 
$W_{s2} = 4\mu_B h - \varepsilon_{s2} (0)$. The spin density $\tilde{m}$ is
a continuous increasing function of $u$ that in the $u\rightarrow 0$ and $u \gg 1$
limits reads,
\begin{equation}
\lim_{u\rightarrow 0}\tilde{m} = 0
\hspace{0.50cm}{\rm and}\hspace{0.50cm}
 \lim_{u\gg 1}\tilde{m} \approx 0.317 \, .
\label{barmu0}
\end{equation}

Momenta involving a related momentum $\tilde{k}$ separate parts of the gapped lower threshold
spectra of $S^{+-} (k,\omega)$, $S^{xx} (k,\omega)$, and $S^{zz} (k,\omega)$ 
that refer to different types of $k$ dependences. At $k=\tilde{k}$
the following relations that define it hold,
\begin{eqnarray}
W_{s2} & = & \varepsilon_{s} (k_{F\uparrow}-\tilde{k}) -  \varepsilon_{s} (k_{F\downarrow}-\tilde{k})
\nonumber \\
& & {\rm for} \hspace{0.20cm}\tilde{k}\geq (k_{F\uparrow} - k_{F\downarrow})\hspace{0.20cm}
{\rm and}\hspace{0.20cm}m\in [0,\tilde{m}]
\nonumber \\
W_{s2} & = & 4\mu_B\,h -\varepsilon_{s2} (\tilde{k}) -  \varepsilon_{s} (k_{F\downarrow}-\tilde{k})
\nonumber \\
& & {\rm for} \hspace{0.20cm}\tilde{k}\leq (k_{F\uparrow} - k_{F\downarrow})\hspace{0.20cm}
{\rm and}\hspace{0.20cm}m\in [\tilde{m},1[ \, .
\label{ktilde}
\end{eqnarray}
The momentum $\tilde{k}$ is given by $\tilde{k}=(k_{F\uparrow} - k_{F\downarrow})$ at $m = \tilde{m}$.

The spectra of the transverse gapped lower thresholds are such that,
\begin{equation}
\Delta^{xx} (k) = \Delta^{+-} (k)\hspace{0.20cm} {\rm for} \hspace{0.20cm}k\in [0,\pi] \, .
\label{GappLT}
\end{equation}
(The equality $\Delta^{-+} (k) = \Delta^{+-} (k)$ also holds, yet as reported above the amount of $S^{-+} (k,\omega)$'s 
spectral weight produced by excited $n$-string states is negligible in the thermodynamic
limit and finite magnetic fields.) The spectrum of the longitudinal gapped lower threshold is also related to $\Delta^{+-} (k)$ as follows,
\begin{equation}
\Delta^{zz} (k) = \Delta^{+-} (\pi - k)\hspace{0.20cm} {\rm for} \hspace{0.20cm}k\in [0,\pi] \, .
\label{GappLTlong}
\end{equation}

For smaller spin densities $m\in [0,\tilde{m}]$, the spectrum $\Delta^{+-} (k)$
is given by,
\begin{eqnarray}
\Delta^{+-} (k) & = & \varepsilon_{s2} (k) \hspace{0.20cm}{\rm for}\hspace{0.20cm}k\in [0,(k_{F\uparrow} - k_{F\downarrow})]
\nonumber \\
& = & 4\mu_B\,h - \varepsilon_{s} (k_{F\uparrow}-k) \hspace{0.20cm}{\rm for}\hspace{0.20cm}
k\in [(k_{F\uparrow} - k_{F\downarrow}),{\tilde{k}}]
\nonumber \\
& = & 4\mu_B\,h - W_{s2} - \varepsilon_{s} (k_{F\downarrow}-k) 
 \hspace{0.20cm}{\rm for}\hspace{0.20cm}k \in [{\tilde{k}},2k_{F\downarrow}]
\nonumber \\
& = & \varepsilon_{s2} (k - 2k_{F\downarrow}) \hspace{0.20cm}{\rm for}\hspace{0.20cm}k\in [2k_{F\downarrow},\pi]  \, .
\label{Dxx03}
\end{eqnarray}

For larger spin densities $m\in [\tilde{m},1[$, that spectrum is slightly different and reads,
\begin{eqnarray}
\Delta^{+-} (k) & = & \varepsilon_{s2} (k) \hspace{0.20cm}{\rm for}\hspace{0.20cm}k\in [0,{\tilde{k}}[
\nonumber \\
& = & 4\mu_B\,h - W_{s2} - \varepsilon_{s} (k_{F\downarrow}-k) 
\nonumber \\
& & {\rm for}\hspace{0.20cm}k \in ]{\tilde{k}},2k_{F\downarrow}]
\nonumber \\
& = & \varepsilon_{s2} (k - 2k_{F\downarrow}) \hspace{0.20cm}{\rm for}\hspace{0.20cm}k\in [2k_{F\downarrow},\pi]  \, .
\label{Dxx31}
\end{eqnarray}

The expressions of the previously studied two-parametric transverse gapless spectra 
\cite{Carmelo_16} $\omega^{-+} (k)$ and $\omega^{+-} (k)$, whose superposition 
gives $\omega^{xx} (k)$, and that of the longitudinal gapless spectrum $\omega^{zz} (k)$ 
that [except for $\omega^{-+} (k)$] refer to the lower continua in Figs. \ref{figure1}-\ref{figure6},
are given in Eqs. (\ref{dkEdPxxMP})-(\ref{dkEdPl}). 
The corresponding excited energy eigenstates are described by real Bethe-ansatz rapidities. 
The expressions of the one-parametric spectra of their upper thresholds 
$\omega^{-+}_{ut} (k)$, $\omega^{+-}_{ut} (k)$, $\omega^{xx}_{ut} (k)$, and $\omega^{zz}_{ut} (k)$
and lower thresholds $\omega^{-+}_{lt} (k)$, $\omega^{+-}_{lt} (k)$, $\omega^{xx}_{lt} (k)$,
and $\omega^{zz}_{lt} (k)$ are also provided in Appendix \ref{B}.

We consider the following energy gaps,
\begin{eqnarray}
\Delta_{\rm gap}^{+-} (k) & = & \Delta^{+-} (k) - \omega^{+-}_{ut} (k) \geq 0
\nonumber \\
\Delta_{\rm gap}^{xx} (k) & = & \Delta^{xx} (k) - \omega^{xx}_{ut} (k) 
\nonumber \\
\Delta_{\rm gap}^{zz} (k) & = & \Delta^{zz} (k) - \omega^{zz}_{ut} (k) \geq 0 \, ,
\label{gapPMMP}
\end{eqnarray}
where,
\begin{eqnarray}
\Delta_{\rm gap}^{xx} (k) & = & \Delta^{+-} (k) - \omega^{+-}_{ut} (k)
\hspace{0.20cm}{\rm for}\hspace{0.20cm}k\in [0,k^{xx}_{ut}]
\nonumber \\
\Delta_{\rm gap}^{xx} (k) & = & \Delta^{+-} (k) - \omega^{-+}_{ut} (k)
\hspace{0.20cm}{\rm for}\hspace{0.20cm}k\in [k^{xx}_{ut},\pi] \, ,
\label{gap}
\end{eqnarray}
and
\begin{equation}
\Delta_{\rm gap}^{zz} (k) = \Delta_{\rm gap}^{+-} (\pi - k)  
\hspace{0.20cm}{\rm for}\hspace{0.20cm}k \in [0,\pi] \, .
\label{gapL}
\end{equation}
The momentum $k^{xx}_{ut}>k_{F\uparrow}-k_{F\downarrow}$ in Eq. (\ref{gap}) is that at which the equality 
$\omega^{-+}_{ut} (k^{xx}_{ut}) = \omega^{+-}_{ut} (k^{xx}_{ut})$ holds.
In the thermodynamic limit and for the $k$ intervals for which such energy gaps are positive,
there is a negligible amount of spectral weight in their corresponding $(k,\omega)$-plane 
regions. This justifies why here we named them gaps.

The upper threshold spectra $\omega^{-+}_{ut} (k)$, $\omega^{+-}_{ut} (k)$,
$\omega^{xx}_{ut} (k)$, $\omega^{zz}_{ut} (k)$ in Eqs. (\ref{gapPMMP})-(\ref{gapL})
are given in Eqs. (\ref{Omxxut1})-(\ref{Omlut}).
The spectra $\omega^{+-}_{ut} (k)$, $\omega^{xx}_{ut} (k)$, and $\omega^{zz}_{ut} (k)$ refer to the upper thresholds of
the lower continua in Figs. \ref{figure1} and \ref{figure2}, \ref{figure3} and \ref{figure4}, and \ref{figure5} and \ref{figure6}, respectively.

As confirmed from analysis of Figs. \ref{figure1}-\ref{figure6},
one has that $\Delta_{\rm gap}^{+-} (k)\geq 0$ and $\Delta_{\rm gap}^{zz} (k)\geq 0$, whereas
$\Delta_{\rm gap}^{xx} (k)$ is negative for some $k$ intervals.
Specifically,
\begin{eqnarray}
\Delta_{\rm gap}^{xx} (k) & \leq & 0 \hspace{0.20cm}{\rm for}
\nonumber \\
k & \in & [\bar{k}_0,\pi]\hspace{0.20cm}{\rm for}\hspace{0.20cm}m\in ]0,\bar{m}_0] 
\nonumber \\
k & \in & [\bar{k}_0,\bar{k}_1]\hspace{0.20cm}{\rm for}\hspace{0.20cm}m\in ]\bar{m}_0,\bar{m}] \, .
\label{gapineq}
\end{eqnarray}
The values of the spin densities $\bar{m}_0$ and $\bar{m}> \bar{m}_0$ increase
and decrease upon increasing $u$, their limiting values being,
\begin{eqnarray}
\lim_{u\rightarrow 0}\bar{m}_0 & = & {2\over\pi}\arcsin\left({1\over 3}\right) \approx 0.216
\nonumber \\
\lim_{u\rightarrow 0}\bar{m} & = & {2\over\pi}\arctan\left({1\over 2}\right) \approx 0.295
\nonumber \\
 \lim_{u\gg 1}\bar{m}_0 & \approx & 0.239
\hspace{0.25cm}{\rm and}\hspace{0.25cm}
\lim_{u\gg 1}\bar{m} \approx 0.276 \, .
\label{barmu0}
\end{eqnarray}

The momenta $\bar{k}_0$ and $\bar{k}_1$ also appearing in Eq. (\ref{gapineq}) are such  
$k^{xx}_{ut}\leq \bar{k}_0\leq\bar{k}_1$, and $\bar{k}_0\leq \bar{k}_1\leq\pi$. 
The equality, $\bar{k}_0=\bar{k}_1$, holds at $m=\bar{m}$. At that spin density
the momentum $\bar{k}_0=\bar{k}_1$ is very little $u$ dependent. It is
given by $\bar{k}_0=\bar{k}_1= 2k_{F\downarrow}$ in the $\lim_{u\rightarrow 0}$ limit and for
$u\gg 1$ it reaches a value very near and just above $2k_{F\downarrow}$.

For $m\in ]0,\bar{m}]$ and the $k$ intervals in Eq. (\ref{gapineq}), the $S^{xx} (k,\omega)$'s expressions in 
the vicinity of that factor gapped lower threshold obtained in this paper are not valid because $\Delta_{\rm gap}^{xx} (k) < 0$. 
However, the $S^{+-} (k,\omega)$ and $S^{zz} (k,\omega)$'s expressions in the vicinity of their gapped lower thresholds 
considered in the following are valid for all $k$ intervals, since the energy gaps $\Delta_{\rm gap}^{+-} (k)$ 
and $\Delta_{\rm gap}^{zz} (k)$ are finite and positive for $0<m<1$ and $u>0$. 

In Appendix \ref{C}, limiting values of the energy gaps considered here and their values
at some specific momenta are provided.

\section{The line shape at and near the spin dynamical structure factors's singularities}
\label{SECIV}

The spin dynamical structure factors's singularities studied in this paper
occur at and just above spectral lines that within the dynamical theory of Refs. \onlinecite{Carmelo_16,Carmelo_05} 
are called branch lines. Such lines coincide with well defined $k$ intervals of the
$(k,\omega)$-plane lower thresholds of both the spectra of excited states populated and
not populated by spin $n$-strings, respectively, plotted in Figs. \ref{figure1}-\ref{figure6}. 

In the case of the contribution from spin $n$-string states, the dynamical theory line shape expressions 
are valid provided there is no or nearly no spectral weight just below the corresponding gapped lower 
thresholds. In the present thermodynamic limit, the amount of spectral weight just below such thresholds 
either vanishes or is extremely small. In the latter case, the very weak coupling to it leads to a higher order contribution to the line shape 
expressions given in the following that can be neglected in that limit. 

In the case of the lower $(k,\omega)$-plane spectrum continua in Figs. \ref{figure1}-\ref{figure6}
of excited states not populated by spin $n$-strings
and thus described by real Bethe-ansatz rapidities, there is
no spectral weight below the corresponding lower thresholds. This ensures that the
expressions of the spin dynamical structure factors at and just
above such thresholds are exact.

The momentum interval $k\in [0,\pi]$ of the gapped lower thresholds of spectra of spin $n$-string states 
is divided in several subintervals that refer to a set of branch lines called
$s2$, $\bar{s}$, $\bar{s}'$, and $s2'$ branch line. The corresponding excited states 
are populated by $N_{\downarrow}-2$ $s$ particles and one $s2$ particle. 
The lower thresholds of the spectra associated with excited states 
populated by $N_{\downarrow}$ $s$ particles, either correspond to a single
$s$ branch line or to two sections of such a line.

The $\bar{s}$, $\bar{s}'$, and $s$ branch lines
refer to $k$ ranges corresponding to a maximum $s$ band $q$
interval $q\in[-(k_{F\downarrow}-\delta q_{s}),(k_{F\downarrow}-\delta q_{s})]$ in the
case of $s$ hole creation and to a maximum $s$ band $q$
interval such that $\vert q\vert\in[(k_{F\downarrow}+\delta q_s),k_{F\uparrow}]$
in case of $s$ particle creation. Here $\delta q_{s}$ such that $\delta q_{s}/k_{F\uparrow} \ll 1$ for $0<m<1$
is for the different branch lines either very small or vanishes in the thermodynamic limit. 

In the very small $k$ intervals corresponding to the $s$ band intervals
$q\in[-k_{F\downarrow},-(k_{F\downarrow}-\delta q_{s})]$
and $q\in [(k_{F\downarrow}-\delta q_{s}),k_{F\downarrow}]$ the line shape of
the spin dynamical structure factors is different, as given in Ref. \onlinecite{Carmelo_16}.
(See Eqs. (128)-(133) of Ref. \onlinecite{Carmelo_16}.)

Similarly, in the case of the $(k,\omega)$-plane vicinity of the $s2$ and $s2'$ branch lines,
which are part of the gapped lower thresholds, the line shape expressions obtained in this paper are 
valid in $k$ ranges corresponding to $s2$ band maximum intervals 
$q \in [-(k_{F\uparrow}-k_{F\downarrow}-\delta q_{s2}),0]$ or $q \in [0,(k_{F\uparrow}-k_{F\downarrow}-\delta q_{s2})]$.
Here $\delta q_{s2}$ such that $\delta q_{s2}/(k_{F\uparrow}-k_{F\downarrow}) \ll 1$ is for $0<m<1$
very small and may vanish in the thermodynamic limit. (And again, the spin dynamical structure factors
expressions are different and known for 
$q \in [-(k_{F\uparrow}-k_{F\downarrow}),-(k_{F\uparrow}-k_{F\downarrow}-\delta q_{s2})]$ 
and $q \in [(k_{F\uparrow}-k_{F\downarrow}-\delta q_{s2}),(k_{F\uparrow}-k_{F\downarrow})]$
yet are not of interest for this study.)

In the present thermodynamic limit, the above $s$ band momentum intervals are thus
represented in the following as $q\in ]-k_{F\downarrow},k_{F\downarrow}[$ and
$\vert q\vert\in ]k_{F\downarrow},k_{F\uparrow}]$ and the $s2$ band momentum intervals
by $q \in ]-(k_{F\uparrow}-k_{F\downarrow}),0]$ or
$q \in [0,(k_{F\uparrow}-k_{F\downarrow})[$. 

Around the specific momentum values where along a gapped lower threshold
or a lower threshold two neighboring branch lines or branch line sections
cross, there are small momentum widths where the corresponding lower threshold refers to 
a boundary line that connects the two branch lines or branch line sections under consideration.

In the thermodynamic limit, such momentum intervals are in general
negligible and the corresponding small spectra deviations
are not visible in the spectra plotted in Figs. \ref{figure1}-\ref{figure6}.
In the cases they are small yet more extended, the two branch lines or branch line sections
run very near the lower threshold and there is very little spectral weight
between it and such lines. In this case, the singularities on the two branch lines or branch line sections
remain the dominant spectral feature. 

We again account for such negligible effects
by replacing $[$ and $]$ by $]$ and $[$, respectively, at the $k$ limiting values 
that separate lower thresholds's $k$ intervals associated with two neighboring 
branch lines or branch line sections.

\subsection{The line shape near the $s2$, $\bar{s}$, $\bar{s}'$, and $s2'$
branch lines (gapped lower thresholds)}
\label{SECIVA}

Here we study the line shape at and just above the gapped lower thresholds of the spectra plotted
in Figs. \ref{figure1}-\ref{figure6} of the transverse and longitudinal structure factors. 
In the case of $S^{xx} (k,\omega)$, this refers to $k$ intervals for which
$\Delta_{\rm gap}^{xx} (k)>0$ and thus different from those given in Eq. (\ref{gapineq}).
In Appendix \ref{D}, the number and current number deviations
as well as the spectral functionals that control the expressions of the spin dynamical structure factors 
given below are provided.
\begin{figure}
\begin{center}
\centerline{\includegraphics[width=9cm]{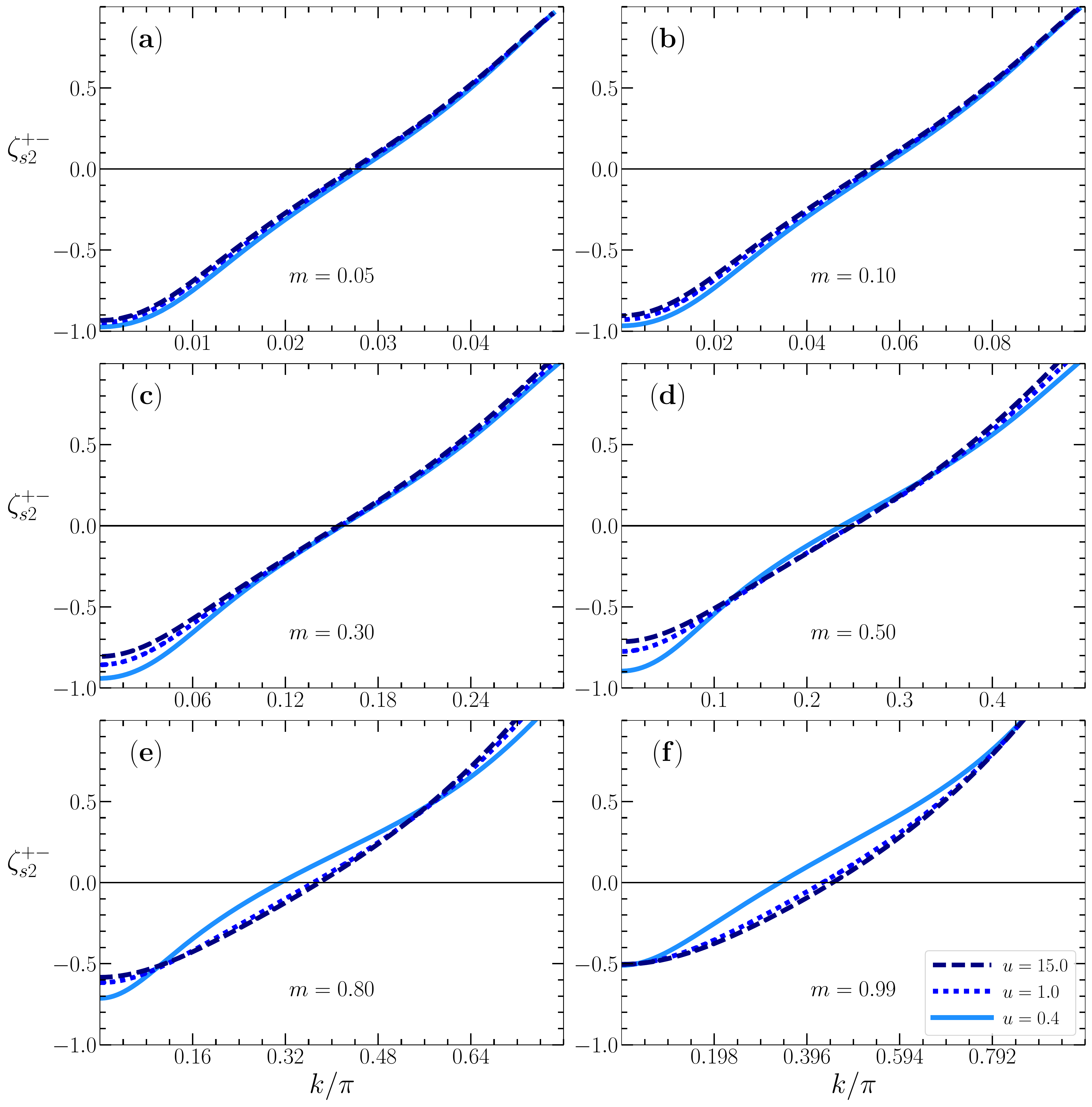}}
\caption{The momentum dependence of the exponent that 
in the $k$ intervals for which it is negative controls the
$S^{+-} (k,\omega)$ line shape near and just above the $s2$ branch
line for spin densities $m$ (a) $0.05$, (b) $0.1$, (c) $0.3$, (d) $0.5$, (e) $0.8$,
and (f) $0.99$ and $u=0.4,1.0,15.0$. The $s2$ branch line is part of the
gapped lower threshold of the spin $n$-strings continuum displayed in
Figs. \ref{figure1} and \ref{figure2}. The same exponent, in the $k$ intervals
for which it is negative, also controls the
$S^{xx} (k,\omega)$'s line shape near and just above the $s2$ branch line 
in the spin $n$-strings continuum displayed in
Figs. \ref{figure3} and \ref{figure4}.}
\label{figure7}
\end{center}
\end{figure}

The line shape near the gapped lower thresholds has the following general form,
\begin{eqnarray}
S^{ab} (k,\omega) & = & C_{ab}^{\Delta}
\Bigl(\omega - \Delta_{\beta}^{ab} (k))\Bigr)^{\zeta_{\beta}^{ab} (k)}  
\nonumber \\
& & {\rm for} \hspace{0.20cm}(\omega - \Delta_{\beta}^{ab} (k)) \geq 0
\hspace{0.20cm}{\rm where}
\nonumber \\
& & \beta = s2,\bar{s},\bar{s}',s2'
\hspace{0.20cm}{\rm and}\hspace{0.20cm}
ab = +-,xx,zz 
\nonumber \\
& & ({\rm valid}\hspace{0.20cm}{\rm when}\hspace{0.20cm}\Delta_{\rm gap}^{ab} > 0) \, .
\label{MPSsFMB}
\end{eqnarray}
Here $C_{ab}^{\Delta}$ is a constant that has a fixed value for the $k$ and $\omega$ ranges associated with 
small values of the energy deviation $(\omega - \Delta_{\beta}^{ab} (k))\geq 0$,
the gapped lower threshold spectra $\Delta_{\beta}^{ab} (k)$ are given in Eqs. (\ref{GappLT})-(\ref{Dxx31}) 
and the index $\beta=s2,\bar{s},\bar{s}',s2'$ labels branch lines or branch line sections 
that are part of the gapped lower thresholds in some specific $k$ intervals defined in the following.
\begin{figure}
\begin{center}
\centerline{\includegraphics[width=6cm]{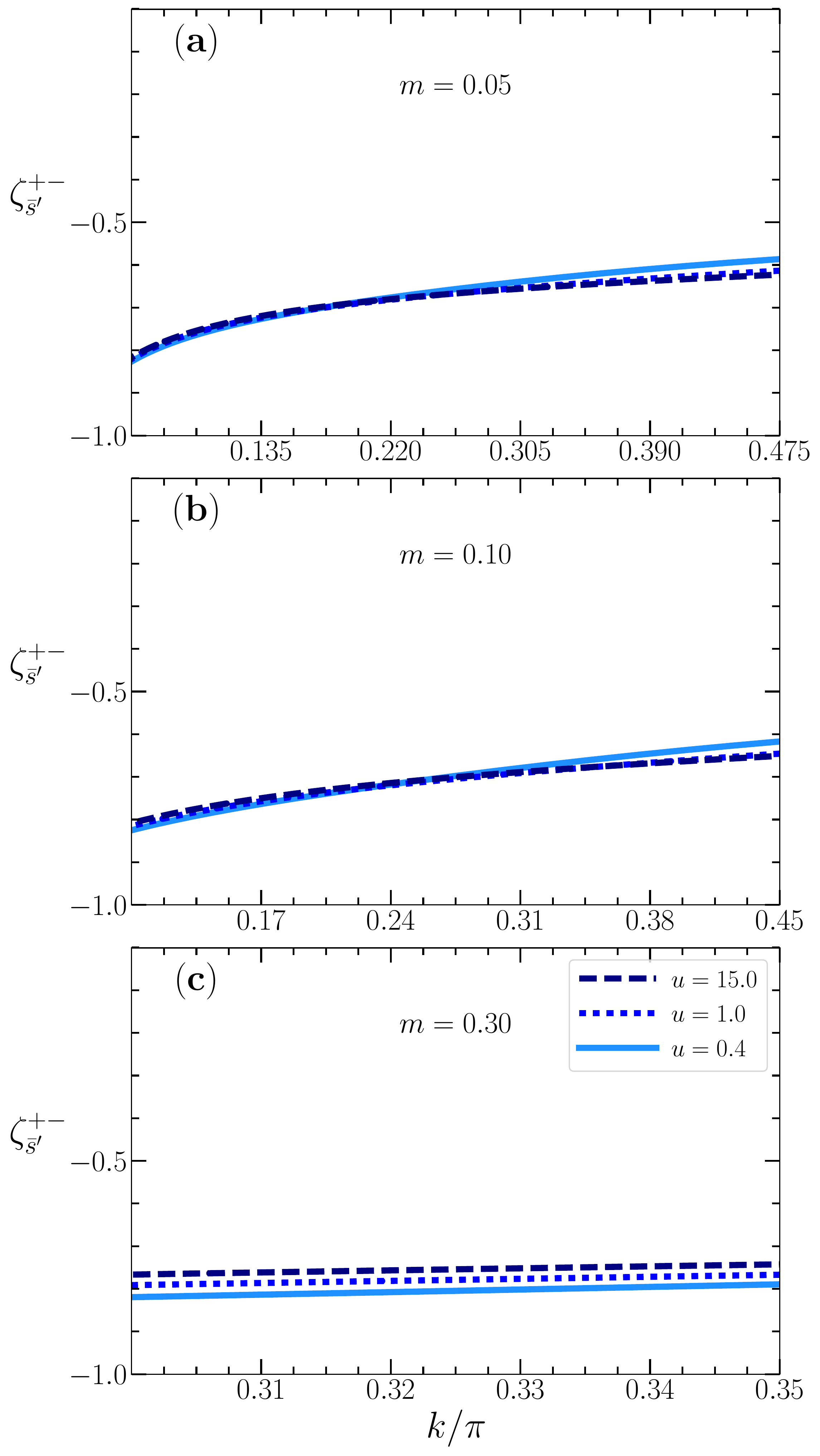}}
\caption{The same as in Fig. \ref{figure7} for the $\bar{s}'$ branch
line. That line coincides with the gapped lower threshold of the spin $n$-strings continuum 
for small $k$ intervals and only for spin densities $0<m<\tilde{m}$ where $\tilde{m}$ continuously increases from 
$\tilde{m}=0$ for $u\rightarrow 0$ to $\tilde{m}\approx 0.317$ for $u\gg 1$. The
corresponding exponent plotted here is negative for such $k$ intervals.}
\label{figure8}
\end{center}
\end{figure}

The branch-line exponents that appear in Eq. (\ref{MPSsFMB}) have the following general form,
\begin{equation}
\zeta^{aa}_{\beta} (k) = -1 + \sum_{\iota =\pm1}\Phi_{\iota}^2 (q) 
\hspace{0.20cm}{\rm for}\hspace{0.20cm} \beta=s2,\bar{s},\bar{s}',s2',s \, ,
\label{expTS}
\end{equation}
where the spectral functionals $\Phi_{\iota} (q)$ suitable to each type
of branch line are given in Eqs. (\ref{Fs2})-(\ref{Fs}). 
[This also includes the $s$ branch lines that define the lower thresholds 
of the lower continua in Figs. \ref{figure1}-\ref{figure6}. Their exponents
are also of form, Eq. (\ref{expTS}), and appear in the 
spin dynamical structure factors's general expression provided
below in Eq. (\ref{MPSsGen}).]
\begin{figure}
\begin{center}
\centerline{\includegraphics[width=9cm]{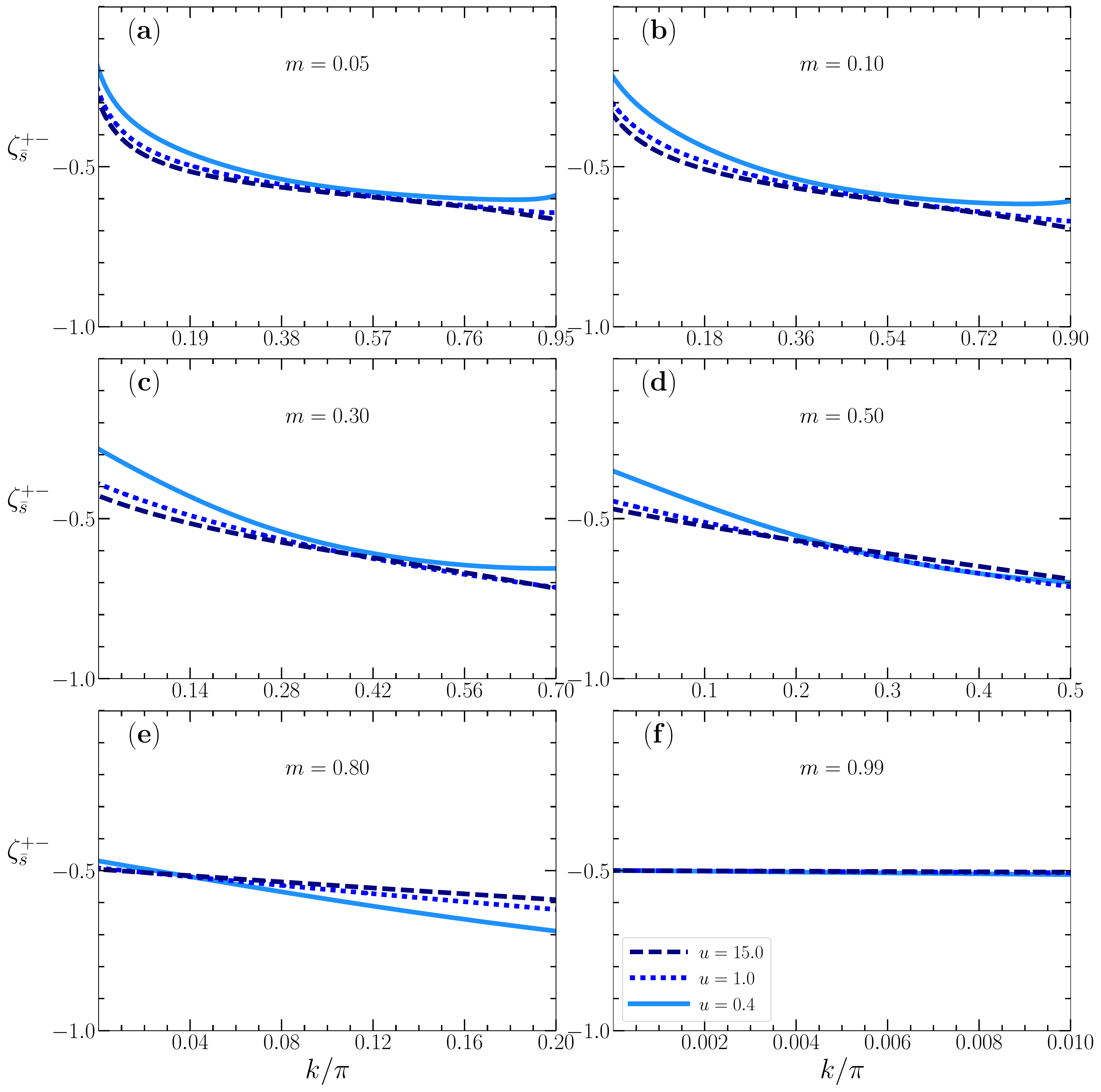}}
\caption{The same as in Fig. \ref{figure7} for the $\bar{s}$ branch
line, which refers to subintervals of the gapped lower threshold
of the spin $n$-string continuum of both $S^{+-} (k,\omega)$
and $S^{xx} (k,\omega)$. In the case of $S^{xx} (k,\omega)$, the momentum dependent
exponent plotted here is valid only for the $k$ intervals of the
$\bar{s}$ branch line in Figs. \ref{figure3} and \ref{figure4} for which 
there is a gap between it and the upper threshold of the lower
continuum.}
\label{figure9}
\end{center}
\end{figure}

As mentioned above, the amount of spectral weight below the gapped thresholds 
either vanishes or is very small. In the latter case, the very weak coupling to it leads to a higher order 
contribution to the line shapes given in Eqs. (\ref{MPSsFMB}) and (\ref{expTS})
that can be neglected in the present 
thermodynamic limit.

The relation of the excitation momentum $k$ to the $s$ band momentum $q$ or
$s2$ band momentum $q$ that appear in the $\Phi_{\iota}$'s argument 
in the general exponent expression,
Eq. (\ref{expTS}), is branch-line dependent. Hence it is useful to revisit the
expressions of the spectra of the gapped lower thresholds,
Eqs. (\ref{GappLT})-(\ref{Dxx31}) and (\ref{GappLTlong}), for each 
of their branch lines or branch line sections, including information on the
relation between the physical excitation momentum $k$ and the
$s$ or $s2$ bands momenta $q$. The corresponding expressions
are given for the $k$ intervals for which the dynamical structure
factor's expression is of the form, Eq. (\ref{MPSsFMB}), which implies
replacements of $[$ and $]$ by $]$ and $[$, respectively, in the limits
of such intervals.

In the case of $S^{+-} (k,\omega)$, the gapped lower threshold spectrum $\Delta^{+-} (k)$ is divided
in the following branch-line intervals,
\begin{eqnarray}
\Delta_{s2}^{+-} (k) & = & \varepsilon_{s2} (k) \hspace{0.20cm}{\rm and}\hspace{0.20cm}k = q
\nonumber \\
{\rm where} & & k\in ]0,(k_{F\uparrow}-k_{F\downarrow})[\hspace{0.20cm}{\rm and}
\nonumber \\
& & q\in ]0,(k_{F\uparrow}-k_{F\downarrow})[
\nonumber \\
& & {\rm for}\hspace{0.20cm}m\in ]0,\tilde{m}]
\nonumber \\
{\rm and} & & k\in ]0,{\tilde{k}}[\hspace{0.20cm}{\rm and}\hspace{0.20cm}
q\in [0,{\tilde{k}}[\hspace{0.20cm}
\nonumber \\
& &{\rm for}\hspace{0.20cm}m\in[\tilde{m},1[ \, ,
\label{Ds2}
\end{eqnarray}
\begin{eqnarray}
\Delta_{\bar{s}'}^{+-} (k) & = & 4\mu_B\,h - \varepsilon_{s} (k_{F\uparrow}-k)
\hspace{0.20cm}{\rm and}
\nonumber \\
&& k = k_{F\uparrow} - q
\nonumber \\
{\rm where} & & k\in ](k_{F\uparrow}-k_{F\downarrow}),\tilde{k}[\hspace{0.20cm}{\rm and}
\nonumber \\
& & q\in ](k_{F\uparrow}-\tilde{k}),k_{F\downarrow}[
\nonumber \\
& & {\rm for}\hspace{0.20cm}m\in ]0,\tilde{m}] \, ,
\label{Dsppp1}
\end{eqnarray}
\begin{eqnarray}
\Delta_{\bar{s}}^{+-} (k) & = & 4\mu_B\,h - W_{s2} - \varepsilon_{s} (k_{F\downarrow}-k) 
\nonumber \\
{\rm and} & & k = k_{F\downarrow}- q 
\nonumber \\
{\rm where} & & k \in [{\tilde{k}},2k_{F\downarrow}[\hspace{0.20cm}{\rm and}
\nonumber \\
& & q\in ]-k_{F\downarrow},(k_{F\downarrow} - {\tilde{k}})[
\nonumber \\
& & {\rm for}\hspace{0.20cm}m\in ]0,\tilde{m}]
\nonumber \\
{\rm and} & & k \in ]{\tilde{k}},2k_{F\downarrow}[\hspace{0.20cm}{\rm and}
\nonumber \\
& & q\in ]-k_{F\downarrow},(k_{F\downarrow} - {\tilde{k}})[
\nonumber \\
& &{\rm for}\hspace{0.20cm}m\in[\tilde{m},1[ \, ,
\label{Dsppp2}
\end{eqnarray}
and
\begin{eqnarray}
\Delta_{s2'}^{+-} (k) & = & \varepsilon_{s2} (k-2k_{F\downarrow}) \hspace{0.20cm}{\rm and}\hspace{0.20cm}
k = 2k_{F\downarrow} + q
\nonumber \\
{\rm where} && k\in ]2k_{F\downarrow},\pi[\hspace{0.20cm}{\rm and}
\nonumber \\
& & q\in ]0,(k_{F\uparrow}-k_{F\downarrow})[
\nonumber \\
&& {\rm for}\hspace{0.20cm}m\in ]0,1[ \, .
\label{Ds2p}
\end{eqnarray}
\begin{figure}
\begin{center}
\centerline{\includegraphics[width=9cm]{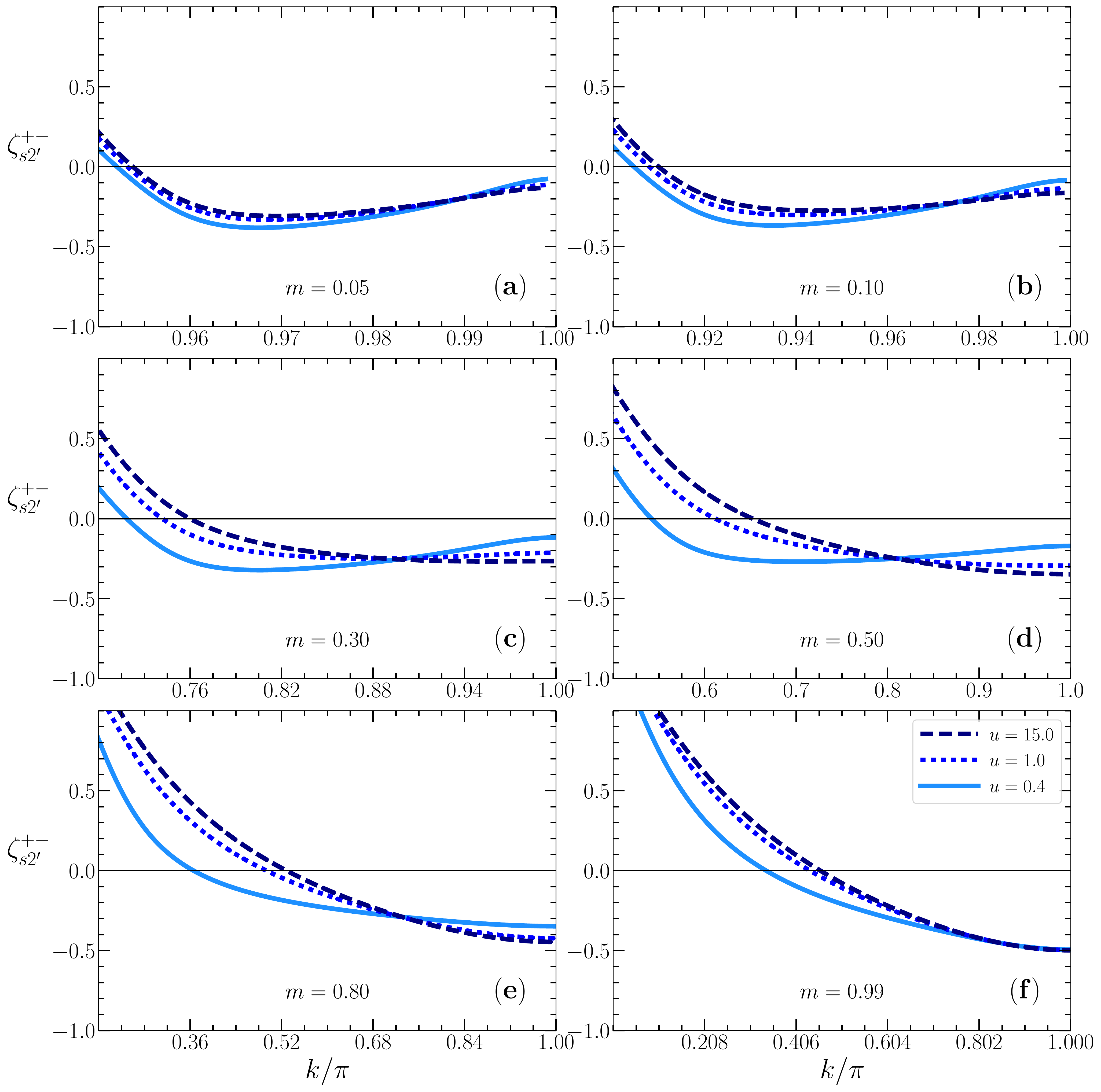}}
\caption{The same as in Fig. \ref{figure7} for the $s2'$ branch line.
As in in Fig. \ref{figure9}, in the case of $S^{xx} (k,\omega)$, the momentum dependent
exponent plotted here is valid only for the $k$ intervals of the
$s2'$ branch line in Figs. \ref{figure3} and \ref{figure4} for which 
there is a gap between it and the upper threshold of the lower
continuum.}
\label{figure10}
\end{center}
\end{figure}

The corresponding $k$ dependent exponents of general form,
Eq. (\ref{expTS}), that appear in the expression, $S^{+-} (k,\omega) = C_{+-}^{\Delta}
(\omega - \Delta_{\beta}^{+-} (k))^{\zeta_{\beta}^{+-} (k)}$,
Eq. (\ref{MPSsFMB}) for $ab=+-$ and $\beta=s2,\bar{s}',\bar{s},s2'$, are given by,
\begin{eqnarray}
\zeta_{s2}^{+-} (k) & = & -1 + \sum_{\iota=\pm 1}\left(- {\iota\over 2\xi_{s\,s}^1} 
+ \Phi_{s,s2}(\iota k_{F\downarrow},q)\right)^2  
\nonumber \\
&& {\rm for}\hspace{0.20cm}q=k
\hspace{0.20cm}{\rm and}
\nonumber \\
k & \in & ]0,(k_{F\uparrow}-k_{F\downarrow})[
\hspace{0.20cm}{\rm for}\hspace{0.20cm}m\in ]0,\tilde{m}]
\nonumber \\
k & \in & ]0,{\tilde{k}}[
\hspace{0.20cm}{\rm for}\hspace{0.20cm}m\in[\tilde{m},1[ 
\nonumber \\
\zeta_{\bar{s}'}^{+-} (k) & = & -1 +
\sum_{\iota=\pm 1}\left(- {\xi_{s\,s}^1\over 2} 
- \Phi_{s,s}(\iota k_{F\downarrow},q)\right)^2
\nonumber \\
&& {\rm for}\hspace{0.20cm}q=k_{F\uparrow}-k
\hspace{0.20cm}{\rm and}
\nonumber \\
k & \in & ](k_{F\uparrow}-k_{F\downarrow}),\tilde{k}[
\hspace{0.20cm}{\rm for}\hspace{0.20cm}m\in ]0,\tilde{m}] 
\nonumber \\
\zeta_{\bar{s}}^{+-} (k) & = & -1 + \sum_{\iota=\pm 1}
\left(\iota {\xi_{s\,s2}^0\over 2} + {\xi_{s\,s}^1\over 2} 
- \Phi_{s,s}(\iota k_{F\downarrow},q)\right)^2
\nonumber \\
&& {\rm for}\hspace{0.20cm}q=k_{F\downarrow}-k
\hspace{0.20cm}{\rm and}
\nonumber \\
k & \in & ]{\tilde{k}},2k_{F\downarrow}[
\hspace{0.20cm}{\rm for}\hspace{0.20cm}m\in ]0,\tilde{m}]
\nonumber \\
k & \in & ]{\tilde{k}},2k_{F\downarrow}[
\hspace{0.20cm}{\rm for}\hspace{0.20cm}m\in ]0,\tilde{m}] 
\nonumber \\
\zeta_{s2'}^{+-} (k) & = & -1 + \sum_{\iota=\pm 1}
\left(- {\iota\over 2\xi_{s\,s}^1} + \xi_{s\,s}^1
+ \Phi_{s,s2}(\iota k_{F\downarrow},q)\right)^2 
\nonumber \\
&& {\rm for}\hspace{0.20cm}q=k-2k_{F\downarrow}
\hspace{0.20cm}{\rm and}\hspace{0.20cm}k\in ]2k_{F\downarrow},\pi[ \, .
\label{expG+-}
\end{eqnarray}

The phase shifts in units of $2\pi$, $\Phi_{s,s}\left(\iota k_{F\downarrow},q\right)$ and 
$\Phi_{s,s2}\left(\iota k_{F\downarrow},q\right)$ where $\iota = \pm 1$, appearing in this equation
and in other exponents's expressions provided in the following are defined by Eqs. (\ref{Phi-barPhi})-(\ref{Phis-all-qq}).
Limiting behaviors of such phase shifts are provided in Eqs. (\ref{Phi-barPhim0})-(\ref{PhiUinfm1qF}).
The phase-shifts related parameters $\xi^{1}_{s\,s}=1/\xi^{0}_{s\,s}$ and
$\xi_{s\,s2}^{0}$ also appearing in the above exponents's expressions are defined 
by Eqs. (\ref{x-aa})-(\ref{xi1Phiss2}) and (\ref{xis20})-(\ref{Limxis20}), respectively.

Physically, $\pm 2\pi\Phi_{s,s}\left(\pm k_{F\downarrow},q\right)$ is the phase shift acquired by a $s$ particle
of momentum $\pm k_{F\downarrow}$ upon creation of one $s$ band hole 
($-2\pi\Phi_{s,s}$) and one $s$ particle ($+2\pi\Phi_{s,s}$) at a momentum $q$ in the $s$ band interval 
$q\in ]-k_{F\downarrow},k_{F\downarrow}[$ and such that $\vert q\vert\in ]k_{F\downarrow},k_{F\uparrow}]$,
respectively. However, $2\pi\Phi_{s,s2}\left(\pm k_{F\downarrow},q\right)$
is the phase shift acquired by a $s$ particle of momentum $\pm k_{F\downarrow}$ 
upon creation of one $s2$ particle at a momentum $q$ in the $s2$ band subinterval 
$q\in [0, (k_{F\uparrow}-k_{F\downarrow}[$ or $q\in ](k_{F\uparrow}-k_{F\downarrow},0]$. 

The three functionals $\Phi_{\iota} (q)$ in the general expression, Eq. (\ref{expTS}),
specific to the exponents given in Eq. (\ref{expG+-}) for the $S^{+-} (k,\omega)$'s 
$s2,s2'$ branch lines, $\bar{s}$ branch line,
and $\bar{s}'$ branch line are provided in Eqs. (\ref{Fs2}), (\ref{Fbars}), and
(\ref{Fbarsl}), respectively. The corresponding suitable specific values
of the number and current number deviations used in such functionals are
for the present branch lines given in Table \ref{table1}.
\begin{table}
\begin{center}
\begin{tabular}{|c|c|c|c|c|c|c|} 
\hline
b. line& $k$ in terms of $q$ & $\delta N_s^F$ & $\delta J_s^F$ & $\delta N_s^{NF}$ & $\delta J_{s2}$ & $\delta N_{s2}$ \\
\hline
$s2$ & $k=q$ & $-1$ & $0$ & $0$ & $0$ & $1$ \\
\hline
$\bar{s}'$ & $k=k_{F\uparrow} - q$ & $0$ & $1/2$ & $-1$ & $1/2$ & $1$ \\
\hline
$\bar{s}$ & $k=k_{F\downarrow} - q$ & $0$ & $1/2$ & $-1$ & $0$ & $1$ \\
\hline
$s2'$ & $k=2k_{F\downarrow} + q$ & $-1$ & $1$ & $0$ & $0$ & $1$ \\
\hline
\end{tabular}
\caption{The momentum $k>0$ and $s$ and $s2$ bands number and current number 
deviations defined in Appendix \ref{D} for $+-$ transverse spin excitations 
populated by one $s2$ particle and thus described by both real and complex nonreal rapidities in the case of
the $s2$ branch line, $\bar{s}'$ branch line, $\bar{s}$ branch line, and $s2'$ branch line that for the momentum
intervals given in the text are part of the corresponding gapped lower threshold.}
\label{table1}
\end{center}
\end{table}

The $S^{+-} (k,\omega)$'s $s2$, $\bar{s}'$, $\bar{s}$, and $s2'$ branch line exponents whose expressions
are given in Eq. (\ref{expG+-}) are plotted as a function of $k$ in Figs. 
\ref{figure7}, \ref{figure8}, \ref{figure9}, and \ref{figure10}, respectively.
In the $k$ intervals of the gapped lower threshold of the spin $n$-string
continuum in Figs. \ref{figure1} and \ref{figure2} for which they are negative, 
which are represented by solid lines in these figures, there are singularities
at and just above the corresponding $\beta=s2,\bar{s}',\bar{s},s2'$ branch lines
in the expression $S^{+-} (k,\omega) = C_{+-}^{\Delta}
(\omega - \Delta_{\beta}^{+-} (k))^{\zeta_{\beta}^{+-} (k)}$,
Eq. (\ref{MPSsFMB}) for $ab=+-$.

The related $S^{xx} (k,\omega)$'s expression, Eq. (\ref{MPSsFMB}) for $ab=xx$, 
in the vicinity and just above the gapped lower threshold of the spin $n$-string
continuum in Figs. \ref{figure3} and \ref{figure4} is similar to that of $S^{+-} (k,\omega)$
for the $k$ intervals for which there is no overlap with the lower continuum spectrum
associated with excited states described by real Bethe-ansatz rapidities. This
thus excludes the low-$m$ $k$ intervals considered in Eq. (\ref{gapineq}).

Concerning again the relation between the physical excitation momentum $k$ and the
$s$ and $s2$ bands momenta $q$, it is useful to provide the $S^{zz} (k,\omega)$'s expressions
of the gapped lower threshold spectrum $\Delta^{zz} (k)$, Eqs. (\ref{gapPMMP}) and (\ref{gapL}), 
for each of its branch-line parts as,
\begin{eqnarray}
\Delta_{s2}^{zz} (k) & = & \varepsilon_{s2} (k - (k_{F\uparrow}-k_{F\downarrow})) \hspace{0.20cm}{\rm and}
\nonumber \\
& & k = (k_{F\uparrow}-k_{F\downarrow}) + q
\nonumber \\
{\rm where} && k\in ]0,(k_{F\uparrow}-k_{F\downarrow})[\hspace{0.20cm}{\rm and}
\nonumber \\
& & q\in ]-(k_{F\uparrow}-k_{F\downarrow}),0[
\nonumber \\
&& {\rm for}\hspace{0.20cm}m\in]0,1[ \, ,
\label{Ds2pL}
\end{eqnarray}
\begin{eqnarray}
\Delta_{\bar{s}}^{zz} (k) & = & 4\mu_B\,h - W_{s2} - \varepsilon_{s} \left(k_{F\uparrow} - k\right)
\hspace{0.20cm}{\rm and}
\nonumber \\
& & k = k_{F\uparrow}- q
\nonumber \\
{\rm where} & & k \in ](k_{F\uparrow}-k_{F\downarrow}),(\pi - {\tilde{k}})[
\hspace{0.20cm}{\rm and}
\nonumber \\
& & q\in ]-(k_{F\downarrow} - {\tilde{k}}),k_{F\downarrow}[
\nonumber \\
& & {\rm for}\hspace{0.20cm}m\in ]0,\tilde{m}]
\nonumber \\
{\rm and} & &k\in ](k_{F\uparrow}-k_{F\downarrow}),(\pi - {\tilde{k}})[\hspace{0.20cm}{\rm and}
\nonumber \\
& & q\in ]-(k_{F\downarrow} - {\tilde{k}}),k_{F\downarrow}[
\nonumber \\
& &{\rm for}\hspace{0.20cm}m\in[\tilde{m},1[ \, ,
\label{Dsppp2L}
\end{eqnarray}
\begin{eqnarray}
\Delta_{\bar{s}'}^{zz} (k) & = & 4\mu_B\,h - \varepsilon_{s} (k_{F\downarrow}-k)
\hspace{0.20cm}{\rm and}\hspace{0.20cm}k = k_{F\downarrow} - q
\nonumber \\
{\rm where} & & k\in ](\pi - {\tilde{k}}),2k_{F\downarrow}[ \hspace{0.20cm}{\rm and}
\nonumber \\
& & q\in ]-k_{F\downarrow},- (k_{F\uparrow}-\tilde{k})[
\nonumber \\
& & {\rm for}\hspace{0.20cm}m\in ]0,\tilde{m}] \, ,
\label{Dsppp1L}
\end{eqnarray}
and
\begin{eqnarray}
\Delta_{s2'}^{zz} (k) & = & \varepsilon_{s2} (k-\pi) \hspace{0.20cm}{\rm and}\hspace{0.20cm}k = \pi + q
\nonumber \\
{\rm where} & & k\in ]2k_{F\downarrow},\pi[\hspace{0.20cm}{\rm and}
\nonumber \\
& & q\in ]-(k_{F\uparrow}-k_{F\downarrow}),0[
\nonumber \\
& & {\rm for}\hspace{0.20cm}m\in ]0,\tilde{m}]
\nonumber \\
{\rm and} & & k\in ](\pi - {\tilde{k}}),\pi[ \hspace{0.20cm}{\rm and}
\nonumber \\
&& q\in ]-{\tilde{k}},0[\hspace{0.20cm}{\rm for}\hspace{0.20cm}m\in[\tilde{m},1[ \, .
\label{Ds2L}
\end{eqnarray}
\begin{figure}
\begin{center}
\centerline{\includegraphics[width=6cm]{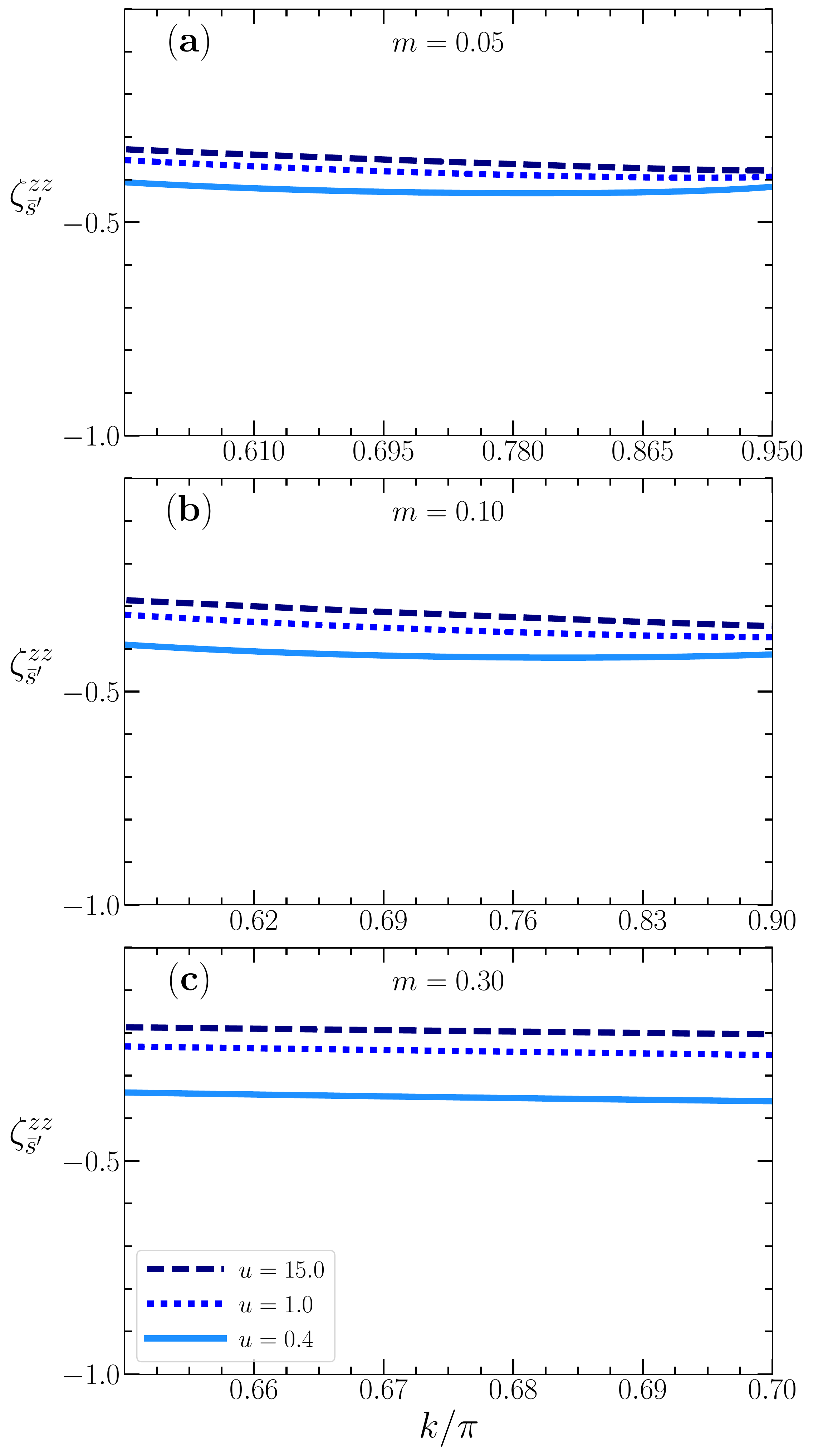}}
\caption{The same as in Fig. \ref{figure8} for the $\bar{s}'$ branch
line of $S^{zz} (k,\omega)$. For that dynamical structure factor, this
exponent is the only that is negative and refers to singularities near
the corresponding small momentum intervals of the
gapped lower threshold of the spin $n$-string continuum in
Figs. \ref{figure5} and \ref{figure6}. Such singularities only emerge in $S^{zz} (k,\omega)$
for spin densities $0<m<\tilde{m}$ where $\tilde{m}=0$ for $u\rightarrow 0$ 
and $\tilde{m}\approx 0.317$ for $u\gg 1$.}
\label{figure11}
\end{center}
\end{figure}

The corresponding $k$ dependent exponents of general form,
Eq. (\ref{expTS}), that appear in the expression, $S^{zz} (k,\omega) = C_{zz}^{\Delta}
(\omega - \Delta_{\beta}^{zz} (k))^{\zeta_{\beta}^{zz} (k)}$,
Eq. (\ref{MPSsFMB}) for $ab=+-$ and $\beta=s2,\bar{s},\bar{s}',s2'$, read,
\begin{eqnarray}
\zeta_{s2}^{zz} (k) & = & -1 + \sum_{\iota=\pm 1}
\left(- {\iota\over \xi_{s\,s}^1} - \xi_{s\,s}^1
+ \Phi_{s,s2}(\iota k_{F\downarrow},q)\right)^2 
\nonumber \\
&& {\rm for}\hspace{0.20cm}q=k-k_{F\uparrow}+k_{F\downarrow}
\hspace{0.20cm}{\rm and}\hspace{0.20cm}k\in ]0,(k_{F\uparrow}-k_{F\downarrow})[
\nonumber \\
\zeta_{\bar{s}}^{zz} (k) & = & -1 
\nonumber \\
& + & \sum_{\iota=\pm 1}
\left(- {\iota\over\xi_{s\,s}^1} 
+ \iota {\xi_{s\,s2}^0\over 2} - {\xi_{s\,s}^1\over 2} 
- \Phi_{s,s}(\iota k_{F\downarrow},q)\right)^2
\nonumber \\
&& {\rm for}\hspace{0.20cm}q=k_{F\uparrow}-k
\hspace{0.20cm}{\rm and}
\nonumber \\
k & \in &  ](k_{F\uparrow}-k_{F\downarrow}),(\pi - {\tilde{k}})[
\hspace{0.20cm}{\rm for}\hspace{0.20cm}m\in ]0,\tilde{m}]
\nonumber \\
k & \in & ](k_{F\uparrow}-k_{F\downarrow}),(\pi - {\tilde{k}})[
\hspace{0.20cm}{\rm for}\hspace{0.20cm}m\in ]0,\tilde{m}] 
\nonumber \\
\zeta_{\bar{s}'}^{zz} (k) & = & -1 + \sum_{\iota=\pm 1}
\left({\iota\over 2\xi_{s\,s}^1} + {\xi_{s\,s}^1\over 2} 
- \Phi_{s,s}(\iota k_{F\downarrow},q)\right)^2 
\nonumber \\
&& {\rm for}\hspace{0.20cm}q=k_{F\downarrow}-k
\hspace{0.20cm}{\rm and}
\nonumber \\
k & \in & ](\pi - {\tilde{k}}),2k_{F\downarrow}[
\hspace{0.20cm}{\rm for}\hspace{0.20cm}m\in ]0,\tilde{m}] 
\nonumber \\
\zeta_{s2'}^{zz} (k) & = & -1 + \sum_{\iota=\pm 1}\left(- {\iota\over\xi_{s\,s}^1} 
+ \Phi_{s,s2}(\iota k_{F\downarrow},q)\right)^2 
\nonumber \\
&& {\rm for}\hspace{0.20cm}q=k-\pi
\hspace{0.20cm}{\rm and}
\nonumber \\
k & \in & ]2k_{F\downarrow},\pi[
\hspace{0.20cm}{\rm for}\hspace{0.20cm}m\in ]0,\tilde{m}]
\nonumber \\
k & \in & ](\pi - {\tilde{k}}),\pi[
\hspace{0.20cm}{\rm for}\hspace{0.20cm}m\in[\tilde{m},1[ \, .
\label{exps2pL}
\end{eqnarray}
Also in the present case of $S^{zz} (k,\omega)$, the three functionals $\Phi_{\iota} (q)$ in the general
expression, Eq. (\ref{expTS}), specific to the $s2,s2'$ branch lines, $\bar{s}$ branch line,
and $\bar{s}'$ branch line are provided in Eqs. (\ref{Fs2}), (\ref{Fbars}), and
(\ref{Fbarsl}), respectively. The corresponding suitable values
of the number and current number deviations used in such functionals
are though different for the present branch lines. They are given in Table \ref{table2}.
\begin{table}
\begin{center}
\begin{tabular}{|c|c|c|c|c|c|c|} 
\hline
b. line & $k$ in terms of $q$ & $\delta N_s^F$ & $\delta J_s^F$ & $\delta N_s^{NF}$ & $\delta J_{s2}$ & $\delta N_{s2}$ \\
\hline
$s2$ & $k=k_{F\uparrow} - k_{F\downarrow} + q$ & $-2$ & $-1$ & $0$ & $0$ & $1$ \\
\hline
$\bar{s}$ & $k=k_{F\uparrow} - q$ & $-2$ & $-1/2$ & $-1$ & $0$ & $1$ \\
\hline
$\bar{s}'$ & $k=k_{F\uparrow} - q$ & $1$ & $-1/2$ & $-1$ & $-1/2$ & $1$ \\
\hline
$s2'$ & $k=\pi + q$ & $-2$ & $0$ & $0$ & $0$ & $1$ \\
\hline
\end{tabular}
\caption{The momentum $k>0$ and $s$ and $s2$ bands
number and current number deviations defined in Appendix \ref{D}
for longitudinal spin excitations 
populated by one $s2$ particle and thus
described both real and complex nonreal rapidities in the case of
the $s2$ branch line, $\bar{s}$ branch line,
$\bar{s}'$ branch line, and
$s2'$ branch line that for the momentum
intervals given in the text are part of the corresponding gapped lower threshold.}
\label{table2}
\end{center}
\end{table} 

The behaviors of the spin dynamical structure factor $S^{zz} (k,\omega)$ are actually
qualitatively different from those of $S^{+-} (k,\omega)$. Except for $\zeta_{\bar{s}'}^{zz} (k)$, the exponents in Eq. (\ref{exps2pL}) 
are positive for all their $k$ intervals. That $\bar{s}'$ branch line exponent
is plotted as a function of $k$ in Fig. \ref{figure11}. It is negative for its whole $k$ subinterval, which
is part of the $k$ interval of the gapped lower threshold in Fig. \ref{figure5}.
The $\bar{s}'$ branch line's $m$-dependent subinterval is either small or 
that line is not part of the $S^{zz} (k,\omega)$'s gapped lower threshold at all. Its momentum width decreases upon increasing 
$m$ up to the spin density $\tilde{m}$. As mentioned above, this spin density decreases upon 
decreasing $u$, having the limiting values $\tilde{m}=0$ for $u\rightarrow 0$ 
and $\tilde{m}\approx 0.317$ for $u\gg 1$. For $\tilde{m}<m<1$, the $\bar{s}'$ branch line
is not part of the $S^{zz} (k,\omega)$'s gapped lower threshold spectrum.
This is why for $m=0.5>\tilde{m}$ and $m=0.8>\tilde{m}$ that line does not appear in the
gapped lower threshold plotted in Fig. \ref{figure6}.

Hence gapped lower threshold's singularities only emerge in $S^{zz} (k,\omega)$
for spin densities $0<m<\tilde{m}$ at and just above the $\bar{s}'$ branch line,
the corresponding line shape reading, $S^{zz} (k,\omega) = C_{zz}^{\Delta} (\omega - \Delta_{\bar{s}'}^{zz} (k))^{\zeta_{\bar{s}'}^{+-} (k)}$.
That branch line $k$ subinterval width though strongly decreases upon increasing $m$ up to $\tilde{m}$. 
 
These behaviors are consistent with the $S^{zz} (k,\omega)$'s spectral 
weight stemming from spin $n$-string states decreasing upon increasing the spin density,
being negligible for $\tilde{m}<m<1$. Consistent with the $u$ dependence of the spin density 
$\tilde{m}$, this spectral weight suppression becomes stronger upon decreasing $u$. Hence increasing the spin
density $m$ within the interval $m\in ]0,\tilde{m}]$ and lowering the $u$ value tends
to suppress the contribution of spin $n$-string states to $S^{zz} (k,\omega)$.

\subsection{The line shape near the lower thresholds}
\label{SECIVB}

To provide an overall physical picture that accounts for all gapped lower threshold's singularities and
lower threshold's singularities in the spin dynamical structure factors, here we shortly revisit their line shape
behavior at and just above the lower thresholds of the lower continua in Figs. \ref{figure1}-\ref{figure6}. 
The corresponding contributions are from excited states described by real Bethe-ansatz rapidities. Such 
lower continua contain most spectral weight of the corresponding spin dynamical structure factors.
\begin{figure}
\begin{center}
\centerline{\includegraphics[width=9cm]{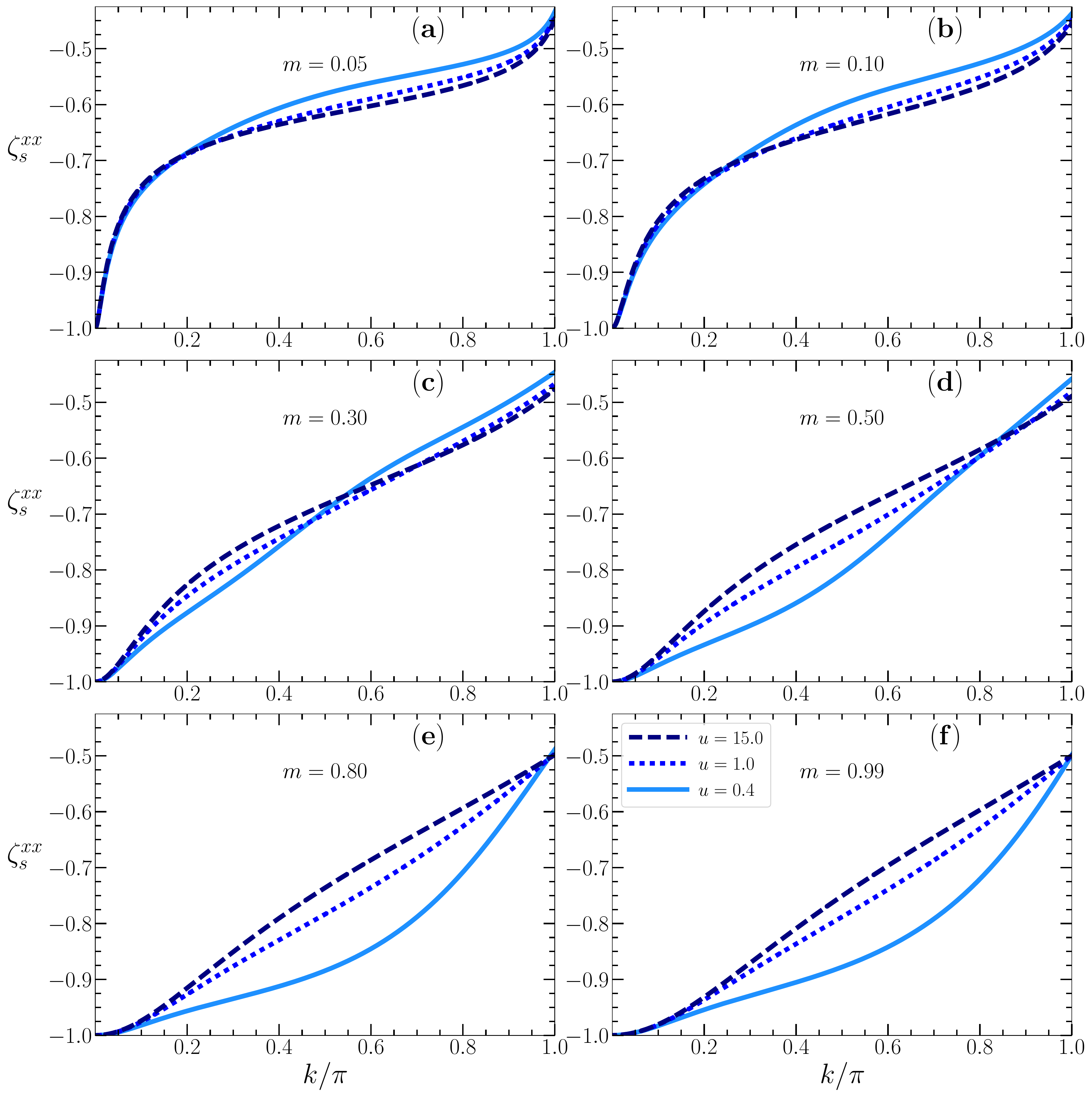}}
\caption{The momentum dependence of the exponent that controls the
$S^{xx} (k,\omega)$'s line shape near and just above the lower
threshold of the lower continuum in Figs. \ref{figure3} and \ref{figure4}
for spin densities $m$ (a) $0.05$, (b) $0.1$, (c) $0.3$, (d) $0.5$, (e) $0.8$,
and (f) $0.99$ and $u=0.4,1.0,15.0$. For $k\in ]0,(k_{F\uparrow}-k_{F\downarrow})[$ and 
$k\in ](k_{F\uparrow}-k_{F\downarrow}),\pi[$ that exponent corresponds
to that of $S^{+-} (k,\omega)$ and $S^{-+} (k,\omega)$, respectively.}
\label{figure12}
\end{center}
\end{figure}

In the case of the transverse dynamical structure factor,
$S^{xx} (k,\omega) = {1\over 4}\left(S^{+-} (k,\omega)+S^{-+} (k,\omega)\right)$,
we consider the transitions to excited states that determine the line shape in the vicinity
of the lower thresholds of both $S^{+-} (k,\omega)$ and $S^{-+} (k,\omega)$, respectively. 
The spectrum of $S^{xx} (k,\omega)$ at and just above its
lower threshold, refers to a superposition of the lower threshold spectra $\omega^{+-} (k)$ and $\omega^{-+} (k)$,
Eqs. (\ref{OkPMRs}) and (\ref{OkMPRs})-(\ref{OkMPRs2}), respectively.
The $(k,\omega)$-plane lower continuum that results from such a spectra superposition is 
represented in Figs. \ref{figure3} and \ref{figure4}.

Similarly to Eq. (\ref{MPSsFMB}), for spin densities $0<m<1$, $u>0$, and $k\in ]0,\pi[$ the 
line shape of the spin dynamical structure factors $S^{ab} (k,\omega)$ where $ab = +-,-+,xx,zz$ 
near and just above their lower thresholds has the following general form,
\begin{eqnarray}
S^{ab} (k,\omega) & = & C_{ab} 
\Bigl(\omega - \omega^{ab}_{lt} (k)\Bigr)^{\zeta_{s}^{ab} (k)}  
\nonumber \\
& & {\rm for}\hspace{0.20cm} (\omega - \omega^{ab}_{lt} (k)) \geq 0  
\nonumber \\
& & {\rm where}\hspace{0.20cm} ab = +-,-+,xx,zz \, .
\label{MPSs}
\end{eqnarray}
In the case of $S^{xx} (k,\omega)$, this expression can be expressed as
\begin{eqnarray}
&& S^{xx} (k,\omega) = S^{+-} (k,\omega)\hspace{0.20cm} {\rm for}\hspace{0.20cm} 
k\in [0,(k_{F\uparrow}-k_{F\downarrow})[
\nonumber \\
&& \hspace{1.25cm} = S^{-+} (k,\omega)\hspace{0.20cm} {\rm for}
\hspace{0.20cm}k \in ](k_{F\uparrow}-k_{F\downarrow}),\pi[ \, .
\label{MPSsGen}
\end{eqnarray}

The lower thresholds under consideration refer to a single $s$ branch line that except for $S^{-+} (k,\omega)$
has two sections. In Eq. (\ref{MPSs}), $C_{ab}$ are constants that have a fixed value 
for the $k$ and $\omega$ ranges corresponding to small values of the energy deviation 
$(\omega - \omega^{ab}_{lt} (k))\geq 0$. The $ab=+-,-+,zz$ lower threshold spectra 
$\omega^{+-} (k)$, $\omega^{-+} (k)$, and $\omega^{zz} (k)$ in that deviation
are given in Eqs. (\ref{OkPMRs}), (\ref{OkMPRs})-(\ref{OkMPRs2}), 
and (\ref{OkPMRsL})-(\ref{OkMPRsL}), respectively.
\begin{figure}
\begin{center}
\centerline{\includegraphics[width=9cm]{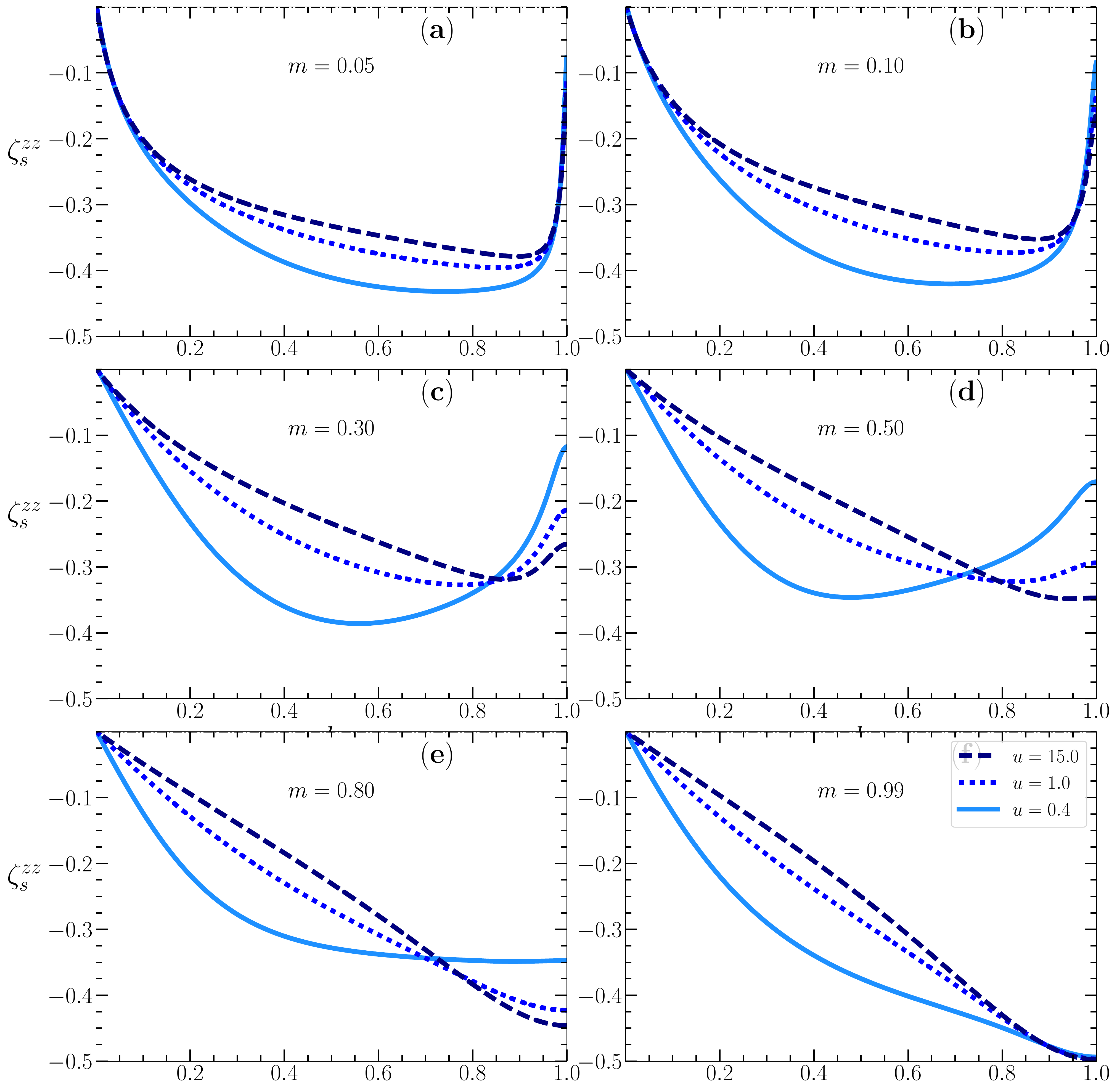}}
\caption{The momentum dependence of the exponent that controls the
$S^{zz} (k,\omega)$ line shape near and just above the lower
threshold of the lower continuum in Figs. \ref{figure5} and \ref{figure6}
for spin densities $m$ (a) $0.05$, (b) $0.1$, (c) $0.3$, (d) $0.5$, (e) $0.8$,
and (f) $0.99$ and $u=0.4,1.0,15.0$.}
\label{figure13}
\end{center}
\end{figure} 

The $k$ dependent exponents appearing in the spin dynamical factors's
expression, Eq. (\ref{MPSs}), are also of general form,
Eq. (\ref{expTS}). In the present case, they are given by,
\begin{eqnarray}
\zeta_{s}^{-+} (k) & = & -1 + \sum_{\iota =\pm 1}\left(- {\xi_{s\,s}^1\over 2} 
- \Phi_{s,s}(\iota k_{F\downarrow},q)\right)^2 
\nonumber \\
&& {\rm for}\hspace{0.20cm}q=k_{F\uparrow}-k
\hspace{0.20cm}{\rm and}\hspace{0.20cm}
k\in ](k_{F\uparrow}-k_{F\downarrow}),\pi[
\nonumber \\
\zeta_{s}^{+-} (k) & = & -1 + \sum_{\iota =\pm 1}\left(- {\xi_{s\,s}^1\over 2} 
+ \Phi_{s,s}(\iota k_{F\downarrow},q)\right)^2 
\nonumber \\
&& {\rm for}\hspace{0.20cm}q=k-k_{F\uparrow}
\hspace{0.20cm}{\rm and}\hspace{0.20cm}k\in ]0,(k_{F\uparrow}-k_{F\downarrow})[
\nonumber \\
 \zeta_{s}^{+-} (k) & = & -1 + \sum_{\iota =\pm 1}\left({\iota\over\xi_{s\,s}^1} - {\xi_{s\,s}^1\over 2} 
- \Phi_{s,s}(\iota k_{F\downarrow},q)\right)^2 
\nonumber \\
&& {\rm for}\hspace{0.20cm}q=k_{F\uparrow}-k
\hspace{0.20cm}{\rm and}\hspace{0.20cm}k\in ](k_{F\uparrow}-k_{F\downarrow}),\pi[
\nonumber \\
\zeta_{s}^{zz} (k) & = & -1 +
\sum_{\iota =\pm 1}\left({\iota\over 2\xi_{s\,s}^1}  + {\xi_{s\,s}^1\over 2} 
- \Phi_{s,s}(\iota k_{F\downarrow},q)\right)^2
\nonumber \\
&& {\rm for}\hspace{0.20cm}q=k_{F\downarrow}-k
\hspace{0.20cm}{\rm and}\hspace{0.20cm}k\in ]0,2k_{F\downarrow}[
\nonumber \\
\zeta_{s}^{zz} (k) & = & -1 + \sum_{\iota =\pm 1}\left(- {\iota\over 2\xi_{s\,s}^1} + {\xi_{s\,s}^1\over 2} 
+ \Phi_{s,s}(\iota k_{F\downarrow},q)\right)^2 
\nonumber \\
&& {\rm for}\hspace{0.20cm}q=k-k_{F\downarrow}
\hspace{0.20cm}{\rm and}\hspace{0.20cm}k\in ]2k_{F\downarrow},\pi[ \, .
\label{expsPM}
\end{eqnarray}
The functional $\Phi_{\iota} (q)$ in the general exponent expression, Eq. (\ref{expTS}), 
is for the present $s$ branch lines given in Eq. (\ref{Fs}). The 
suitable specific values of the number and current number deviations used in such a functional to
obtain the exponents in Eq. (\ref{expsPM}) are provided in Table \ref{table3}.

As confirmed by the form of the expressions given in Eqs. (\ref{OkPMRs}) and (\ref{OkMPRs2}),
one has that $\omega^{+-}_{lt} (k) = \omega^{-+}_{lt} (k)$ for $k\in ](k_{F\uparrow}-k_{F\downarrow}),\pi[$.
In that $k$ interval, the line shape of 
$S^{xx} (k,\omega) = {1\over 4}\left(S^{+-} (k,\omega)+S^{-+} (k,\omega)\right)$ is
controlled by the smallest of the exponents $\zeta_{s}^{-+} (k)$ and $\zeta_{s}^{+-} (k)$
in Eq. (\ref{expsPM}), which turns out to be $\zeta_{s}^{-+} (k)$. Hence, the exponent $\zeta_{s}^{xx} (k)$
is given by,
\begin{eqnarray}
\zeta_{s}^{xx} (k) & = & -1 + \sum_{\iota =\pm 1}\left(-{\xi_{s\,s}^1\over 2} 
+ \Phi_{s,s}(\iota k_{F\downarrow},q)\right)^2
\hspace{0.20cm}{\rm for}
\nonumber \\
&& q=k-k_{F\uparrow}
\hspace{0.20cm}{\rm and}\hspace{0.20cm}k\in ]0,(k_{F\uparrow}-k_{F\downarrow})[
\nonumber \\
& = & -1 + \sum_{\iota =\pm 1}\left(-{\xi_{s\,s}^1\over 2} 
- \Phi_{s,s}(\iota k_{F\downarrow},q)\right)^2 
\hspace{0.20cm}{\rm for}
\nonumber \\
&& q=k_{F\uparrow}-k
\hspace{0.20cm}{\rm and}\hspace{0.20cm}k\in ](k_{F\uparrow}-k_{F\downarrow}),\pi[ \, .
\label{expszz}
\end{eqnarray}

This exponent is plotted as a function of $k$ in Fig. \ref{figure12}.
The $s$ branch line exponent $\zeta_{s}^{zz} (k)$ whose expression is given in Eqs. (\ref{expsPM}) 
is also plotted as a function of momentum in Fig. \ref{figure13}.
\begin{table}
\begin{center}
\begin{tabular}{|c|c|c|c|c|c|} 
\hline
$s$ & $k=k(q)$ intervals & $\delta N_s^F$ & $\delta J_s^F$ & $\delta N_s^{NF}$ \\
\hline
$-+$ & $k=k_{F\uparrow} - q\in ](k_{F\uparrow}-k_{F\downarrow}),\pi[$ & $0$ & $-1/2$ & $-1$ \\
\hline
$+-$ & $k=k_{F\uparrow} + q\in [0,(k_{F\uparrow}-k_{F\downarrow})[$ & $0$ & $-1/2$ & $1$ \\
\hline
$+-$ & $k=k_{F\uparrow} - q\in ](k_{F\uparrow}-k_{F\downarrow}),\pi[$ & $2$ & $-1/2$ & $-1$ \\
\hline
$zz$ & $k=k_{F\downarrow} - q\in ]0,2k_{F\downarrow}[$ & $1$ & $1/2$ & $-1$ \\
\hline
$zz$ & $k=k_{F\downarrow} + q k\in ]2k_{F\downarrow},\pi] $ & $-1$  & $1/2$ & $1$ \\
\hline
\end{tabular}
\caption{The momentum $k>0$ intervals and $s$ band number and current number 
deviations defined in Appendix \ref{D} for the $s$ branch lines that coincide
with the lower thresholds of the $-+$, $+-$, and $zz$ dynamical structure factors
lower continua. In the case of $S^{+-} (k,\omega)$ and $S^{zz} (k,\omega)$,
such lower continua appear in Figs. \ref{figure1} and \ref{figure2} and \ref{figure5} and \ref{figure6}, respectively.
The lower continua of $S^{xx} (k,\omega)$ displayed in Figs. \ref{figure3} and \ref{figure4}
are a superposition of those of $S^{+-} (k,\omega)$ and $S^{-,+} (k,\omega)$.}
\label{table3}
\end{center}
\end{table}

Both such exponents are negative in the whole momentum interval $k\in ]0,\pi[$ for spin densities
$0<m<1$ and $u>0$. It follows that there are singularities at and just above the corresponding 
lower thresholds. (Due to a sign error, the minus sign in the quantity $-\xi_{s\,s}^1/2$ appearing
in Eq. (\ref{expszz}) was missed in Ref. \onlinecite{Carmelo_16} where
the exponent $\zeta_{1}^{xx} (k)$ is named $\xi^t$. Its momentum dependence
plotted in Fig. \ref{figure12} corrects that plotted in Fig. 5 of Ref. \onlinecite{Carmelo_16}.)
 
\section{Limiting behaviors of the spin dynamical structure factors}
\label{SECV}

Consistent with the relation, Eq. (\ref{SPMMPmm}),
the spin dynamical structure factor $S^{-+} (k,\omega)$ is at $m=0$ that obtained in the $m\rightarrow 0$
limit from $m>0$ values whereas $S^{+-} (k,\omega)$ is at $m=0$ that obtained in the $m\rightarrow 0$
limit from $m<0$ values. One then confirms that $S^{-+} (k,\omega)=S^{+-} (k,\omega)$ at $m=0$. 
However, in the $m\rightarrow 0$
limit from $m>0$ values, the $S^{+-} (k,\omega)$ gapped continuum, Eq. (\ref{dkEdPPM}), becomes 
a gapless line that coincides with both its $\bar{s}$ and $\bar{s}'$ branch lines and
the lower threshold of $S^{-+} (k,\omega)=S^{+-} (k,\omega)$ at $m=0$.

In the case of the initial ground state referring to $h=0$ and thus $m=0$, one
has in addition that $S^{zz} (k,\omega) = S^{xx} (k,\omega)$. 
The selection rules in Eq. (\ref{SRh0}) impose that the longitudinal dynamical structure factor
is fully controlled by transitions from the $S =S^z=0$ ground state to spin triplet excited states with
spin numbers $S=1$ and $S^z=0$. This is different from the case 
when the initial ground state refers to $h\neq 0$ and $m\neq 0$.
Then according to the selection rules, Eq. (\ref{SRhfinite}), the longitudinal dynamical structure factor
$S^{zz} (k,\omega) \neq S^{xx} (k,\omega)$ is controlled by transitions from the ground state 
with spin numbers $S^z = S$ or $S^z = -S$ to excited states with
the same spin numbers $S^z = S$ or $S^z = -S$, respectively.

In the case of the $h=0$ and $m=0$ initial ground state, (i) $S^{zz} (k,\omega)$ and
(ii) $S^{+-} (k,\omega)$ and $S^{-+} (k,\omega)$ are fully controlled by transitions to spin triplet 
$S=1$ excited states with (i) $S^z=0$ and (ii) $S^z=\pm 1$, respectively. Their $s$ band two-hole spectrum is obtained 
in the $m\rightarrow 0$ limit from that of $S^{+-} (k,\omega)$
for $m<0$ and from that of $S^{-+} (k,\omega)$ for $m>0$ and thus reads,
\begin{eqnarray}
\omega^{xx} (k) & = & \omega^{zz} (k) = - \varepsilon_{s} (q_1) - \varepsilon_{s} (q_2)
\nonumber \\
& &{\rm where}\hspace{0.20cm} k = \iota\pi - q_1 - q_2 \hspace{0.20cm}
{\rm and}\hspace{0.20cm}\iota = \pm 1
\nonumber \\
& & {\rm for}\hspace{0.20cm}q_1 \in [-\pi/2,\pi/2]
\nonumber \\
& & {\rm and}\hspace{0.20cm}q_2 \in [-\pi/2,\pi/2] \, .
\label{spectram0}
\end{eqnarray}
Consistent, spin $SU (2)$ symmetry implies that the triplet $S=1$ and $S^z=0$ excited states 
that control $S^{zz} (k,\omega)$ have exactly the same 
spectrum, Eq. (\ref{spectram0}), as the triplet $S=1$ and $S^z=\pm 1$ excited states
that control $S^{+-} (k,\omega)$ and $S^{-+} (k,\omega)$.

In spite of the singular behavior concerning the class of excited states that
control the longitudinal dynamical structure factor for $m=0$ and $m>0$
initial ground states, respectively, one confirms in the following that
the same line shape near the spin dynamical structure factors's 
lower thresholds is obtained at $m=0$ and in the $m\rightarrow 0$ limit,
respectively.

\subsection{Behaviors of the spin dynamical structure factors in the $m\rightarrow 0$ limit}
\label{SECVA}

In the $m\rightarrow 0$ limit from $m>0$ values, the transverse spin structure factor
$S^{-+} (k,\omega)$ lower threshold spectrum, Eq. (\ref{OkPMRs}), expands to the 
whole $k\in [0,\pi]$ interval. The corresponding
line shape near the $s$ branch line is then valid for $k\in]0,\pi[$. 
Since a similar spectrum is obtained for the lower threshold of $S^{-+} (k,\omega)$ 
in the $m\rightarrow 0$ limit from $m<0$ values, one finds,
\begin{eqnarray}
\omega^{xx}_{lt} (k) & = & - \varepsilon_s (k_{F}-k) \hspace{0.2cm}{\rm where}
\nonumber \\
k & = & {\pi\over 2} - q\in]0,\pi[
\hspace{0.2cm}{\rm for}
\nonumber \\
q & \in & ]-\pi/2,\pi/2[ \, .
\label{OkPMRs0}
\end{eqnarray}

As reported above, in the $m\rightarrow 0$ limit from $m>0$ values
the $S^{+-} (k,\omega)$'s gapped continuum associated with the spectrum, Eq. (\ref{dkEdPPM}), becomes 
a gapless line that coincides with both the spectra in Eqs. (\ref{Dsppp1}) and
(\ref{Dsppp2}) of its $\bar{s}$ and $\bar{s}'$ branch lines, respectively, and
the lower threshold of $S^{-+} (k,\omega)=S^{+-} (k,\omega)$ at $m=0$.
(In the $m\rightarrow 0$ limit from $m<0$ values, the $\bar{s}$ and $\bar{s}'$ branch lines
rather stem from $S^{-+} (k,\omega)$.) Hence the spectra
$\Delta_{\bar{s}'}^{+-} (k) = \Delta_{\bar{s}}^{+-} (k)$ read in that limit,
\begin{eqnarray}
\Delta_{\bar{s}'}^{+-} (k) & = & \Delta_{\bar{s}}^{+-} (k)
\nonumber \\
& = & - \varepsilon_{s} (\pi/2-k)
\hspace{0.20cm}{\rm where}\hspace{0.20cm}k = {\pi\over 2} - q
\nonumber \\
{\rm for} & & k\in ]0,\pi[
\hspace{0.2cm}{\rm for}\hspace{0.2cm}q \in ]-\pi/2,\pi/2[ \, .
\label{Dsppp10}
\end{eqnarray}

It then turns out that the corresponding exponents $\zeta_{\bar{s}'}^{+-} (k)$ and $\zeta_{\bar{s}}^{+-} (k)$, Eq. (\ref{expG+-}),
have in the $m\rightarrow 0$ limit exactly the same value. In addition, that value is 
the same as that of $\zeta_{s}^{xx} (k)$, Eq. (\ref{expszz}), reached in that limit.
Indeed, by use of the limiting behaviors $\lim_{m\rightarrow 0}\Phi_{s,s}\left(\pm k_{F\downarrow},q\right) = \pm 1/(2\sqrt{2})$ for
$q\neq \pm k_{F\downarrow}$, $\lim_{m\rightarrow 0}\Phi_{s,s2}\left(\pm k_{F},0\right) = \pm 1/\sqrt{2}$,
and $\lim_{m\rightarrow 0}\xi_{s\,s}^1 = 1/\sqrt{2}$
reported in Eqs. (\ref{Phis-all-qq-0}), (\ref{Phis2-all-qq-0}), and (\ref{Limxiss}),
one finds that,
\begin{eqnarray}
\zeta_{s}^{xx} (k) & = & -1 + \sum_{\iota =\pm 1}\left(- {\xi_{s\,s}^1\over 2} 
- \Phi_{s,s}(\iota\pi/2,q)\right)^2 
\nonumber \\
& = & - {1\over 2} 
\nonumber \\
\zeta_{\bar{s}'}^{+-} (k) & = & -1 + \sum_{\iota=\pm 1}\left(- {\xi_{s\,s}^1\over 2} 
- \Phi_{s,s}(\iota \pi/2,q)\right)^2  
\nonumber \\
& = & - {1\over 2} 
\nonumber \\
\zeta_{\bar{s}}^{+-} (k) & = & -1 + \sum_{\iota=\pm 1}\left(\iota {\xi_{s\,s2}^0\over 2}
+ {\xi_{s\,s}^1\over 2} 
- \Phi_{s,s}(\iota \pi/2,q)\right)^2 
\nonumber \\
& = & - {1\over 2} \, .
\label{expsPM0}
\end{eqnarray}

The spin $SU(2)$ symmetry obliges as well that at $m=0$ the results should be similar
for the transverse and longitudinal spin structure factors, respectively.
In the $m\rightarrow 0$ limit, the longitudinal spin structure factor
lower threshold spectrum, Eq. (\ref{OkPMRsL}), expands to the whole $k\in ]0,\pi[$ 
interval and indeed is similar to that in Eq. (\ref{OkPMRs0}), as it reads,
\begin{eqnarray}
\omega^{zz}_{lt} (k) & = & \omega_{s}^{zz} (k) = - \varepsilon_s (\pi/2 - k) \hspace{0.2cm}{\rm where}
\nonumber \\
k & = & k_{F} - q \in]0,\pi[
\hspace{0.2cm}{\rm for}\hspace{0.2cm}
q \in ]-\pi/2,\pi/2[ \, .
\label{OkPMRsL0}
\end{eqnarray}

In spite of such a similarity, the longitudinal dynamical structure factor
is at $m=0$ fully controlled by transitions from the ground state to excited states with
spin numbers $S=1$ and $S^z=0$. The line shape obtained from such spin triplet excited states
is though exactly the same as that obtained in the $m\rightarrow 0$ limit from 
the $S^z=S$ or $S^z=-S$ and $S>0$ excited states.

However, in the $m\rightarrow 0$ limit the $S^{zz} (k,\omega)$'s gapped 
$\bar{s}$ and $\bar{s}'$ branch line spectra in Eqs. (\ref{Dsppp2L}) and (\ref{Dsppp1L}), respectively, 
become gapless and coincide with both each other and with the lower threshold of the longitudinal spin structure factor,
Eq. (\ref{OkPMRsL0}), for whole $k\in ]0,\pi[$ interval,
\begin{eqnarray}
\Delta_{\bar{s}'}^{zz} (k) & = & - \varepsilon_{s} (\pi/2-k)
\hspace{0.20cm}{\rm and}\hspace{0.20cm}k = {\pi\over 2} - q
\nonumber \\
{\rm where} & & k\in ]0, \pi[ 
\hspace{0.2cm}{\rm for}
\nonumber \\
q & \in & ]-\pi/2,\pi/2[ \, .
\label{Dsppp1L0}
\end{eqnarray}

One then finds that in such a limit, $\zeta_{\bar{s}'}^{zz} (k)< \zeta_{\bar{s}}^{zz} (k)$.
Here $\zeta_{\bar{s}'}^{zz} (k)$ and  $\zeta_{\bar{s}}^{zz} (k)$ are
the corresponding branch line exponents given in Eq. (\ref{exps2pL}). Such an inequality 
implies that the line shape is controlled by the exponents $\zeta_{\bar{s}'}^{zz} (k)$ and $\zeta_{s}^{zz} (k)$ 
such that $\zeta_{\bar{s}'}^{zz} (k)=\zeta_{s}^{zz} (k)$ in the $m\rightarrow 0$ limit, as given
below. Here $\zeta_{\bar{s}'}^{zz} (k)$ is the exponent associated with the spectrum
in Eq. (\ref{Dsppp1L}). 

The use of the limiting behaviors reported in Eqs. (\ref{Phis-all-qq-0}) and (\ref{Limxiss}),
confirms that the exponent $\zeta_{\bar{s}'}^{zz} (k)$, Eq. (\ref{exps2pL}), equals 
both the exponent $\zeta_{s}^{zz} (k)$, Eq. (\ref{expsPM}), and those given in Eq. (\ref{expsPM0}).
The former two exponents are found to be given by,
\begin{eqnarray}
\zeta_{s}^{zz} (k) & = & -1 + \sum_{\iota =\pm 1}\left({\iota\over 2\xi_{s\,s}^1}  + {\xi_{s\,s}^1\over 2} 
- \Phi_{s,s}(\iota \pi/2,q)\right)^2
\nonumber \\
& = & - {1\over 2} 
\nonumber \\
\zeta_{\bar{s}'}^{zz} (k) & = & -1 + \sum_{\iota=\pm 1}\left({\iota\over 2\xi_{s\,s}^1} + {\xi_{s\,s}^1\over 2} 
- \Phi_{s,s}(\iota \pi/2,q)\right)^2 
\nonumber \\
& = & - {1\over 2} \, .
\label{expsppp1L0}
\end{eqnarray}

Again and in spite of such similarities, the two classes of excited states described by real and complex
nonreal rapidities, respectively, that at $m=0$ contribute to the longitudinal dynamical structure factor have
rather spin numbers $S=1$ and $S^z=0$. The line shape associated with such spin triplet 
excited states is though exactly the same as that obtained in the $m\rightarrow 0$ limit from 
the above excited states.

One then concludes that for $u>0$ and in the $m\rightarrow 0$ limit the line shape
at and just above the lower threshold of the spin structure factor is of the form,
\begin{eqnarray}
&& S^{aa} (k,\omega) = C\,(\omega - \omega (k))^{-1/2}\hspace{0.20cm}{\rm where}
\nonumber \\
&& \omega (k) = 2t\int_0^{\infty}d\omega\,{\cos \left(\omega\,\Lambda_s \left({\pi\over 2} - k\right)\right)\over\omega\cosh \omega}
\,J_1 (\omega) \, ,
\label{DSF-BL-m0}
\end{eqnarray}
for $]0,\pi[$ and $aa = xx,yy,zz$ where  $C$ is a constant that has a fixed value for the $k$ and $\omega$ ranges corresponding 
to small values of the energy deviation $(\omega - \omega (k))$, $J_1 (\omega)$ is a Bessel function, and the 
$s$ band rapidity function $\Lambda_s (q)$ is defined in terms of its inverse function $q = q_s (\Lambda)$ 
in Eq. (\ref{qLambdam0}). The exponent $-1/2$ is indeed that known to control the line shape at and 
just above the lower threshold of $\omega (k)$ \cite{Essler_99}.

\subsection{Behaviors of the spin dynamical structure factors in the $m\rightarrow 1$ limit}
\label{SECVB}

The sum rules, Eq. (\ref{SRDSF}), imply that  $\lim_{m\rightarrow 1}S^{-+} (k,\omega)=0$ and
$\lim_{m\rightarrow 1}S^{zz} (k,\omega)=0$. It follows that as $m\rightarrow 1$ and thus
$h\rightarrow h_c$, the spin dynamical structure factor is dominated by $S^{xx} (k,\omega)$. 
Here $h_c$ is the critical field associated with the spin energy scale $2\mu_B\,h_c$, Eq. (\ref{hc}),
at which fully polarized ferromagnetism is achieved. 

At $h=h_c$ the power-law expressions of the present dynamical theory involving $k$ dependent exponents 
are not valid, being replaced by a $\delta$-function like distribution,
\begin{eqnarray}
&& S^{xx} (k,\omega) = {\pi\over 2} \delta \left(\omega - \omega^{xx}_{lt} (k)\right) \hspace{0.20cm}{\rm where}
\nonumber \\
&& \omega^{xx}_{lt} (k) = 4t\,\left(\sqrt{1+u^2}-u\right) 
\nonumber \\
&& - {2t\over \pi}\int_{-\pi}^{\pi}d k \sin k
\arctan \left({\sin k - \Lambda_s (\pi -k)\over u}\right) ,
\label{Sxxm1}
\end{eqnarray}
for $[0,\pi]$. Here the $s$ band rapidity function $\Lambda_s (q)$ is defined in terms of its inverse function 
$q = q_s (\Lambda)$ in Eq. (\ref{qLs0}).

\section{Discussion and concluding remarks}
\label{SECVI}

\subsection{Discussion of the results}
\label{SECVIA}

Our results provide important information about how in 1D Mott-Hubbard insulators electron itinerancy 
associated in the present model with the transfer integral $t$ affects 
the spin dynamics: The main effect of increasing $t$ at constant $U$ and thus
decreasing the ratio $u=U/4t$ is on the energy bandwidth of the corresponding relevant spectra.

Physically, this is justified by the interplay of kinetic energy and spin fluctuations.
However, the matrix elements that control the spectral weights and
the related momentum-dependent exponents in the dynamical structure factors's
expressions studied in this paper are little affected by decreasing the ratio $u=U/4t$.

The internal degrees of freedom of the $s$ and $s2$ particles refer to
one unbound singlet pair of spins $1/2$ and two bound singlet pairs of such spins.
The spins $1/2$ in such pairs refer to rotated electrons that singly occupy sites.
In the $u\rightarrow 0$ limit, the corresponding $s$ and $s2$ energy dispersion's 
bandwidths reach their maximum values, $\lim_{u\rightarrow 0}W_{s} = 2t\left(1 + \sin\left({\pi\over 2}\,m\right)\right)$
and $\lim_{u\rightarrow 0}W_{s2} = 4t\sin\left({\pi\over 2}\,m\right)$, respectively,
whereas $\lim_{u\rightarrow\infty}W_{s} = \lim_{u\rightarrow\infty}W_{s2} = 0$, 
as given in Eq. (\ref{W2u0}). Indeed,
for small, intermediate, and large yet finite $u$ values the $s$ particles for all 
spin densities $m$ and the $s2$ particles for $m>0$,
along with the two and four spins $1/2$ within them, respectively, contribute to the kinetic energy 
associated with electron itinerancy. However, in the $u\rightarrow\infty$ limit
{\it all} spin configurations become degenerate and the spins $1/2$ within
the $s$ and $s2$ particles become localized. 

Consistently, the kinetic energy, $E_{\rm kin}=t\,\partial\langle\hat{H}\rangle/\partial t$, of all 
Mott-Hubbard insulator's states decreases
from a maximum value reached in the $u\rightarrow 0$ limit to zero for $u\rightarrow\infty$.
Intermediate $u$ values refer to a crossover
between these two limiting behaviors. While this applies to all spin densities,
for further information on the interplay of kinetic energy
and spin fluctuations at $m=0$, see for instance Sec. IV of Ref. \onlinecite{Carmelo_88}
for electronic density $n=1$.

The dynamical theory used in the studies of this paper refers to a specific case of the general 
dynamical theory considered in Ref. \onlinecite{Carmelo_16}. The former theory
refers to the Hamiltonian, Eq. (\ref{H}), acting onto a subspace that includes spin $n$-string states.
It has specific values for the spectral parameters that control the momentum dependent exponents in the
spin dynamical structure factors's expressions that have been obtained in this paper
for $(k,\omega)$-plane regions at and near well-defined types of spectral features. 

As mentioned in Sec. \ref{SECI}, the issue of how the branch-line cusp 
singularities stem from the behavior of matrix elements between the $m>0$ ground states 
and specific classes of excited states is shortly discussed in Appendix \ref{D}. 
The dynamical theory refers to the thermodynamic limit, in which the matrix elements squares
$\vert\langle \nu\vert\hat{S}^{a}_k\vert GS\rangle\vert^2$ in Eq. (\ref{SDSF})
have in terms of the relative weights $a (m_{+1},\,m_{-1})$ and lowest peak weights $A^{(0,0)}$
defined in that Appendix the general form given in its Eq. (\ref{ME}).
The theory provides in Eq. (\ref{Aamm}) the dependence of such weights on
the $\iota =\pm1$ functionals $\Phi_{\iota}^2$ that control
the cusp singularities exponents.

Unfortunately, it does not provide the precise values 
of the $u$ and $m$ dependent constant $0<B_s\leq 1$ and 
$u$ dependent constants $0<f_l<1$ where $l=0,2,4$
in the $A^{(0,0)}$ expression under consideration. Those contribute to the
coefficients $C_{ab}^{\Delta}$ and $C_{ab}$, respectively, in 
the spin dynamical structure factors's 
analytical expressions, Eqs. (\ref{MPSsFMB}) and (\ref{MPSs}), which
are determined by the lowest peaks spectral weights.
In spite of this limitation, our results provide important
physical information on such factors.

The possible alternative use of form factors of the $\sigma =\uparrow,\downarrow$ 
electron creation and annihilation operators involved in the dynamical
structure factors studied in this paper remains an unsolved problem for 
the present 1D Hubbard model. 

When $\zeta^{ab}_{\beta} (k) = -1 + \sum_{\iota =\pm1}\Phi_{\iota}^2<0$, Eq. 
(\ref{expTS}), there are cusp singularities at and just above the corresponding 
$\beta$ branch lines. The form of the matrix elements expression, Eq. (\ref{ME}),
reveals both that the occurrence of cusp singularities is controlled
by the matrix elements $\langle \nu\vert\hat{S}^{a}_k\vert GS\rangle$ and
that $\vert\langle \nu\vert\hat{S}^{a}_k\vert GS\rangle\vert^2$ also diverges
in the case of the excited states that generate such singularities. This confirms that
there is a direct relation between the negativity of the exponents
$\zeta^{ab}_{\beta} (k)$ and the amount of spectral weight at
and just above the corresponding $\beta$ branch lines.

For simplicity, in this paper we have not provided further details of the dynamical theory 
that are common to those already given in Ref. \onlinecite{Carmelo_16}.
The form of both the relative weights and the lowest peak weights considered in the studies
of Ref. \onlinecite{Karlo_97} for the charge degrees of
freedom of the 1D Hubbard model for electronic
densities $n_e \in [0,1]$ at spin density $m=0$
is similar to that of the present relative weights $a (m_{+1},\,m_{-1})$ 
and lowest peak weights $A^{(0,0)}$ for the spin degrees of
freedom of the same model for spin densities $m \in [0,1]$ at electronic density $n_e = 1$.
Such studies consider the $u\rightarrow\infty$
limit in which for the dynamical correlation function 
under consideration the values of the lowest peak weights can
be calculated. The results of that reference confirm that
the cusp singularities correspond to $(k,\omega)$-plane regions with a larger
amount of spectral weight.

That the momentum-dependent exponents in  Eqs. (\ref{MPSsFMB}) and (\ref{MPSs})
and thus the corresponding matrix elements that control the spectral weights, Eq. (\ref{ME}), 
are little affected by decreasing the ratio $u=U/4t$ reveals
that in the present case of the spin dynamical structure factors
of the 1D Hubbard model's Mott-Hubbard insulating phase the relative spectral-weight contributions of different
types of excited energy eigenstates is little $u$ dependent.
This means that concerning that issue, results for the most known limit 
of small yet finite $t^2/U$ and thus large $u$ in which the present
quantum problem is equivalent to the spin-$1/2$ $XXX$ chain \cite{Kohno_09,Muller}
also apply to small and intermediate $u$ values. This applies
to the analysis presented in Sec. \ref{SECIII}, concerning the spectral weight in 
the gap regions being negligible in the present thermodynamic limit

Our results have focused on the contribution from spin $n$-string states. This refers 
to the line shape at and just above the $(k,\omega)$-plane
gapped lower threshold's spectra $\Delta_{\beta}^{ab} (k)$ where $ab = +-,xx,zz$
and $\beta$ refers to different branch lines. In well-defined $m$-dependent $k$ subintervals, 
Eqs. (\ref{Ds2})-(\ref{Ds2p}) and (\ref{Ds2pL})-(\ref{Ds2L}), such branch lines coincide 
with the gapped lower thresholds under consideration.
In these physically important $(k,\omega)$-plane regions, the spin dynamical 
structure factors $S^{ab} (k,\omega)$ have the general analytical expression provided in 
Eq. (\ref{MPSsFMB}). In the case of $S^{+-} (k,\omega)$ and $S^{xx} (k,\omega)$, such gapped 
lower thresholds refer to the $n$-string states's upper continua shown in the $(k,\omega)$-plane 
in Figs. \ref{figure1} and \ref{figure2} and \ref{figure3} and \ref{figure4}, respectively.

That as justified in Sec. \ref{SECIII} the spectral weight in the gap regions is negligible
in the present thermodynamic limit, is consistent with the amount of that weight existing 
just below the $(k,\omega)$-plane gapped lower thresholds of the $n$-string states's spectra shown in 
Figs. \ref{figure1}-\ref{figure6} being vanishingly small or negligible.
This is actually behind the validity at finite magnetic fields $0<h<h_c$ and
in the thermodynamic limit of the analytical expressions of the spin dynamical structure factors 
of general form, Eq. (\ref{MPSsFMB}), obtained in this paper. 
	
The momentum dependent exponents that control the spin dynamical structure factors's
line-shape in such expressions are given in Eq. (\ref{expG+-}) for $S^{+-} (k,\omega)$
and $S^{xx} (k,\omega)$ and in Eq. (\ref{exps2pL}) for $S^{zz} (k,\omega)$. In the former case,
the exponents associated with the $(k,\omega)$-plane vicinity of
the $s2-$, $\bar{s}'-$, $\bar{s}-$, and $s2'$-branch lines are plotted in Figs. \ref{figure7}-\ref{figure10}. 
Such lines refer to different $k$ intervals of the gapped lower threshold of the $n$-string states's spectra of 
$S^{+-} (k,\omega)$ and $S^{xx} (k,\omega)$. The solid lines in Figs. \ref{figure1} and \ref{figure2} and 
\ref{figure3} and \ref{figure4} that belong to that gapped lower threshold correspond to $k$ intervals for which the exponents
are negative. In them, singularities occur in the spin dynamical structure factors's expression,
Eq. (\ref{MPSsFMB}), at and above the gapped lower thresholds.

In the case of $S^{xx} (k,\omega)$, the expression given in that equation does not apply for
small spin densities in the ranges and corresponding $k$ intervals given in Eqs. \ref{gapineq} and \ref{barmu0}. 
For these spin-density ranges and momentum intervals, there is overlap between the lower 
continuum and upper $n$-string states's continuum, as shown in
Figs. \ref{figure3} (a-c). 

However, consistently with the perturbative arguments provided in Appendix \ref{D}
in terms of the number of elementary processes associated with annihilation of one $s$ particle,
the contribution to $S^{zz} (k,\omega)$ from excited states populated by $n$-strings
is much weaker than for $S^{+-} (k,\omega)$ and $S^{xx} (k,\omega)$ and
is negligible in the case of $S^{-+} (k,\omega)$. In the case of $S^{zz} (k,\omega)$
it does not lead to a $(k,\omega)$-plane continuum. The gapped lower threshold of such states
is shown in Figs. \ref{figure5} and \ref{figure6}. There the $k$ subinterval associated with the $\beta=\bar{s}'$ branch line 
is the only one at and above which there are singularities. We have found that
out of the four branch-line's exponents whose expressions are provided in Eq. (\ref{exps2pL}),
only that of the $\beta=\bar{s}'$ branch line is indeed negative. That line is represented in the gapped lower threshold 
of $S^{zz} (k,\omega)$ shown in Figs. \ref{figure5} (a) - \ref{figure5} (c) by a solid (green) line. The 
corresponding exponent is plotted in Fig. \ref{figure11}. 

That line's $k$ subinterval is though small. 
Its momentum width decreases upon decreasing $u$ and/or increasing 
the spin density within the range $0<m\leq\tilde{m}$. Here $\tilde{m}$ increases from
$\tilde{m}=0$ for $u\rightarrow 0$ to $\tilde{m}\approx 0.317$ for large $u$. 
For spin densities $\tilde{m}\leq m<1$, that line is not part of the gapped lower threshold, so that
the contribution to $S^{zz} (k,\omega)$ from excited states populated by $n$-strings 
becomes negligible. Consistent, in Figs. \ref{figure5} (d) - \ref{figure5} (f) and \ref{figure6} that
line is lacking.

To provide an overall physical picture that includes the
relative $(k,\omega)$-plane location of all spectra with a significant amount
of spectral weight, we also accounted for the contributions from all
types of excited energy eigenstates that lead to gapped and gapless lower threshold singularities in the spin dynamical structure factors.
This includes excited energy eigenstates described only by real Bethe-ansatz rapidities
and thus without $n$-strings, which are known to lead to most spectral weight
of the sum rules, Eq. (\ref{SRDSF}). Their contribution to 
$S^{+-} (k,\omega)$, $S^{xx} (k,\omega)$, and  $S^{zz} (k,\omega)$ leads to
the $(k,\omega)$-plane lower continua shown in 
Figs. \ref{figure1} and \ref{figure2}, \ref{figure3} and \ref{figure4}, and \ref{figure5} and \ref{figure6}, respectively.

\subsection{Concluding remarks}
\label{SECVIB}

Spin $n$-strings have been identified in experimental studies of CuCl$_2$$\cdot$2N(C$_5$D$_5$) 
and Cu(C$_4$H$_4$N$_2$)(NO$_3$)$_2$ \cite{Kohno_09,Stone_03,Heilmann_78}. 
In this paper the contribution of spin $n$-strings to the spin dynamical structure factors of 
the 1D fermionic Hubbard model with one electron per site in a magnetic field 
has been studied. That model describes a 1D Mott-Hubbard insulator. 

1D Mott-Hubbard insulators are a paradigm for the importance of strong 
correlations and are known to exhibit a wide variety of unusual physical phenomena. 
For instance, while in the 1D Hubbard metallic phase increasing the onsite repulsion $U$
reduces the lattice distortion, in its Mott-Hubbard insulating phase Coulomb correlations enhance 
the lattice dimerization \cite{Baeriswyl}. 1D Mott-Hubbard insulators can be studied within 
condensed matter by inelastic neutron scattering in spin chains such as for instance chain cuprates, as well as a 
number of quasi-1D organic compounds \cite{Kohno_09,Stone_03,Pollet}. 

The theoretical description of the spin degrees of freedom of some of such condensed-matter systems is commonly 
modeled by the spin-$1/2$ $XXX$ antiferromagnet \cite{Kohno_09,Stone_03}. As justified in the following,
our study indicates that the 1D Hubbard model with one electron per site can alternatively be used to describe 
the spin dynamical properties of such systems. 

Analysis of the spin dynamical structure factors spectra plotted in
the $(k,\omega)$ plane in Figs. \ref{figure1}-\ref{figure6}, reveals that the only effect of decreasing
the ratio $u=U/4t$ is to increase such spectra energy bandwidths. (Within the isotropic spin-$1/2$ $XXX$ chain,
this can be achieved by increasing the exchange integral $J$.)

It is somehow surprising that the 1D Hubbard model with one electron per site for $u=15$, which is
equivalent to a isotropic spin-$1/2$ $XXX$ chain with $J=4t^2/U$, and the former model for $u=0.4$ and $u=1.0$, 
lead to spin dynamical structure factors's spectra that except for their energy bandwidth have
basically the same form.

However, the type of momentum dependences of the exponents plotted in Figs. \ref{figure7}-\ref{figure13} 
that control the $(k,\omega)$-plane line shape of the spin dynamical structure factors in the vicinity of the 
singularities located in the gapped lower thresholds of the spin $n$-string states's spectra and lower thresholds 
of the lower continua represented in Figs. \ref{figure1}-\ref{figure6} is not affected by decreasing $u$. 

That as found in this paper the main effect of increasing $t$ at constant $U$ and thus
decreasing the ratio $u=U/4t$ is on the energy bandwidth of the corresponding relevant spectra
is an important information about how in 1D Mott-Hubbard insulators electron itinerancy 
associated in the present model with the transfer integral $t$ affects the spin dynamics.

This seems to confirm that concerning the 
spin dynamical properties of spin chain compounds in a magnetic field, both 
the 1D Hubbard model with one electron per site and the spin-$1/2$ $XXX$ antiferromagnet
are suitable model candidates. Consistent, for general Mott-Hubbard insulating materials there is no reason for the
on-site repulsion to be much stronger than the electron hopping amplitude $t$. This situation is realized in the Bechgaard 
salts \cite{Pollet}. 

Since the dynamical theory used in our study for the whole $u>0$ range and the thermodynamic limit
only provides the line shape at and near the cusp singularities located
at the gapped lower thresholds and lower thresholds, it cannot access
other possible peaks, as for instance those due to the Brillouin-zone folding effect.
However and as discussed in Sec. \ref{SECVIA}, results for the most known limit 
of small yet finite $t^2/U$ and thus large $u$ in which the present
quantum problem is equivalent to the spin-$1/2$ $XXX$ chain \cite{Kohno_09}
also apply to small and intermediate $u$ values provided that the 
spectral features energy bandwidths are suitably rescaled. Hence one can at least
confirm that the cusp singularities located at the gapped lower thresholds and lower 
thresholds predicted by the half-filled 1D Hubbard model are observable
in neutron scattering experiments.

In such experiments, the quantity that is observed is proportional to,
\begin{equation}
S^{av} (k,\omega) = {1\over 6}\left(S^{-+} (k,\omega) + S^{+-} (k,\omega)  + 4S^{zz} (k,\omega)\right) \, .
\label{Sav}
\end{equation}
Upon superposition of the spectra of the spin dynamical structure factors on the right-hand
side of this equation, we have checked that {\it all} cusp singularities at and near
both the gapped lower thresholds and lower thresholds found in this paper for the
1D Hubbard model at any of the $u$ values $u=0.4$, $u=0.1$, and $u = 15.0$ correspond
to peaks shown in Fig. 4 of Ref. \onlinecite{Kohno_09} for CuCl$_2$$\cdot$2N(C$_5$D$_5$) 
and in Fig. 5 of that reference for Cu(C$_4$H$_4$N$_2$)(NO$_3$)$_2$ at the finite values
of the magnetic field considered in these figures and suitable transfer integral $t$ values,
to reach agreement with the corresponding energy bandwidths. This should obviously apply to $u=15.0$
at which large $u$ value the spin degrees of freedom of the present model are described  
by the spin-$1/2$ $XXX$ chain (with exchange integral $J=4t^2/U=t/u$) 
used in the studies of Ref. \onlinecite{Kohno_09} to theoretically access the cusp singularities
under consideration.

That such a correspondence also applies to $u=0.4$ and $u=1.0$ is justified by the results of this paper
according to which: The dependence on $u$ of the momentum dependence 
of the negative exponents that control the spin dynamical structure factors's line shape 
is rather weak; The main effect of decreasing $u$ on
such factors's spectra is merely to increase their energy bandwidth.

The dynamical theory used in our study provides analytical
expressions of the spin dynamical structure factors at and just above the
$(k,\omega)$-plane gapped lower thresholds and lower thresholds of their spectra
with more spectral weight. The use of other methods such as
the time-dependent density matrix renormalization group \cite{White,Schollwock,Moreno}
to obtain the line shape of such dynamical functions over other
$(k,\omega)$-plane regions would provide valuable complementary information. 

In the case of 1D Mott-Hubbard insulators, the apparent independence on the $u$ values of the spin dynamics
found in this paper, suggests that the suitable values of the interaction for such systems are rather 
settled by the agreement with experimental results
on the charge dynamics and one-particle spectral function at energy scales above the Mott-Hubbard gap. 
 
\acknowledgements
J. M. P. C. thanks the Boston University's Condensed Matter Theory Visitors Program for support and
Boston University for hospitality during the initial period of this research. He acknowledges the support from
FCT through the Grants No. PTDC/FIS-MAC/29291/2017 and No. SFRH/BSAB/142925/2018.
J. M. P. C. and T. \v{C}. thank Pedro D. Sacramento for illuminating discussions and they
acknowledge the support from FCT through the Grant No. UID/FIS/04650/2013.
T. \v{C}. gratefully acknowledges the support by the Institute for Basic Science in Korea (Project No. IBS-R024-D1).
J. M. P. C. and T. \v{C}. contributed equally to this work.


\appendix

\section{Useful selection rules and sum rules}
\label{A}

Let $\vert S,\alpha\rangle$, $\vert S^z,\beta\rangle$, and $\vert S,S^z,\gamma\rangle$
denote energy eigenstates where $S \in [0,N/2]$ is their spin, $S^z$ their spin projection, and
$\alpha$, $\beta$ and $\gamma$ represent all other quantum numbers needed to uniquely specify these 
states, respectively. The selection rules given in the following are derived from the properties
of the operators $\hat{S}^{z}_k$ and $\hat{S}^{\pm}_k$ by straightforward manipulations involving
their operator algebra \cite{Muller}.

At vanishing magnetic field, $h=0$, the following selection rules hold in the thermodynamic limit,
\begin{eqnarray}
\langle S,\alpha\vert\hat{S}^a_k\vert S'\alpha'\rangle & = & 0 
\hspace{0.20cm}{\rm for}\hspace{0.20cm}S=S'=0\hspace{0.20cm}{\rm and}\hspace{0.20cm}
a = z, \pm
\nonumber \\
\langle S,\alpha\vert\hat{S}^a_k\vert S'\alpha'\rangle & = & 0 
\hspace{0.20cm}{\rm for}\hspace{0.20cm}\vert S-S'\vert \neq 0,1 \hspace{0.20cm}{\rm and}\hspace{0.20cm}
a = z, \pm
\nonumber \\
\langle S^z,\beta\vert\hat{S}^{\pm}_k\vert S^{z'},\beta'\rangle & = & 0 
\hspace{0.20cm}{\rm for}\hspace{0.20cm}S^{z'}\neq S^z \pm 1
\nonumber \\
\langle S^z,\beta\vert\hat{S}^z_k\vert S^{z'},\beta'\rangle & = & 0 
\hspace{0.20cm}{\rm for}\hspace{0.20cm}S^{z'}\neq S^z \, .
\label{SRh0}
\end{eqnarray}

However, for finite magnetic fields $0<h<h_c$ the following selection rules 
are valid in that limit,
\begin{eqnarray}
\langle S,S,\gamma\vert\hat{S}^{\pm}_k\vert S',S^{z'},\gamma'\rangle & = & 0 
\nonumber \\
{\rm for} && S'\neq S\pm 1 \hspace{0.20cm}{\rm and}\hspace{0.20cm}S^{z'}\neq S\pm 1
\nonumber \\
\langle S,S,\gamma\vert\hat{S}^z_k\vert S',S^{z'},\gamma'\rangle & = & 0 
\nonumber \\
{\rm for} && S'\neq S\hspace{0.20cm}{\rm and}\hspace{0.20cm}S^{z'}\neq S \, .
\label{SRhfinite}
\end{eqnarray}

Finally, the dynamical structure factors satisfy the following sum rules,
\begin{eqnarray}
{1\over 2\pi^2}\int_{-\pi}^{\pi}dk\int_{0}^{\infty}d\omega
\,S^{+-} (k,\omega) & = & (1+m)
\nonumber \\
{1\over 2\pi^2}\int_{-\pi}^{\pi}dk\int_{0}^{\infty}d\omega
\,S^{-+} (k,\omega) & = & (1-m) 
\nonumber \\
{1\over 2\pi^2}\int_{-\pi}^{\pi}dk\int_{0}^{\infty}d\omega
\,S^{zz} (k,\omega) & = & {1\over 2}(1-m^2) \,  .
\label{SRDSF}
\end{eqnarray}

\section{Gapless transverse and longitudinal continuum spectra}
\label{B}

Within a $k$ extended zone scheme, the $S^{-+} (k,\omega)$'s spectrum $\omega^{-+} (k)$ and 
the $S^{+-} (k,\omega)$'s spectrum $\omega^{+-} (k)$ associated with the lower continuum in 
Figs. \ref{figure1} and \ref{figure2} read,
\begin{eqnarray}
& & \omega^{-+} (k) = - \varepsilon_{s} (q_1) - \varepsilon_{s} (q_2)
\nonumber \\
& & {\rm where}\hspace{0.20cm} k = \iota\pi - q_1 - q_2 \hspace{0.20cm}
{\rm and}\hspace{0.20cm}\iota = \pm 1
\nonumber \\
& & {\rm for}\hspace{0.20cm}q_1 \in [-k_{F\downarrow},k_{F\downarrow}]
\hspace{0.20cm}{\rm and}\hspace{0.20cm}q_2 \in [-k_{F\downarrow},k_{F\downarrow}] \, ,
\label{dkEdPxxMP}
\end{eqnarray}
and
\begin{eqnarray}
& & \omega^{+-} (k) = \varepsilon_{s} (q_1) - \varepsilon_{s} (q_2) 
\nonumber \\
& & {\rm where}\hspace{0.20cm} k = \iota\pi + q_1 - q_2 \hspace{0.20cm}
{\rm and}\hspace{0.20cm}\iota = \pm 1
\nonumber \\
& & {\rm for}\hspace{0.20cm}\vert q_1\vert \in [k_{F\downarrow},k_{F\uparrow}] 
\hspace{0.20cm}{\rm and}\hspace{0.20cm}q_2 \in [-k_{F\downarrow},k_{F\downarrow}] \, ,
\label{dkEdPxxPM}
\end{eqnarray}
respectively. Here $\varepsilon_{s} (q)$ is the $s$ band energy dispersion given in Eq. (\ref{equA4}).

The spectrum $\omega^{xx} (k)$ of the transverse dynamical structure factor $S^{xx} (k,\omega)$ 
associated with the lower continuum in Figs. \ref{figure3} and \ref{figure4}
results from combination of the two spectra $\omega^{-+} (k)$ and $\omega^{+-} (k)$
in Eqs. (\ref{dkEdPxxMP}) and (\ref{dkEdPxxPM}), respectively.

However, the spectrum $\omega^{zz} (k)$ associated with the lower continuum in 
Figs. \ref{figure5} and \ref{figure6} is given by,
\begin{eqnarray}
& & \omega^{zz} (k) = \varepsilon_{s} (q_1) - \varepsilon_{s} (q_2) 
\nonumber \\
& & {\rm where}\hspace{0.20cm} k = q_1 - q_2
\nonumber \\
& & {\rm for}\hspace{0.20cm}\vert q_1\vert \in [k_{F\downarrow},k_{F\uparrow}] 
\hspace{0.20cm}{\rm and}\hspace{0.20cm}q_2 \in [-k_{F\downarrow},k_{F\downarrow}] \, .
\label{dkEdPl}
\end{eqnarray}

The upper thresholds of the two-parametric spectra, Eqs. (\ref{dkEdPxxMP}) and (\ref{dkEdPxxPM}),
have the following one-parametric spectra for spin densities $m \in ]0,1[$,
\begin{eqnarray}
\omega^{+-}_{ut} (k) & = & 2\mu_B\,h - \varepsilon_{s} (k_{F\downarrow}-k) 
\hspace{0.20cm}{\rm where}\hspace{0.20cm} k = k_{F\downarrow} - q
\nonumber \\
&& {\rm for}\hspace{0.20cm}k\in [0,k_{F\downarrow}] 
\hspace{0.20cm}{\rm and}\hspace{0.20cm}q \in [0,k_{F\downarrow}] \, ,
\nonumber \\
& = & \varepsilon_{s} (q_1) - \varepsilon_{s} (q_2) 
\hspace{0.20cm}{\rm where}\hspace{0.20cm} k = \pi + q_1 - q_2
\nonumber \\
&& {\rm for}\hspace{0.20cm}k\in [k_{F\downarrow},\pi] 
\hspace{0.20cm}{\rm and}\hspace{0.20cm}v_s (q_1) = v_s (q_2)\hspace{0.20cm}
\nonumber \\
&& {\rm with} \hspace{0.20cm}q_1 \in [-k_{F\uparrow},-k_{F\downarrow}]
\nonumber \\
&& {\rm and}\hspace{0.20cm}
q_2 \in [-k_{F\downarrow},0] \, ,
\label{Omxxut1}
\end{eqnarray}
and
\begin{eqnarray}
\omega^{-+}_{ut} (k) & = & -2\varepsilon_{s} \left({\pi - k\over 2}\right) 
\hspace{0.20cm}{\rm where}\hspace{0.20cm} k = \pi - 2q
\nonumber \\
&& {\rm for}\hspace{0.20cm}k\in [(k_{F\uparrow}-k_{F\downarrow}),\pi] 
\nonumber \\
&& {\rm and}\hspace{0.20cm}q \in [-k_{F\downarrow},0] \, ,
\label{Omxxut2}
\end{eqnarray}
respectively. 

The upper threshold spectrum $\omega^{xx}_{ut} (k)$ of the combined spectra, Eqs. (\ref{dkEdPxxMP}) and (\ref{dkEdPxxPM}),
is given by,
\begin{eqnarray}
\omega^{xx}_{ut} (k) & = & \omega^{+-}_{ut} (k)\hspace{0.20cm}{\rm for}\hspace{0.20cm} k \in [0,k^{xx}_{ut}]
\nonumber \\
& = & \omega^{-+}_{ut} (k)\hspace{0.20cm}{\rm for}\hspace{0.20cm} k \in [k^{xx}_{ut},\pi] \, ,
\label{Omxxutxx}
\end{eqnarray}
where the momentum $k^{xx}_{ut}$ is such that $\omega^{+-}_{ut} (k^{xx}_{ut}) = \omega^{-+}_{ut} (k^{xx}_{ut})$.

However, the one-parametric upper threshold spectrum associated
with the two-parametric longitudinal spectrum, Eq. (\ref{dkEdPl}), 
reads for $m \in ]0,1[$,
\begin{eqnarray}
\omega^{zz}_{ut} (k) & = & \varepsilon_{s} (q_1) - \varepsilon_{s} (q_2) 
\hspace{0.20cm}{\rm where}\hspace{0.20cm} k = q_1 - q_2
\nonumber \\
&& {\rm for}\hspace{0.20cm}v_s (q_1) = v_s (q_2)\hspace{0.20cm}
{\rm and}\hspace{0.20cm}k\in [0,k_{F\uparrow}]\hspace{0.20cm}{\rm with}
\nonumber \\
&& q_1 \in [k_{F\downarrow},k_{F\uparrow}]\hspace{0.20cm}{\rm and}\hspace{0.20cm}
q_2 \in [0,k_{F\downarrow}] \, ,
\nonumber \\
& = & 2\mu_B\,h - \varepsilon_{s} (k_{F\uparrow}-k) 
\hspace{0.20cm}{\rm where}\hspace{0.20cm} k = k_{F\uparrow} - q
\nonumber \\
&& {\rm for}\hspace{0.20cm}k\in [k_{F\uparrow},\pi] 
\hspace{0.20cm}{\rm and}\hspace{0.20cm}q \in [-k_{F\downarrow},0] \, .
\label{Omlut}
\end{eqnarray}

At $k=0,k_{F\downarrow},\pi$ and $k=0,k_{F\uparrow}-k_{F\downarrow},\pi$,
the upper threshold spectra, Eqs. (\ref{Omxxut1}) and (\ref{Omxxut2}), respectively, are given by,
\begin{eqnarray}
\omega^{+-}_{ut} (0) & = & W_s^h = 2\mu_B\,h 
\nonumber \\
\omega^{+-}_{ut} (k_{F\downarrow}) & = & W_s = 2\mu_B\,h + W_s^p
\nonumber \\
\omega^{+-}_{ut} (\pi) & = & 0 
\nonumber \\
\omega^{-+}_{ut} (k_{F\uparrow}-k_{F\downarrow}) & = & 0
\nonumber \\
\omega^{-+}_{ut} (\pi) & = & 2W_s^p \, .
\label{Omxxutlim}
\end{eqnarray}

At $k=0,k_{F\uparrow},\pi$ the upper threshold spectrum $\omega^{zz}_{ut} (k)$ reads,
\begin{eqnarray}
\omega^{zz}_{ut} (0) & = & 0
\nonumber \\
\omega^{zz}_{ut} (k_{F\uparrow}) & = & W_s = 2\mu_B\,h + W_s^p
\nonumber \\
\omega^{zz}_{ut} (\pi) & = & W_s^h = 2\mu_B\,h  \, .
\label{Omlutlim}
\end{eqnarray}

The energy scales $W_s=W_s^p+W_s^h$, $W_s^p$, and $W_s^h$ are in the
above equations the $s$ band energy width, the $s$ particle energy bandwidth,
and the $s$ hole energy bandwidth defined by Eqs. (\ref{vares2limits})-(\ref{W2um1}).

The dynamical theory used in our study provides the spin
dynamical structure factors's line shape near the lower thresholds
of the spectra, Eqs. (\ref{dkEdPxxMP}), (\ref{dkEdPxxPM}), and (\ref{dkEdPl}).
In the case of (i) $S^{-+} (k,\omega)$ and (ii) $S^{+-} (k,\omega)$ and $S^{zz} (k,\omega)$
such lower thresholds refer to (i) a single $s$ branch line and 
(ii) two sections of a $s$ branch line, respectively.

These lower thresholds spectra can be expressed in terms of the excitation
momentum $k$ or of the $s$ band momentum $q$ and are given by,
\begin{eqnarray}
\omega^{-+}_{lt} (k) & = & - \varepsilon_s (k_{F\uparrow}-k) \hspace{0.2cm}{\rm and}
\nonumber \\
k & = & k_{F\uparrow} - q\hspace{0.2cm}{\rm where}
\nonumber \\
k & \in & ](k_{F\uparrow}-k_{F\downarrow}),\pi[ 
\hspace{0.2cm}{\rm and}
\nonumber \\
q & \in & ]-k_{F\downarrow},k_{F\downarrow}[ \, ,
\label{OkPMRs}
\end{eqnarray}
\begin{eqnarray}
\omega^{+-}_{lt} (k) & = & \varepsilon_s (k - k_{F\uparrow}) \hspace{0.2cm}{\rm and}
\nonumber \\
k & = & k_{F\uparrow} + q\hspace{0.2cm}{\rm where}
\nonumber \\
k & \in & ]0, (k_{F\uparrow}-k_{F\downarrow})[ 
\hspace{0.2cm}{\rm and}
\nonumber \\
q & \in & ]-k_{F\uparrow},-k_{F\downarrow}[ \, ,
\label{OkMPRs}
\end{eqnarray}
\begin{eqnarray}
\omega^{+-}_{lt} (k) & = & - \varepsilon_s (k_{F\uparrow}-k) \hspace{0.2cm}{\rm and}
\nonumber \\
k & = & k_{F\uparrow} - q\hspace{0.2cm}{\rm where}
\nonumber \\
k & \in & ](k_{F\uparrow}-k_{F\downarrow}),\pi[ 
\hspace{0.2cm}{\rm and}
\nonumber \\
q & \in & ]-k_{F\downarrow},k_{F\downarrow}[ \, ,
\label{OkMPRs2}
\end{eqnarray}
\begin{eqnarray}
\omega^{zz}_{lt} (k) & = & - \varepsilon_s -k_{F\downarrow}[_{F\downarrow} - k) \hspace{0.2cm}{\rm and}
\nonumber \\
k & = & k_{F\downarrow} - q 
\hspace{0.2cm}{\rm where}
\nonumber \\
k & \in &]0,2k_{F\downarrow}[\hspace{0.2cm}{\rm and}
\nonumber \\
q & \in & ]-k_{F\downarrow},k_{F\downarrow}[ \, ,
\label{OkPMRsL}
\end{eqnarray}
\begin{eqnarray}
\omega^{zz}_{lt} (k) & = & \varepsilon_s (k - k_{F\downarrow}) \hspace{0.2cm}{\rm and}
\nonumber \\
k & = & k_{F\downarrow} + q \hspace{0.2cm}{\rm where}
\nonumber \\
k & \in &]2k_{F\downarrow}),\pi[
\hspace{0.2cm}{\rm and}
\nonumber \\
q & \in &]k_{F\downarrow},k_{F\uparrow}[ \, .
\label{OkMPRsL}
\end{eqnarray}

\section{Energy gaps's expressions and limiting values}
\label{C}

In this Appendix, the expressions in terms of the $s$ and $s2$ bands energy dispersions and
limiting values of the energy gaps $\Delta_{\rm gap}^{+-} (k)$, Eq. (\ref{gapPMMP}),
$\Delta_{\rm gap}^{xx} (k)$, Eq. (\ref{gap}), and
$\Delta_{\rm gap}^{zz} (k)$, Eq. (\ref{gapL}), and their values
at some specific momenta are provided.

For $m\in [0,\tilde{m}]$ the energy gap $\Delta_{\rm gap}^{+-} (k)$ reads,
\begin{eqnarray}
\Delta_{\rm gap}^{+-} (k) & = & - 2\mu_B\,h + \varepsilon_{s2} (k) + \varepsilon_{s} (k_{F\downarrow}-k) 
\nonumber \\
& & {\rm for}\hspace{0.20cm}k\in ]0,(k_{F\uparrow}-k_{F\downarrow})[
\nonumber \\
\Delta_{\rm gap}^{+-} (k) & = & 2\mu_B\,h - \varepsilon_{s} (k_{F\uparrow}-k) + \varepsilon_{s} (k_{F\downarrow}-k) 
\nonumber \\
& & {\rm for}\hspace{0.20cm}](k_{F\uparrow} - k_{F\downarrow}),{\tilde{k}}[
\nonumber \\
\Delta_{\rm gap}^{+-} (k) & = & 2\mu_B\,h - W_{s2}\hspace{0.20cm}
{\rm for}\hspace{0.20cm}k\in ]{\tilde{k}},k_{F\downarrow}[
\nonumber \\
\Delta_{\rm gap}^{+-} (k) & = &  4\mu_B\,h - W_{s2} - \varepsilon_{s} (k_{F\downarrow}-k) 
\nonumber \\
& & + \varepsilon_{s} (q) - \varepsilon_{s} (k + q - \pi) 
\nonumber \\
& & {\rm for}\hspace{0.20cm}k\in ]k_{F\downarrow},2k_{F\downarrow}[ \hspace{0.20cm}{\rm and}
\nonumber \\
& & q \in ]-(k_{\bullet} - k_{F\uparrow} + k_{F\downarrow}),0[
\nonumber \\
& & q_1 =  k + q - \pi \in ]-k_{F\uparrow},-k_{\bullet}[
\nonumber \\
\Delta_{\rm gap}^{+-} (k) & = & \varepsilon_{s2} (k - 2k_{F\downarrow}) + \varepsilon_{s} (q) - \varepsilon_{s} (k + q - \pi) 
\nonumber \\
& & {\rm for}\hspace{0.20cm}k\in ]2k_{F\downarrow},\pi[ \hspace{0.20cm} {\rm and}
\nonumber \\
& & q \in ]-k_{F\downarrow},- (k_{\bullet} - k_{F\uparrow} + k_{F\downarrow})[
\nonumber \\
& & q_1 =  k + q - \pi \in ]- k_{\bullet},-k_{F\downarrow}[
\nonumber \\
& & {\rm for}\hspace{0.20cm}{\rm spin}\hspace{0.20cm}{\rm densities}\hspace{0.20cm}m\in [0,\tilde{m}] \, .
\label{gapEXPm03}
\end{eqnarray}
For spin densities $m\in [\tilde{m},1[$ its expression is,
\begin{eqnarray}
\Delta_{\rm gap}^{+-} (k) & = & - 2\mu_B\,h + \varepsilon_{s2} (k)  + \varepsilon_{s} (k_{F\downarrow}-k) 
\nonumber \\
& & {\rm for}\hspace{0.20cm}k\in ]0,{\tilde{k}}[
\nonumber \\
\Delta_{\rm gap}^{+-} (k) & = & 2\mu_B\,h - W_{s2}\hspace{0.20cm}
{\rm for}\hspace{0.20cm}k\in ]{\tilde{k}},k_{F\downarrow}[
\nonumber \\
\Delta_{\rm gap}^{+-} (k) & = &  4\mu_B\,h - W_{s2} - \varepsilon_{s} (k_{F\downarrow}-k) 
\nonumber \\
& & + \varepsilon_{s} (q) - \varepsilon_{s} (k + q - \pi) 
\nonumber \\
& & {\rm for}\hspace{0.20cm}k\in ]k_{F\downarrow},2k_{F\downarrow}[ \hspace{0.20cm}{\rm and}
\nonumber \\
& & q \in ]-(k_{\bullet} - k_{F\uparrow} + k_{F\downarrow}),0[
\nonumber \\
& & q_1 =  k + q - \pi \in ]-k_{F\uparrow},-k_{\bullet}[
\nonumber \\
\Delta_{\rm gap}^{+-} (k) & = & \varepsilon_{s2} (k - 2k_{F\downarrow}) + \varepsilon_{s} (q) - \varepsilon_{s} (k + q - \pi) 
\nonumber \\
& & {\rm for}\hspace{0.20cm}k\in ]2k_{F\downarrow},\pi[ \hspace{0.20cm} {\rm and}
\nonumber \\
& & q \in ]-k_{F\downarrow},- (k_{\bullet} - k_{F\uparrow} + k_{F\downarrow})[
\nonumber \\
& & q_1 =  k + q - \pi \in ]-k_{\bullet},-k_{F\downarrow}[
\nonumber \\
& & {\rm for}\hspace{0.20cm}{\rm spin}\hspace{0.20cm}{\rm densities}\hspace{0.20cm}m\in [\tilde{m},1[ \, .
\label{gapEXPm31}
\end{eqnarray}

The momentum $k_{\bullet}$ appearing in the above equations satisfies the following equation,
expressed in terms of the $s$ band group velocity defined in Eq. (\ref{equA4B}),
\begin{equation}
v_s (k_{\bullet}) = v_s (k_{\bullet} - k_{F\uparrow} + k_{F\downarrow})
\hspace{0.20cm}{\rm where}\hspace{0.20cm}k_{\bullet}>k_{F\downarrow} \, .
\label{kbullet}
\end{equation}
(The limiting behaviors of the $s$ band group velocity are given in Eqs. (\ref{vsu0}), (\ref{vvu0}), (\ref{varepsilonsulm0}), (\ref{varepsilonsulm1}), and (\ref{vvm1}).)

The energy gap $\Delta_{\rm gap}^{+-} (k)$ is given by $2\mu_B\,h - W_{s2}$ 
for the following $k$ values and spin densities,
\begin{eqnarray}
\Delta_{\rm gap}^{+-} (k) & = & 2\mu_B\,h - W_{s2}
\nonumber \\
k & = & 0 \hspace{0.20cm}{\rm for}\hspace{0.20cm}m\in ]0,1[
\nonumber \\
k & = & k_{F\uparrow} - k_{F\downarrow}  \hspace{0.20cm}{\rm for}\hspace{0.20cm}m\in [0,1/3]
\nonumber \\
k & \in & ]{\tilde{k}},k_{F\downarrow}[ \hspace{0.20cm}{\rm for}\hspace{0.20cm}m\in [0,\tilde{m}]
\nonumber \\
k & \in & ]{\tilde{k}},k_{F\downarrow}[ \hspace{0.20cm}{\rm for}\hspace{0.20cm}m\in [\tilde{m},1[ \, .
\label{gapEXPm03LIM}
\end{eqnarray}
Here $W_{s2}$ is the $s2$ band energy width. From the use of results given in Appendix \ref{E}, one finds that
the energy scale $2\mu_B\,h - W_{s2}\geq 0$ in Eq. (\ref{gapEXPm03LIM}) has the following limiting values,
\begin{eqnarray}
\lim_{u\rightarrow 0}\,(2\mu_B\,h - W_{s2}) & = & 0\hspace{0.20cm}{\rm for}\hspace{0.20cm}m\in ]0,1[
\nonumber \\
\lim_{m\rightarrow 0}\,(2\mu_B\,h - W_{s2}) & = & 0\hspace{0.20cm}{\rm for}\hspace{0.20cm}u>0
\nonumber \\
\lim_{m\rightarrow 1}\,(2\mu_B\,h - W_{s2}) & = & U - (\sqrt{(4t)^2+(2U)^2} 
\nonumber \\
&& - \sqrt{(4t)^2+U^2})>0
\hspace{0.20cm}{\rm for}\hspace{0.20cm}u>0
\nonumber \\
& \approx & U\hspace{0.20cm}{\rm for}\hspace{0.20cm}u\ll 1
\nonumber \\
& \approx & {t\over u} = {4t^2\over U}\hspace{0.20cm}{\rm for}\hspace{0.20cm}u\gg 1 \, .
\label{gapOtherLIM}
\end{eqnarray}

At $k=\pi$ (that in the spectra expressions means the $k\rightarrow \pi$ limit) the present gap reads,
\begin{eqnarray}
\Delta_{\rm gap}^{+-} (\pi) & = & 4\mu_B\,h
\nonumber \\
& & {\rm for}\hspace{0.20cm}m\in ]0,1[\hspace{0.20cm}{\rm and}\hspace{0.20cm}u>0 \, .
\label{gapEXPm03LIMB}
\end{eqnarray}
This expression has the following limiting values,
\begin{eqnarray}
\lim_{u\rightarrow 0}\,\Delta_{\rm gap}^{+-} (\pi) & = & 8t\sin\left({\pi\over 2}m\right)\hspace{0.20cm}{\rm for}\hspace{0.20cm}m\in ]0,1[
\nonumber \\
\lim_{m\rightarrow 0}\,\Delta_{\rm gap}^{+-} (\pi) & = & 0\hspace{0.20cm}{\rm for}\hspace{0.20cm}u>0
\nonumber \\
\lim_{m\rightarrow 1}\,\Delta_{\rm gap}^{+-} (\pi) & = & \sqrt{(4t)^2+(U)^2} - U >0
\hspace{0.20cm}{\rm for}\hspace{0.20cm}u>0
\nonumber \\
& \approx & 4t - U\hspace{0.20cm}{\rm for}\hspace{0.20cm}u\ll 1
\nonumber \\
& \approx & {2t\over u} = {4t^2\over U}\hspace{0.20cm}{\rm for}\hspace{0.20cm}u\gg 1 \, .
\label{gappiOtherLIM}
\end{eqnarray}

The energy gap $\Delta_{\rm gap}^{xx} (k)$, Eqs. (\ref{gapPMMP}) and (\ref{gap}), 
can be expressed as,
\begin{eqnarray}
\Delta_{\rm gap}^{xx} (k) & = & \Delta_{\rm gap}^{+-} (k)
\hspace{0.20cm}{\rm for}\hspace{0.20cm}k\in ]0,k^{xx}_{ut}[
\nonumber \\
\Delta_{\rm gap}^{xx} (k) & = & \Delta_{\rm gap}^{-+} (k)
\hspace{0.20cm}{\rm for}\hspace{0.20cm}k\in ]k^{xx}_{ut},\pi[ \, ,
\label{gapxx}
\end{eqnarray}
where $k^{xx}_{ut}>0$ is the $k$ value at which 
$\omega^{-+}_{ut} (k^{xx}_{ut}) = \omega^{+-}_{ut} (k^{xx}_{ut})$ and,
\begin{equation}
\Delta_{\rm gap}^{-+} (k) = \Delta^{-+} (k) - \omega^{-+}_{ut} (k) \, .
\label{gap-+}
\end{equation}
The gapped lower threshold spectrum $\Delta^{-+} (k)$ in this expression obeys
the equality $\Delta^{-+} (k)=\Delta^{+-} (k)$, where
$\Delta^{+-} (k)$ is given in Eqs. (\ref{Dxx03})-(\ref{Dxx31}).

For spin densities  $m\in [0,\tilde{m}]$, the energy gap
$\Delta_{\rm gap}^{-+} (k)$, Eq. (\ref{gap-+}), reads,
\begin{eqnarray}
\Delta_{\rm gap}^{-+} (k) & = & \varepsilon_{s2} (k) 
\hspace{0.20cm}{\rm for}\hspace{0.20cm}k\in ]0,(k_{F\uparrow} - k_{F\downarrow})[
\nonumber \\
\Delta_{\rm gap}^{-+} (k) & = & 4\mu_B\,h - \varepsilon_{s} (k_{F\uparrow}-k) 
+ 2\varepsilon_{s} \left({\pi - k\over 2}\right)
\nonumber \\
& & {\rm for}\hspace{0.20cm}
k\in ](k_{F\uparrow} - k_{F\downarrow}),{\tilde{k}}[
\nonumber \\
\Delta_{\rm gap}^{-+} (k) & = & 4\mu_B\,h - W_{s2} - \varepsilon_{s} (k_{F\downarrow}-k) 
+ 2\varepsilon_{s} \left({\pi - k\over 2}\right)
\nonumber \\
& & {\rm for}\hspace{0.20cm}k \in ]{\tilde{k}},2k_{F\downarrow}[
\nonumber \\
\Delta_{\rm gap}^{-+} (k) & = & \varepsilon_{s2} (k - 2k_{F\downarrow}) 
+ 2\varepsilon_{s} \left({\pi - k\over 2}\right)
\nonumber \\
& & {\rm for}\hspace{0.20cm}k\in ]2k_{F\downarrow},\pi[  \, ,
\label{GgapSPM03}
\end{eqnarray}
whereas for $m\in [\tilde{m},1[$ it is given by,
\begin{eqnarray}
\Delta_{\rm gap}^{-+} (k) & = & \varepsilon_{s2} (k) \hspace{0.20cm}{\rm for}\hspace{0.20cm}k\in [0,{\tilde{k}}-\delta {\tilde{k}}_1]
\nonumber \\
\Delta_{\rm gap}^{-+} (k) & = & 4\mu_B\,h - W_{s2} - \varepsilon_{s} (k_{F\downarrow}-k) 
\nonumber \\
& & {\rm for}\hspace{0.20cm}k \in ]{\tilde{k}},(k_{F\uparrow} - k_{F\downarrow})[
\nonumber \\
\Delta_{\rm gap}^{-+} (k) & = & 4\mu_B\,h - W_{s2} - \varepsilon_{s} (k_{F\downarrow}-k) 
+ 2\varepsilon_{s} \left({\pi - k\over 2}\right)
\nonumber \\
& & {\rm for}\hspace{0.20cm}k \in ](k_{F\uparrow} - k_{F\downarrow}),2k_{F\downarrow}[
\nonumber \\
\Delta_{\rm gap}^{-+} (k) & = & \varepsilon_{s2} (k - 2k_{F\downarrow}) 
+ 2\varepsilon_{s} \left({\pi - k\over 2}\right)
\nonumber \\
& & {\rm for}\hspace{0.20cm}k\in ]2k_{F\downarrow},\pi[  \, .
\label{GgapSPM3pi}
\end{eqnarray}

At $k=0,k_{F\uparrow}-k_{F\downarrow},\pi$ the
energy gap $\Delta_{\rm gap}^{-+} (k)$ is given by,
\begin{eqnarray}
\Delta_{\rm gap}^{-+} (0) & = & 4\mu_B\,h - W_{s2}
\hspace{0.20cm}{\rm for}\hspace{0.20cm}m\in ]0,1[
\nonumber \\
\Delta_{\rm gap}^{-+} (k_{F\uparrow}-k_{F\downarrow}) & = & 4\mu_B\,h 
\hspace{0.20cm}{\rm for}\hspace{0.20cm}m\in [0,\tilde{m}]
\nonumber \\
\Delta_{\rm gap}^{-+} (\pi) & = & 4\mu_B\,h - 2W_s^p 
\nonumber \\
&& {\rm for}\hspace{0.20cm}m\in ]0,1[ \, .
\label{gapMPkkk}
\end{eqnarray}

In the $k$ intervals $k\in ]\bar{k}_0,\pi[$ and $k\in ]\bar{k}_0,\bar{k}_1[$,
Eq. (\ref{gapineq}), for spin densities $m\in ]0,\bar{m}_0]$ and
$m\in ]0,\bar{m}]$, respectively, one has that $\Delta_{\rm gap}^{xx} (k) = \Delta_{\rm gap}^{-+} (k)<0$. 
For instance, at $k=\pi$ (and in the $k\rightarrow \pi$ limit in the spectra expressions) and for spin densities $m\in ]0,1[$, 
the energy gap $\Delta_{\rm gap}^{xx} (\pi)= \Delta_{\rm gap}^{-+} (\pi)$
is in the $u\rightarrow 0$ limit and for $u\gg 1$ given by,
\begin{eqnarray}
\Delta_{\rm gap}^{xx} (\pi) & = & 12t\left(\sin\left({\pi\over 2}m\right) - {1\over 3}\right)
\nonumber \\
& = & -4t\hspace{0.20cm}{\rm for}\hspace{0.20cm}m\rightarrow 0 
\nonumber \\
& = & 0\hspace{0.20cm}{\rm for}\hspace{0.20cm}m = \bar{m}_0 = {2\over\pi}\arcsin\left({1\over 3}\right)
\approx 0.216
\nonumber \\
& = & 8t\hspace{0.20cm}{\rm for}\hspace{0.20cm}m\rightarrow 1 \, ,
\label{gapEXPm03LIMu0MP}
\end{eqnarray}
and
\begin{eqnarray}
\Delta_{\rm gap}^{xx} (\pi) & = &  -{\pi t\over u}= -{4\pi t^2\over U}\hspace{0.20cm}{\rm for}\hspace{0.20cm}m\rightarrow 0
\nonumber \\
& = & 0\hspace{0.20cm}{\rm for}\hspace{0.20cm}m = \bar{m}_0 \approx 0.239
\nonumber \\
& = & {4t\over u} = {16t^2\over U} \hspace{0.20cm}{\rm for}\hspace{0.20cm}m\rightarrow 1 \, ,
\label{gaMPpuL}
\end{eqnarray}
respectively.

Finally, the energy gap $\Delta_{\rm gap}^{zz} (k)$, Eqs. (\ref{gapPMMP}) and (\ref{gapL}), is
for spin densities $m\in ]0,\tilde{m}]$ and $m \in [\tilde{m},1[$ given by,
\begin{eqnarray}
\Delta_{\rm gap}^{zz} (k) & = & \varepsilon_{s2} (k - (k_{F\uparrow}-k_{F\downarrow}))
\nonumber \\
& & {\rm for}\hspace{0.20cm}k\in ]0,(k_{F\uparrow}-k_{F\downarrow})[
\nonumber \\
& = & 4\mu_B\,h - W_{s2} - \varepsilon_{s} \left(k_{F\uparrow} - k\right)
\nonumber \\
& & {\rm for}\hspace{0.20cm}k\in ](k_{F\uparrow}-k_{F\downarrow}),(\pi - {\tilde{k}})[
\nonumber \\
& = & 4\mu_B\,h - \varepsilon_{s} (k_{F\downarrow}-k)
\nonumber \\
& & {\rm for}\hspace{0.20cm}k\in ](\pi - {\tilde{k}}),2k_{F\downarrow}[
\nonumber \\
& = & \varepsilon_{s2} (k-\pi) \hspace{0.20cm}{\rm for}\hspace{0.20cm}k\in ]2k_{F\downarrow},\pi[
\nonumber \\
& & {\rm when}\hspace{0.20cm}m\in [0,\tilde{m}] \, ,
\label{Dxx03L}
\end{eqnarray}
and
\begin{eqnarray}
\Delta_{\rm gap}^{zz} (k) & = & \varepsilon_{s2} (k - (k_{F\uparrow}-k_{F\downarrow})) 
\hspace{0.20cm}{\rm for}\hspace{0.20cm}k\in [0,(k_{F\uparrow}-k_{F\downarrow})]
\nonumber \\
& = & 4\mu_B\,h - W_{s2} - \varepsilon_{s} \left(k_{F\uparrow} - k\right)
\nonumber \\
& & {\rm for}\hspace{0.20cm}k\in ](k_{F\uparrow}-k_{F\downarrow}),(\pi - {\tilde{k}})[
\nonumber \\
& = & \varepsilon_{s2} (k-\pi)  \hspace{0.20cm}{\rm for}\hspace{0.20cm}
k\in ](\pi - {\tilde{k}}),\pi[ 
\nonumber \\
& & {\rm when}\hspace{0.20cm}m\in [\tilde{m},1[ \, ,
\label{Dxx31L}
\end{eqnarray}
respectively.

\section{Matrix elements functionals and cusp singularities}
\label{D}

The ground state at a given spin density $m$ is populated by $N_s = N_{\downarrow}$ $s$ particles that 
fill a $s$ band Fermi sea $q\in[-k_{F\downarrow},k_{F\downarrow}]$
where $k_{F\downarrow}$ is given in Eq. (\ref{kkk}) 
and by a full $c$ band $q\in [-\pi,\pi]$ populated by $N_c = N$ $c$ particles that
do not participate in the spin dynamical properties. Within the
present thermodynamic limit, we have here ignored corrections
of $1/L$ order to these bands momentum limiting values. There are no $s2$ particles in the ground state. 

However, the following number and current number deviations under 
transitions from the ground state to excited energy eigenstates are associated with 
momentum deviations $1/L$ corrections that must be accounted for
even in the thermodynamic limit,
\begin{eqnarray}
& & \delta N_{s,\iota}^F \hspace{0.20cm}{\rm for}\hspace{0.20cm}\iota =1,-1
\hspace{0.20cm}{\rm (right,left)}\hspace{0.20cm}s\hspace{0.10cm}{\rm particles}
\nonumber \\
& & \delta N_{s}^F = \sum_{\iota = \pm 1}\delta N_{s,\iota}^F
\hspace{0.20cm}{\rm and}\hspace{0.20cm}\delta J_{s}^F =  {1\over 2}\sum_{\iota = \pm 1}\iota\,\delta N_{s,\iota}^F
\nonumber \\
& & \delta J_{s2} = {\iota\over 2}\delta N_{s2} (q)\vert_{q=\iota\,(k_{F\uparrow}-k_{F\downarrow})}
\nonumber \\
& & \delta J_{c}^F =  {1\over 2}\sum_{\iota = \pm 1}\iota\,\delta N_{c,\iota}^F 
\hspace{0.20cm}{\rm where}
\nonumber \\
& & \delta N_{c,\iota}^F = - \delta N_{c,-\iota}^F \, .
\label{NcFNcFJcFJsF}
\end{eqnarray}

Under the transitions from the ground state to the excited energy eigenstates
that span the spin subspaces of the quantum problem studied in this paper, the number of $s$ particles may 
change. This leads to number deviations $\delta N_{s}$. The specific number deviations $\delta N_{s,\iota}^F$ 
in Eq. (\ref{NcFNcFJcFJsF}) refer only to changes of the $s$ particles numbers at the left $(\iota =-1)$ or right $(\iota =1)$
$s$ band Fermi points. The same information is contained in the two Fermi points 
number deviations $\delta N_{s,\iota}^F$ and in the corresponding Fermi points 
number deviations $\delta N_{s}^F = \sum_{\iota = \pm 1}\delta N_{s,\iota}^F$ and
current number deviations $\delta J_{s}^F =  {1\over 2}\sum_{\iota = \pm 1}\iota\,\delta N_{s,\iota}^F$.

The overall $s$ particles number deviation $\delta N_{s}$ can be expressed as,
\begin{equation}
\delta N_{s} = \delta N_{s}^{F}+\delta N_s^{NF} \, .
 \label{dNdNFsNNF}
\end{equation}
Here $\delta N_s^{NF}$ refers to changes in the number of $s$ particles at $s$ band momenta
other than at the Fermi points.

For the spin subspaces under consideration, the $s2$ band number 
deviations only read $\delta N_{s2}=0$ or $\delta N_{s2}=1$.
In the latter case, the $s2$ particle can be created at any $s2$ band momentum 
$q \in [-(k_{F\uparrow}-k_{F\downarrow}),(k_{F\uparrow}-k_{F\downarrow})]$. Only 
when the $s2$ particle is created at the $s2$ band limiting values $q= -(k_{F\uparrow}-k_{F\downarrow})$ or
$q= (k_{F\uparrow}-k_{F\downarrow})$ that process leads to a current number deviation
$\delta J_{s2} = -1/2$ and  $\delta J_{s2} = 1/2$, respectively.

The dynamical structure factors are within the dynamical theory used in the studies
of this paper expressed as a sum of $s$-band spectral function terms $B_{s} (k,\omega)$
(denoted by $B_{Q} (k,\omega)$ in Ref. \onlinecite{Carmelo_16}), each associated with 
a reference energy eigenstate whose $s$-band Fermi sea changes from occupancy one to zero
at the $\iota =+1$ right and $\iota =-1$ left Fermi points $q_{Fs,\iota} = q_{Fs,\iota}^0+\pi\delta N_{s,\iota}^F/L$.
Here $q_{Fs,\iota}^0$ refers to the ground state and the $\iota = \pm 1$ number deviations $\delta N_{s,\iota}^F$ are
those in Eq. (\ref{NcFNcFJcFJsF}). 

In the subspaces of our study, that reference state corresponds to fixed $\iota =\pm 1$ deviations $\delta N_{s,\iota}^F$ and can have 
no holes within the $s$-band Fermi sea, one hole at a fixed $s$-band momentum $q$, or two holes at fixed $s$-band momenta 
$q$ an $q'$, all inside that Fermi sea and away from the Fermi points. In addition, that state
can have no $s$ particles or a single $s$ particle at a fixed $s$-band momentum $q$ 
outside the $s$-band Fermi sea and away from its Fermi points. It can also have
no $s2$ particles or one $s2$ particle at a fixed momentum 
$q''\in [-(k_{F\uparrow}-k_{F\downarrow}),(k_{F\uparrow}-k_{F\downarrow})]$. 

Besides the reference state, each term $B_{s} (k,\omega)$ 
involves sums that run over $m_{\iota}=1,2,3,...$ elementary particle-hole processes of 
$\iota = \pm 1$ momenta $\iota 2\pi/L$
around the corresponding Fermi points $q_{Fs,\iota}$ that generate a tower of excited states
upon that reference state. It reads \cite{Carmelo_16}, 
\begin{eqnarray}
B_{s} (k,\omega) & = & {L\over 2\pi}\sum_{m_{+1};m_{-1}}\,A^{(0,0)}\,a (m_{+1},\,m_{-1})
\nonumber \\
& \times & \delta \Bigl(\omega - \epsilon - {2\pi\over L}\,v_{s}\sum_{\iota =\pm1} (m_{\iota}+\Phi_{\iota}^2/4)\Bigr)
\nonumber \\
& \times & 
\delta \Bigl(k - {2\pi\over L}\,\sum_{\iota =\pm1}\iota\,(m_{\iota}+\Phi_{\iota}^2/4)\Bigr) \, .
\label{BQ-gen}
\end{eqnarray}
Here $v_s = v_s (k_{F\downarrow})$ where
$v_s (q)$ is the $s$-band group velocity, Eq. (\ref{equA4B}),
and the {\it lowest peak weight} $A^{(0,0)}$ and the weights $A^{(0,0)}\,a (m_{+1},\,m_{-1})$ 
refer to the matrix elements square $\vert\langle \nu\vert\hat{S}^{a}_k\vert GS\rangle\vert^2$
in Eq. (\ref{SDSF}) between the ground state and the $m_{\iota}=0$ reference excited state
and the corresponding $m_{\iota}>0$ tower excited states. 
For the present subspaces, the $\iota =\pm 1$ functionals 
$\Phi_{\iota}$ and the spectrum $\epsilon$ in Eq. (\ref{BQ-gen}) have the general form,
\begin{eqnarray}
&& \Phi_{\iota} = {\iota\,\delta N^F_{s}\over 2\xi_{s\,s}^{1}} + \xi_{s\,s}^{1}\,(\delta J^F_{s} - 2\delta J_{s2}) 
\nonumber \\
&& +\,\Phi_{s,s}(\iota k_{F\downarrow},q)\delta N_s (q) + \Phi_{s,s}(\iota k_{F\downarrow},q')\delta N_s (q')
\nonumber \\
&& +\,(1-\delta_{\vert q''\vert,(k_{F\uparrow}-k_{F\downarrow})})\,\Phi_{s,s2}(\iota k_{F\downarrow},q'')\delta N_{s2} (q'') 
\nonumber \\
&& \epsilon = \varepsilon_s (q)\delta N_s (q) +  \varepsilon_s (q')\delta N_s (q') + \varepsilon_{s2} (q'')\delta N_{s2} (q'')
\nonumber \\
&& {\rm where}
\nonumber \\
&& \delta N_s (q) = 0, \pm 1 \, ; \hspace{0.20cm}
\delta N_s (q') = 0, -1 \hspace{0.20cm}{\rm and}
\nonumber \\
&& \delta N_{s2} (q'') = 0, 1 \, .
\label{functional}
\end{eqnarray}
Here the deviations $\delta N^F_{s}$, $\delta J^F_{s}$, and $\delta J_{s2}$
are given in Eq. (\ref{NcFNcFJcFJsF}),
the $\iota = \pm 1$ phase shifts $\Phi_{s,s}\left(\iota k_{F\downarrow},q\right)$ and 
$\Phi_{s,s2}\left(\iota k_{F\downarrow},q\right)$ in units of $2\pi$ are defined by Eq. (\ref{Phis-all-qq}),
the phase-shift related parameter $\xi_{s\,s}^{1}$ is defined in Eq. (\ref{xi1all}),
and the energy dispersions $\varepsilon_{s} (q)$ and $\varepsilon_{s2} (q)$ are given in
Eqs. (\ref{equA4}) and (\ref{vares2}), respectively.

The relative weights $a (m_{+1},\,m_{-1})$ in Eq. (\ref{BQ-gen})
can be expressed in terms of the gamma function as \cite{Carmelo_16},
\begin{eqnarray}
a (m_{+1},m_{-1}) & = & \prod_{\iota =\pm 1}
a_{\iota}(m_{\iota})\hspace{0.20cm}{\rm where}
\nonumber \\
a_{\iota}(m_{\iota}) & = & \frac{\Gamma (m_{\iota} +
\Phi_{\iota}^2)}{\Gamma (m_{\iota}+1)\,
\Gamma (\Phi_{\iota}^2)}  \, .
\label{amm}
\end{eqnarray}

In the present thermodynamic limit, the matrix elements weights have 
the following asymptotic behavior \cite{Carmelo_16},
\begin{eqnarray}
A^{(0,0)} & = & \left({1\over L\,B_s}\right)^{-1+\sum_{\iota =\pm1}\Phi_{\iota}^2}
\nonumber \\
& \times & \prod_{\iota =\pm 1}e^{-f_0 + f_2\left(2{\tilde{\Phi}}_{\iota}\right)^2 - f_4\left(2{\tilde{\Phi}}_{\iota}\right)^4} 
\nonumber \\
a (m_{+1},m_{-1}) & = & \prod_{\iota =\pm 1}{
(m_{\iota}+\Phi_{\iota}^2/4)^{-1+ \Phi_{\iota}^2}\over\Gamma (\Phi_{\iota}^2)} \, .
\label{Aamm}
\end{eqnarray}
Here ${\tilde{\Phi}}_{\iota} = \Phi_{\iota} -\iota\delta N_{s,\iota}^F$,
the constant $0<B_s\leq 1$ and the constants $0<f_l<1$ where $l=0,2,4$ depends on $u$ and $m$ and
depend only on $u$, respectively, and are independent of $L$. Importantly, in that limit the matrix elements square
in Eq. (\ref{SDSF}) then read,
\begin{eqnarray}
&& \vert\langle \nu\vert\hat{S}^{a}_k\vert GS\rangle\vert^2 =
\left({1\over L\,B_s}\right)^{-1+\sum_{\iota =\pm1}\Phi_{\iota}^2}
\nonumber \\
&& \times 
\prod_{\iota =\pm 1}{e^{-f_0 + f_2\left(2{\tilde{\Phi}}_{\iota}\right)^2 - f_4\left(2{\tilde{\Phi}}_{\iota}\right)^4}  \over\Gamma (\Phi_{\iota}^2)} 
\left(m_{\iota}+\Phi_{\iota}^2/4\right)^{-1+ \Phi_{\iota}^2}
\nonumber \\
&& = \left({1\over L\,B_s}\right)^{-1+\sum_{\iota =\pm1}\Phi_{\iota}^2}
\prod_{\iota =\pm 1}{e^{-f_0 + f_2\left(2{\tilde{\Phi}}_{\iota}\right)^2 - f_4\left(2{\tilde{\Phi}}_{\iota}\right)^4}  \over\Gamma (\Phi_{\iota}^2)} 
\nonumber \\
&& \times \left({L\over 4\pi\,v_{s}}(\omega - \epsilon +\iota\,v_{s}\,k)\right)^{-1+ \Phi_{\iota}^2} \, .
\label{ME}
\end{eqnarray}
Here the equality $m_{\iota} = {L\over 4\pi\,v_{s}}(\omega - \epsilon +\iota\,v_{s}\,k)-\Phi_{\iota}^2/4$
imposed by the $\delta$-functions in Eq. (\ref{BQ-gen}) has been used. 

In the general case in which the two $\iota =\pm 1$ functionals 
$\Phi_{\iota}$ are finite the $s$-particle spectral function $B_{s} (k,\omega)$, Eq. (\ref{BQ-gen}), can
be written as \cite{Carmelo_16},
\begin{eqnarray}
&& B_{s} (k,\omega) = {1\over 4\pi\,B_s\,v_{s}}\,
\prod_{\iota =\pm 1}\,\Theta (\omega - \epsilon + \iota\,v_{s}\,k)
\nonumber \\
&& {e^{-f_0 + f_2\left(2{\tilde{\Phi}}_{\iota}\right)^2 - f_4\left(2{\tilde{\Phi}}_{\iota}\right)^4} \over \Gamma (\Phi_{\iota}^2)}
 \Bigl({\omega - \epsilon +\iota\,v_{s}\,k\over 4\pi\,B_s\,v_{s}}\Bigr)^{-1+\Phi_{\iota}^2} \, .
\label{B-J-i-sum-GG}
\end{eqnarray}
To reach this expression, which in the thermodynamic limit is exact, Eqs. 
(\ref{BQ-gen}), (\ref{Aamm}), and (\ref{ME}) were used. 

The summation of the terms $B_{s} (k,\omega)$ that lead to 
expressions for the dynamical structure factors
can be performed and reach several kinds of contributions.

When $\delta N_s (q) = \delta N_s (q') = 0$ and $\delta N_{s2} (q'') = 0$ or
$\delta N_{s2} (q'') = 1$ at $q''=0$ in Eq. (\ref{functional}), such summations lead to
$S^{ab} (k,\omega) \propto \Bigl(\omega -\omega_0\Bigr)^{\zeta^{ab}}$
for $(\omega - \omega_0) \neq \pm v_{s}\,(k-k_0)$
where $\omega_0 = 0$ and $\omega_0 = 4\mu_B\,h$
for $\delta N_{s2} (q'') = 0$ and $\delta N_{s2} (0) = 1$, respectively,
$k_0 = 2k_{F\downarrow}\,\delta J_s^F$, and $\zeta^{ab} = -2+\sum_{\iota =\pm 1}\Phi_{\iota}^2$.
Moreover, they lead to an alternative behavior
$B (k,\omega) \propto \Bigl(\omega - \omega_0 \mp v_{s}\,(k-k_0)\Bigr)^{\zeta^{ab}_{\pm}}$
for $(\omega - \omega_0) \approx \pm v_{s}\,(k-k_0)$ where $\zeta^{ab}_{\pm} = -1 + \Phi_{\pm}^2$.
These behaviors are only valid in very small $(k,\omega)$-plane regions
associated with very small values of $\omega$ or $(\omega - 4\mu_B\,h)$ 
and of $(k-k_0)$ and lead to cusp singularities 
when $\zeta^{ab}<0$ and/or $\zeta^{ab}_{\pm}<0$ \cite{Carmelo_16}.

When {\it only one} of the deviations $\delta N_s (q)$, $\delta N_s (q')$, and $\delta N_{s2} (q'')$ 
in Eq. (\ref{functional}) reads $1$ (or $-1$) the summation of terms $B_{s} (k,\omega)$ 
gives the line shape of the dynamical structure factors in the $(k,\omega)$-plane vicinity 
of branch lines associated with the lower thresholds, Eqs. (\ref{MPSsFMB}) and (\ref{MPSs}). 
The form of the exponents $\zeta^{ab}_{\beta} (k) = -1 + \sum_{\iota =\pm1}\Phi_{\iota}^2$, Eq. 
(\ref{expTS}), in these expressions is fully determined by the square matrix elements,
Eq. (\ref{ME}). 

When several of the deviations $\delta N_s (q)$, $\delta N_s (q')$, and $\delta N_{s2} (q'')$ 
in Eq. (\ref{functional}) are given by $1$ (or $-1$), the summation of terms $B_{s} (k,\omega)$ 
leads to a line shape without cusp singularities. 

The results of this paper focus on the line shape near the branch lines associated with the lower thresholds, 
Eqs. (\ref{MPSsFMB}) and (\ref{MPSs}). They rely on the specific form that the functional,
Eq. (\ref{functional}), has for the $s2,s2'$ branch lines, $\bar{s}$ branch lines, and $\bar{s}'$ branch lines
that are part of the gapped lower thresholds.

In the case of the $s2$ and $s2'$ branch lines, 
that spectral functional's form is,
\begin{eqnarray}
\Phi_{\iota} (q) 
& = & \iota\,\xi_{s\,s}^{0}{\delta N^F_{s}\over 2} + \xi_{s\,s}^{1}\,\delta J^F_{s}+ \Phi_{s,s2}(\iota k_{F\downarrow},q) 
\nonumber \\
& = & {\iota\,\delta N^F_{s}\over 2\xi_{s\,s}^{1}} + \xi_{s\,s}^{1}\,\delta J^F_{s}+ \Phi_{s,s2}(\iota k_{F\downarrow},q) 
\nonumber \\
& & {\rm for}\hspace{0.20cm}s2\hspace{0.20cm}{\rm and}\hspace{0.20cm}s2'
\hspace{0.20cm}{\rm branch}\hspace{0.20cm}{\rm lines} \, .
\label{Fs2}
\end{eqnarray}
For the excited energy eigenstates that contribute to the singularities
at and above the $s2$ and $s2'$ branch lines, 
the maximum interval of the $s2$ band momentum $q$ in Eq. (\ref{Fs2})
is $q\in [0,(k_{F\uparrow}-k_{F\downarrow})[$ or
$q\in ]-(k_{F\uparrow}-k_{F\downarrow}),0]$.

For $\bar{s}$ and $\bar{s}'$ branch lines, the spectral functionals are different and have the form,
\begin{eqnarray}
& & \Phi_{\iota} (q) = 
\nonumber \\
& = & \iota\,\xi_{s\,s}^{0}{\delta N^F_{s}\over 2} + {\iota\,\xi_{s\,s2}^{0}\over 2} + \xi_{s\,s}^{1}\,\delta J^F_{s} 
- \Phi_{s,s}(\iota k_{F\downarrow},q)
\nonumber \\
& = & {\iota\,\delta N^F_{s}\over 2\xi_{s\,s}^{1}} + {\iota\,\xi_{s\,s2}^{0}\over 2} + \xi_{s\,s}^{1}\,\delta J^F_{s} 
- \Phi_{s,s}(\iota k_{F\downarrow},q)
\nonumber \\
& & {\rm for}\hspace{0.20cm}\bar{s}\hspace{0.20cm}{\rm branch}\hspace{0.20cm}{\rm lines} \, ,
\label{Fbars}
\end{eqnarray}
and
\begin{eqnarray}
& & \Phi_{\iota} (q) = 
\nonumber \\
& = & \iota\,\xi_{s\,s}^{0}{\delta N^F_{s}\over 2} + \xi_{s\,s}^{1}\,(\delta J^F_{s} - 2\delta J_{s2}) - 
\Phi_{s,s}(\iota k_{F\downarrow},q)
\nonumber \\
& = & {\iota\,\delta N^F_{s}\over 2\xi_{s\,s}^{1}} + \xi_{s\,s}^{1}\,(\delta J^F_{s} - 2\delta J_{s2}) - 
\Phi_{s,s}(\iota k_{F\downarrow},q) 
\nonumber \\
& & {\rm for}\hspace{0.20cm}\bar{s}'\hspace{0.20cm}{\rm branch}\hspace{0.20cm}{\rm lines} \, ,
\label{Fbarsl}
\end{eqnarray}
respectively. Here the maximum interval of the $s$ band momentum
is $q\in ]-k_{F\downarrow},k_{F\downarrow}[$,
$\xi_{s\,s}^{0}=1/\xi_{s\,s}^{1}$ at one electron per site and we accounted for the phase shift
$\Phi_{s,s2}(\iota k_{F\downarrow},\pm (k_{F\uparrow}-k_{F\downarrow}))$ reading 
$\mp\xi_{s\,s}^{1}$ [see Eq. (\ref{xi1Phiss2})].

The values of the $s$ and $s2$ bands number and current number 
deviations that in the case of the transverse and longitudinal spin excitations 
are used in Eqs. (\ref{Fs2})-(\ref{Fbarsl}) are provided in Tables \ref{table1} and \ref{table2},
respectively.

Finally, the momentum dependent exponents that control the line shape near the $s$ branch lines that refer to parts of the 
lower thresholds of the combined spectra, Eqs. (\ref{dkEdPxxMP}) and (\ref{dkEdPxxPM}), and 
of the spectrum, Eq. (\ref{dkEdPl}), involve spectral functionals of general form,
\begin{eqnarray}
\Phi_{\iota} (q) & = & 
{\iota\,\delta N^F_{s}\over 2\xi_{s\,s}^{1}} + \xi_{s\,s}^{1}\,\delta J^F_{s} \mp\Phi_{s,s}(\iota k_{F\downarrow},q)
\nonumber \\
& & {\rm where} 
\nonumber \\
- & \rightarrow & \hspace{0.20cm}{\rm maximum}\hspace{0.20cm}{\rm interval}\hspace{0.20cm}
q\in ]-k_{F\downarrow},k_{F\downarrow}[
\nonumber \\
+ & \rightarrow & \hspace{0.20cm}{\rm maximum}\hspace{0.20cm}{\rm interval}\hspace{0.20cm}
\vert q\vert \in ]k_{F\downarrow},k_{F\uparrow}] 
\nonumber \\
& & {\rm for}\hspace{0.20cm}s\hspace{0.20cm}{\rm branch}\hspace{0.20cm}{\rm lines} \, .
\label{Fs}
\end{eqnarray}
Here $-$ and $+$ is the phase-shift sign in $\mp\Phi_{s,s}(\iota k_{F\downarrow},q)$ suitable
to $s$ branch lines involving $s$ band hole and $s$ particle creation, respectively, at a $q$
belonging to the given maximum intervals. 

The values of the $s$ band number and current number deviations that are used 
in Eq. (\ref{Fs}) are provided in Table \ref{table3}.

In terms of many-electron processes, the quantum problem studied in this paper is not
perturbative. However, in terms of the fractionalized particles
that naturally emerge from the rotated-electrons degrees of freedom separation
it is perturbative. (In the subspace of the present quantum problem,
rotated-electron operators are expressed in terms of corresponding 
fractionalized particles operators as given in Eq. (80) of Ref. \onlinecite{Carmelo_16}.)

The case of most interest for the studies of this paper
refers to the gapped excited energy eigenstates populated by
one $s2$ particle. For the $+-$, $xx$, and $zz$ spin dynamical structure factors,
such states are behind the $(k,\omega)$-plane spectral weight located
above the gapped lower thresholds shown in Figs. \ref{figure1}-\ref{figure6}.
For such $+-$, $zz$, and $-+$ factors
the $s$-particle number deviations, $\delta N_{s} = \delta N_{s}^{F}+\delta N_s^{NF}$, Eq. (\ref{dNdNFsNNF}), 
are given by $\delta N_{s} = -1$, $\delta N_{s} = -2$, and $\delta N_{s} = -3$,
respectively. That $\sum_{\iota =\pm1}\Phi_{\iota}^2 (q)$ increases upon increasing $\vert\delta N_{s}\vert$
is behind both a decreasing amount of spectral weight above the 
corresponding gapped lower threshold and an
increase of the momentum-dependent exponents, Eqs. (\ref{MPSsFMB}) and (\ref{MPSs}).

\section{Some useful quantities}
\label{E}

In this Appendix a set o quantities needed for our study are defined and
corresponding useful limiting behaviors are provided.

The quantum problem described by the 1D Hubbard model with one electron per site in a magnetic field
acting in the spin subspaces considered in 
this paper involves a subset of Bethe ansatz equations. 

The equation associated with the $s$ band of the classes of excited energy eigenstates 
that span such spin subspaces is given by, 
\begin{eqnarray}
&& q_j = {2\over L} \sum_{j'=1}^{L}\,\arctan\left({\Lambda_s (q_j)-\sin k (q_{j'})\over u}\right)
\nonumber \\
& & - {2\over L}\sum_{j'=1}^{N_{\uparrow}}\, N_{s}(q_{j'})\arctan
\left({\Lambda_s (q_j)-\Lambda_s (q_{j'})\over 2u}\right) 
\nonumber \\
& & - {2\over L}\sum_{j'=1}^{N_{\uparrow}-N_{\downarrow}+N_{s2}}\, N_{s2}(q_{j'})
\{\arctan\left({\Lambda_s (q_j)-\Lambda_{s2} (q_{j'})\over u}\right) 
\nonumber \\
& & + \arctan\left({\Lambda_s (q_j)-\Lambda_{s2} (q_{j'})\over 3u}\right)\}
\nonumber \\
&& {\rm where} \hspace{0.5cm}  j = 1,...,N_{\uparrow} \, . 
\label{Taps}
\end{eqnarray}
That associated with the $s2$ band reads,
\begin{eqnarray}
q_j & = & {2\over L} \sum_{j'=1}^{L}\,\arctan\left({\Lambda_{s2} (q_j)-\sin k (q_{j'})\over 2u}\right)
\nonumber \\
& - & {2\over L}\sum_{j'=1}^{N_{\uparrow}}\,N_{s}(q_{j'})\{\arctan\left({\Lambda_{s2} (q_j)-\Lambda_s (q_{j'})\over u}\right)
\nonumber \\
& + &\arctan\left({\Lambda_{s2} (q_j)-\Lambda_s (q_{j'})\over 3u}\right)\}
\nonumber \\
&& {\rm where} \hspace{0.5cm}  j = 1,...,N_{\uparrow}-N_{\downarrow}+N_{s2} 
\nonumber \\
&& {\rm and}\hspace{0.5cm}N_{s2} = 0,1 \, .
\label{Taps2}
\end{eqnarray}
In these equations, $N_{s}(q_{j'})=1$ and $N_{s2}(q_{j'})=1$ for occupied $q_{j'}$ and
$N_{s}(q_{j'})=0$ and $N_{s2}(q_{j'})=0$ for unoccupied $q_{j'}$.

For the spin subspaces spanned by excited states populated by $N_s = N_{\downarrow}-2$
$s$ particles and one $s2$ particle, the Bethe-ansatz equation, Eq. (\ref{Taps2}), does not include the third term that involves
the spin rapidity differences $\Lambda_{s2} (q_j)-\Lambda_{s2} (q_{j'})$. Indeed,
it vanishes for $q_j = q_{j'}$.

The $s$ band Bethe ansatz rapidity is real and associated with the rapidity function
$\Lambda_{s} (q_j)$. The $s2$ band rapidity function $\Lambda_{s2} (q_j)$ that
appears in Eqs. (\ref{Taps}) and (\ref{Taps2}) is the real part of the following two Bethe ansatz complex rapidities
associated with a spin $n$-string of length $n=2$,
\begin{equation}
\Lambda_{s2} (q_j) \pm i\,u \, .
\label{L2}
\end{equation} 

The rapidity function $k (q_j)$ that appears in the above equations is associated with the $c$
band that in the present subspaces is full with a constant occupancy of $N$ $c$ particles
and thus is not dynamically active. That function is defined by the following equation,
\begin{eqnarray}
&& k (q_j) = q_j 
- {2\over L}\sum_{j'=1}^{N_{\uparrow}}\,N_{s}(q_{j'})\arctan\left({\sin k (q_j)-\Lambda (q_{j'}) \over u}\right)
\nonumber \\
& & - {2\over L}\sum_{j'=1}^{N_{\uparrow}-N_{\downarrow}+N_{s2}}\, N_{s2}(q_{j'})
\arctan\left({\sin k (q_j)-\Lambda_{s2} (q_{j'})\over 2u}\right) 
\nonumber \\
&& {\rm where} \hspace{0.5cm} j = 1,...,N \, .
\label{Tapc}
\end{eqnarray}

In the above equations,
\begin{equation}
q_j = {2\pi\over L}\,I^{\beta}_j \hspace{0.20cm}{\rm for}\hspace{0.20cm}
\beta = c,s,s2 \, ,
\label{q-j}
\end{equation} 
where the quantum numbers $I^{\beta}_j$ are either integers or half-odd integers according to the 
following boundary conditions \cite{Takahashi},
\begin{eqnarray}
I_j^{c} & = & 0,\pm 1,\pm 2,... \hspace{0.50cm}{\rm for}\hspace{0.15cm}N_s + N_{s2}\hspace{0.15cm}{\rm even} 
\nonumber \\
& = & \pm 1/2,\pm 3/2,\pm 5/2,... \hspace{0.50cm}{\rm for}\hspace{0.15cm}N_s + N_{s2}\hspace{0.15cm}{\rm odd} 
\nonumber \\
I_j^{s} & = & 0,\pm 1,\pm 2,... \hspace{0.50cm}{\rm for}\hspace{0.15cm}N_{\uparrow}\hspace{0.15cm}{\rm odd} 
\nonumber \\
& = & \pm 1/2,\pm 3/2,\pm 5/2,... \hspace{0.50cm}{\rm for}\hspace{0.15cm}N_{\uparrow}\hspace{0.15cm}{\rm even} 
\nonumber \\
I_j^{s2} & = & 0,\pm 1,\pm 2,... \hspace{0.50cm}{\rm for}\hspace{0.15cm}N_{s2} = 1 \, .
\label{Ic-an}
\end{eqnarray}

In the thermodynamic limit, we often use continuous momentum variables $q$ that
replace the discrete $s$ and $s2$ bands momenta $q_j$ such that $q_{j+1} - q_j=2\pi/L$.
They read $q\in[-k_{F\uparrow},k_{F\uparrow}]$ and $q \in [-(k_{F\uparrow}-k_{F\downarrow}),(k_{F\uparrow}-k_{F\downarrow})]$,
respectively. In that limit the momenta $k_{F\downarrow}$ and $k_{F\uparrow}$ rare
given by,
\begin{equation}
k_{F\downarrow} = {\pi\over 2}(1-m)\, ;\hspace{0.20cm}k_{F\uparrow} = {\pi\over 2}(1+m)
\, ;\hspace{0.20cm}k_F = {\pi\over 2} \, ,
\label{kkk}
\end{equation}
for the spin-density interval, $m\in ]0,1[$ where $k_F = \lim_{m\rightarrow 0}k_{F\downarrow} = 
\lim_{m\rightarrow 0}k_{F\uparrow}$.

The energy dispersions $\varepsilon_s (q)$ and $\varepsilon_{s2} (q)$ that appear in the spectra of
the spin excitations are defined as follows,
\begin{eqnarray}
\varepsilon_{s} (q) & = & {\bar{\varepsilon}_{s}} (\Lambda_s (q)) 
\hspace{0.20cm}{\rm for}\hspace{0.20cm}q \in [-k_{F\uparrow},k_{F\uparrow}] 
\hspace{0.20cm}{\rm where}
\nonumber \\
{\bar{\varepsilon}_{s}} (\Lambda) & = & \int_{B}^{\Lambda}d\Lambda^{\prime}\,2t\,\eta_{s} (\Lambda^{\prime}) \, ,
\label{equA4}
\end{eqnarray}
and
\begin{eqnarray}
\varepsilon_{s2} (q) & = & 4\mu_B\,h + \varepsilon_{s2}^0 (q)
\hspace{0.20cm}{\rm for}
\nonumber \\
q & \in & [-(k_{F\uparrow} - k_{F\downarrow}),(k_{F\uparrow} - k_{F\downarrow})]
\hspace{0.20cm}{\rm where}
\nonumber \\
\varepsilon_{s2}^0 (q) & = & {\bar{\varepsilon}}_{s2}^0 (\Lambda_{s2} (q)) \hspace{0.20cm}{\rm and}
\nonumber \\
{\bar{\varepsilon}}_{s2}^0 (\Lambda) & = & \int_{\infty}^{\Lambda}d\Lambda^{\prime}\,2t\,\eta_{s2} (\Lambda^{\prime}) \, ,
\label{vares2}
\end{eqnarray}
respectively.

The corresponding $s$ and $s2$ bands group velocities are given by,
\begin{equation}
v_s (q) = {\partial\varepsilon_s (q)\over\partial q} 
\hspace{0.20cm}{\rm and}\hspace{0.20cm}
v_{s2} (q) = {\partial\varepsilon_{s2} (q)\over\partial q} \, .
\label{equA4B}
\end{equation}

The distribution $2t\,\eta_{s} (\Lambda)$ appearing in Eq. (\ref{equA4}) is coupled to a 
distribution $2t\,\eta_c (k)$ through the following integral equations,
\begin{equation}
2t\,\eta_c (k) = 2t\sin k + \frac{\cos k}{\pi\,u} \int_{-B}^{B}d\Lambda\,{2t\,\eta_{s} (\Lambda)\over 1 
+  \left({\sin k - \Lambda\over u}\right)^2} \, ,
\label{equA5}
\end{equation}
and
\begin{eqnarray}
2t\,\eta_{s} (\Lambda) & = & {1\over\pi\,u}\int_{-\pi}^{\pi}dk\,{2t\,\eta_c (k)\over 1 +  \left({\Lambda-\sin k\over u}\right)^2} 
\nonumber \\
& - & \frac{1}{2\pi\,u} \int_{-B}^{B}d\Lambda^{\prime}\,{2t\,\eta_{s} (\Lambda^{\prime})\over 1 +  \left({\Lambda -
\Lambda^{\prime}\over 2u}\right)^2} \, .
\label{equA6}
\end{eqnarray}

The distribution $2t\,\eta_{s2} (\Lambda)$ appearing in Eq. (\ref{vares2}) is given by,
\begin{eqnarray}
2t\,\eta_{s2} (\Lambda) & = & {1\over 2\pi\,u}\int_{-\pi}^{\pi}dk\,{2t\,\eta_c (k)\over 1 +  \left({\Lambda-\sin k\over 2u}\right)^2} 
\nonumber \\
& - & \frac{1}{\pi\,u} \int_{-B}^{B}d\Lambda^{\prime}\,
{2t\,\eta_{s} (\Lambda^{\prime})\over 1 +  \left({\Lambda - \Lambda^{\prime}\over u}\right)^2}
\nonumber \\
& - & \frac{1}{3\pi\,u} \int_{-B}^{B}d\Lambda^{\prime}\,
{2t\,\eta_{s} (\Lambda^{\prime})\over 1 +  \left({\Lambda - \Lambda^{\prime}\over 3u}\right)^2} \, ,
\label{etas2}
\end{eqnarray}
where the distributions $2t\,\eta_c (k)$ and $2t\,\eta_{s} (\Lambda)$ are the solutions of Eqs.  (\ref{equA5}) and  (\ref{equA6}).

The rapidity distribution function $\Lambda_s (q)$ where $q \in [-k_{F\uparrow},k_{F\uparrow}]$ in the argument of 
the auxiliary dispersion ${\bar{\varepsilon}_{s}}$ in Eq. (\ref{equA4}) is defined in terms of the $s$ band inverse 
function $q = q_s (\Lambda)$ where $\Lambda \in [-\infty,\infty]$. The latter is defined by the equation,
\begin{eqnarray}
q = q_s (\Lambda) & = & {1\over\pi}\int_{-\pi}^{\pi}dk\,2\pi\rho (k)\, \arctan \left({\Lambda-\sin k\over u}\right) 
\nonumber \\
& - & \frac{1}{\pi} \int_{-B}^{B}d\Lambda^{\prime}\,2\pi\sigma (\Lambda^{\prime})\, \arctan \left({\Lambda -
\Lambda^{\prime}\over 2u}\right) \nonumber \\
& & {\rm for}\hspace{0.20cm} \Lambda \in [-\infty,\infty] \, .
\label{equA7}
\end{eqnarray}

The rapidity distribution function $\Lambda_{s2} (q)$ where
$q\in [-(k_{F\uparrow} - k_{F\downarrow}),(k_{F\uparrow} - k_{F\downarrow})]$
is also defined in terms of the $s2$ band inverse function $q = q_{s2} (\Lambda)$ where 
$\Lambda \in [-\infty,\infty]$ as follows,
\begin{eqnarray}
q = q_{s2} (\Lambda) & = & {1\over\pi}\int_{-\pi}^{\pi}dk\,2\pi\rho (k)\, \arctan \left({\Lambda-\sin k\over 2u}\right) 
\nonumber \\
& - & \frac{1}{\pi} \int_{-B}^{B}d\Lambda^{\prime}\,2\pi\sigma (\Lambda^{\prime})
\arctan \left({\Lambda - \Lambda^{\prime}\over u}\right) 
\nonumber \\
& - & \frac{1}{\pi} \int_{-B}^{B}d\Lambda^{\prime}\,2\pi\sigma (\Lambda^{\prime})
\arctan \left({\Lambda - \Lambda^{\prime}\over 3u}\right)
\nonumber \\
& & {\rm for}\hspace{0.20cm} \Lambda \in [-\infty,\infty] \, .
\label{qtwoprime}
\end{eqnarray}
Here the distributions $2\pi\rho (k)$ and $2\pi\sigma (\Lambda)$ are the solution of 
the following coupled integral equations,
\begin{equation}
2\pi\rho (k) = 1 + \frac{\cos k}{\pi\,u} \int_{-B}^{B}d\Lambda\,{2\pi\sigma (\Lambda)\over 1 +  \left({\sin k - \Lambda\over u}\right)^2} \, ,
\label{equA9}
\end{equation}
and
\begin{eqnarray}
2\pi\sigma (\Lambda) & = & {1\over\pi\,u}\int_{-\pi}^{\pi}dk\,{2\pi\rho (k)\over 1 +  \left({\Lambda-\sin k\over u}\right) ^2} 
\nonumber \\
& - & \frac{1}{2\pi\,u} \int_{-B}^{B}d\Lambda^{\prime}\,{2\pi\sigma (\Lambda^{\prime})\over 1 +  \left({\Lambda -
\Lambda^{\prime}\over 2u}\right)^2} \, .
\label{equA10}
\end{eqnarray}
Such distributions obey the sum rules,
\begin{equation}
{1\over\pi}\int_{-\pi}^{\pi}dk\,2\pi\rho (k) = 2
\hspace{0.20cm}{\rm and}\hspace{0.20cm}
\frac{1}{\pi} \int_{-B}^{B}d\Lambda\,2\pi\sigma (\Lambda) = (1-m) \, .
\label{equA10B}
\end{equation}

The parameter $B = \Lambda_s (k_{F\downarrow})$ appearing in the above equations 
has the limiting behaviors,
\begin{eqnarray}
B & = & \Lambda_s (k_{F\downarrow}) \hspace{0.20cm}{\rm with}
\nonumber \\
\lim_{m\rightarrow 0} B & = & \infty\hspace{0.20cm}{\rm and}\hspace{0.20cm}
\lim_{m\rightarrow 1} B = 0 \, .
\label{QB-r0rs}
\end{eqnarray}

Other $\Lambda_s (q)$ and $\Lambda_{s2} (q)$ values are,
\begin{eqnarray}
\Lambda_s (0) & = & 0\hspace{0.20cm}{\rm and}\hspace{0.20cm}
\Lambda_s (\pm k_{F\uparrow}) = \pm\infty
\nonumber \\
\Lambda_{s2} (0) & = & 0\hspace{0.20cm}{\rm and}\hspace{0.20cm}
\Lambda_{s2} (\pm (k_{F\uparrow} - k_{F\downarrow})) = \pm\infty \, .
\label{qtwoprimelimits}
\end{eqnarray}

The $s$ band dispersion,
\begin{eqnarray}
\varepsilon_{s}^0 (q) & = & {\bar{\varepsilon}_{s}}^0 (\Lambda_s (q))\hspace{0.20cm}{\rm where}
\nonumber \\
{\bar{\varepsilon}_{s}}^0 (\Lambda) & = & \int_{\infty}^{\Lambda}d\Lambda^{\prime}\,2t\,\eta_{s} (\Lambda^{\prime}) \, .
\label{equA11}
\end{eqnarray}
whose zero-energy level is for $0<m<1$ shifted relative to that of $\varepsilon_{s} (q)$ 
defines the spin density curve, as given in Eq. (\ref{hc}). 

In the $m\rightarrow 0$ limit, the $s2$ band does not exist in the ground state. In
that limit, it reduces to $q=0$ with $\varepsilon_{s2} (0)=0$ when $N_{s2}=1$. 
In the same limit, the $s$ band energy dispersion can be written as,
\begin{eqnarray}
\varepsilon_{s} (q) & = & {\bar{\varepsilon}_{s}} (\Lambda_s (q)) 
\hspace{0.20cm}{\rm for}\hspace{0.20cm}q \in \left[-{\pi\over 2},{\pi\over 2}\right] 
\hspace{0.20cm}{\rm where}
\nonumber \\
{\bar{\varepsilon}_{s}} (\Lambda) & = & 
- 2t\int_0^{\infty}d\omega\,{\cos (\omega\,\Lambda)\over\omega\cosh (\omega\,u)}\,J_1 (\omega) \, ,
\label{varesm0}
\end{eqnarray}
and the rapidity function $\Lambda_s (q)$ is defined in terms of its inverse function 
$q = q_s (\Lambda)$ where $\Lambda\in [-\infty,\infty]$ as,
\begin{equation}
q = q_s (\Lambda) = \int_0^{\infty}d\omega\,{\sin (\omega\,\Lambda)\over\omega\cosh (\omega\,u)}\,J_0 (\omega) \, .
\label{qLambdam0}
\end{equation}
In these equations $J_0 (\omega)$ and $J_1 (\omega)$ are Bessel functions.

The $s$ and $s2$ band energy dispersions $\varepsilon_{s} (q)$ and $\varepsilon_{s2} (q)$, 
Eqs. (\ref{equA4}) and (\ref{vares2}), respectively, have limiting values,
\begin{eqnarray}
& & \varepsilon_{s} (0) = - W_{s}^p 
\nonumber \\
& & \varepsilon_{s} (\pm k_{F\downarrow}) = 0
\nonumber \\
& & \varepsilon_{s} (\pm k_{F\uparrow}) = W_{s}^h = 2\mu_B\,h
\nonumber \\
& & \varepsilon_{s2} (0) = 4\mu_B\,h - W_{s2}
\nonumber \\
& & \varepsilon_{s2} (\pm (k_{F\uparrow} - k_{F\downarrow})) = 4\mu_B\,h \, ,
\label{vares2limits}
\end{eqnarray}
where, 
\begin{eqnarray}
\lim_{u\rightarrow 0}W_{s}^p & = & 2t\left(1 - \sin\left({\pi\over 2}\,m\right)\right)
\nonumber \\
\lim_{u\rightarrow 0}W_{s}^h & = & \lim_{u\rightarrow 0}2\mu_B\,h = 4t\sin\left({\pi\over 2}\,m\right)
\nonumber \\
\lim_{u\rightarrow 0}W_{s} & = & W_s^p + W_s ^h = 2t\left(1 + \sin\left({\pi\over 2}\,m\right)\right)
\nonumber \\
\lim_{u\rightarrow\infty}W_{s} & = & W_s^p + W_s ^h = 0
\nonumber \\
\lim_{u\rightarrow 0}W_{s2} & = & 4t\sin\left({\pi\over 2}\,m\right) 
\nonumber \\
\lim_{u\rightarrow\infty}W_{s2} & = & 0 \, ,
\label{W2u0}
\end{eqnarray}
for spin densities $m\in ]0,1[$ and,
\begin{eqnarray}
\lim_{m\rightarrow 1}W_{s} & = & W_s^h = 2\mu_B\,h_c = \sqrt{(4t)^2+U^2} - U
\nonumber \\
\lim_{m\rightarrow 1}W_{s2} & = & \sqrt{(4t)^2+(2U)^2} - 2U \, ,
\label{W2um1}
\end{eqnarray}
for all $u>0$ values.

In the $u\rightarrow 0$ limit, the $s$ band energy dispersions have for
spin densities $m\in ]0,1[$ the following expressions,
\begin{eqnarray}
\varepsilon_{s} (q) & = & \varepsilon_{s}^0 (q) - \varepsilon_{s}^0 (k_{F\downarrow})
= -2t\left(\cos q - \cos k_{F\downarrow}\right) 
\nonumber \\
& = & 2t\sin\left({\pi\over 2}\,m\right) - 2t\cos q
\nonumber \\
\varepsilon_{s}^0 (q) & = & -2t\left(\cos q - \cos k_{F\uparrow}\right) 
\nonumber \\
& = & - 2t\sin\left({\pi\over 2}\,m\right) - 2t\cos q
\nonumber \\
& & {\rm for}\hspace{0.20cm}q \in [-k_{F\uparrow},k_{F\uparrow}] \, .
\label{varepsilonsu0}
\end{eqnarray}

The $s2$ band energy dispersions have for $u\rightarrow 0$ and
spin densities $0<m<1$ the following expressions,
\begin{eqnarray}
\varepsilon_{s2} (q) & = & 4\mu_B\,h - 2t\left(\cos (\vert q \vert + k_{F\downarrow}) - \cos k_{F\uparrow}\right) 
\nonumber \\
& = & 8t\sin\left({\pi\over 2}\,m\right) - 2t\left(\cos (\vert q \vert + k_{F\downarrow}) + \sin\left({\pi\over 2}\,m\right)\right) 
\nonumber \\
& = & 6t\sin\left({\pi\over 2}\,m\right) - 2t\cos (\vert q \vert + k_{F\downarrow}) 
\nonumber \\
\varepsilon_{s2}^0 (q) & = & - 2t\left(\cos (\vert q \vert + k_{F\downarrow}) - \cos k_{F\uparrow}\right) 
\nonumber \\
& = & - 2t\sin\left({\pi\over 2}\,m\right) - 2t\cos (\vert q \vert + k_{F\downarrow}) 
\nonumber \\
{\rm for} & & q\in [-(k_{F\uparrow}-k_{F\downarrow}),(k_{F\uparrow}-k_{F\downarrow})] \, .
\label{varepsilons2u0}
\end{eqnarray}

In the $u\rightarrow 0$ limit, the corresponding group velocities, Eq. (\ref{equA4B}), read,
\begin{eqnarray}
v_{s} (q) & = & 2t\sin q\hspace{0.20cm}
{\rm for}\hspace{0.20cm}q \in [-k_{F\uparrow},k_{F\uparrow}] 
\nonumber \\
v_{s2} (q) & = & {\rm sgn}\{q\}\,2t\sin (\vert q \vert + k_{F\downarrow})\hspace{0.20cm}{\rm for}
\nonumber \\
&& q\in [-(k_{F\uparrow}-k_{F\downarrow}),(k_{F\uparrow}-k_{F\downarrow})] \, ,
\label{vsu0}
\end{eqnarray}
respectively, so that,
\begin{equation}
v_{s} (k_{F\downarrow}) = v_{s2} (k_{F\uparrow}-k_{F\downarrow}) = 2t\cos\left({\pi\over 2}m\right) \, .
\label{vvu0}
\end{equation}

In the $m\rightarrow 1$ spin density limit, the $s$ band energy dispersions are for all $u>0$
values given by the following integrals,
\begin{eqnarray}
\varepsilon_{s} (q) & = &
- {2t\over \pi}\int_{-\pi}^{\pi}d k \sin k
\arctan \left({\sin k - \Lambda_s (q)\over u}\right) 
\nonumber \\
& + & \sqrt{(4t)^2+U^2}-U 
\nonumber \\
\varepsilon_{s}^0 (q) & = &
- {2t\over \pi}\int_{-\pi}^{\pi}d k \sin k
\arctan \left({\sin k - \Lambda_s (q)\over u}\right) 
\nonumber \\
{\rm for} & & q\in [-\pi,\pi] \, ,
\label{band-fullyP}
\end{eqnarray}
where the rapidity function $\Lambda_s (q)$ is defined by its inverse function as,
\begin{equation}
q = q_s (\Lambda) = {1\over \pi}\int_{-\pi}^{\pi}d k  
\arctan \left({\Lambda - \sin k\over u}\right) \, .
\label{qLs0}
\end{equation}

In the same $m\rightarrow 1$ limit, the $s2$ band energy dispersions are for all $u>0$
values given by the integrals,
\begin{eqnarray}
\varepsilon_{s2} (q) & = &
- {2t\over \pi}\int_{-\pi}^{\pi}d k \sin k
\arctan \left({\sin k - \Lambda_s (q)\over 2u}\right) 
\nonumber \\
& + & \sqrt{(8t)^2+(2U)^2} - 2U 
\nonumber \\
\varepsilon_{s2}^0 (q) & = &
- {2t\over \pi}\int_{-\pi}^{\pi}d k \sin k
\arctan \left({\sin k - \Lambda_{s2} (q)\over 2u}\right) 
\nonumber \\
{\rm for} & & q\in [-\pi,\pi] \, ,
\label{s2band-fullyP}
\end{eqnarray}
where the rapidity function $\Lambda_{s2} (q)$ is again defined by its inverse function as,
\begin{equation}
q = q_{s2} (\Lambda) =  {1\over \pi}\int_{-\pi}^{\pi}d k  
\arctan \left({\Lambda - \sin k\over 2u}\right) \, .
\label{qLs20}
\end{equation}

For $u\gg 1$, one can derive analytical expressions for the $s$ and $s2$ band energy dispersions 
and the corresponding group velocities, Eq. (\ref{equA4B}), for spin densities $m$ in the limits $m\rightarrow 0$ and $(1-m)\ll 1$. 
For $u\gg 1$ and in the $m\rightarrow 0$ limit, the behaviors of the $s$ band energy dispersions and group velocity are,
\begin{eqnarray}
\varepsilon_{s} (q) & = & - {\pi\,t\over 2u} \cos q
\hspace{0.20cm}{\rm and}\hspace{0.20cm}
\varepsilon_{s}^0 (q) = \varepsilon_{s} (q)
\nonumber \\
v_{s} (q) & = & {\pi\,t\over 2u} \sin q 
\nonumber \\
& & {\rm for}\hspace{0.20cm}q \in [-\pi/2,\pi/2]\hspace{0.20cm}{\rm and}\hspace{0.20cm}m\rightarrow 0 \, .
\label{varepsilonsulm0}
\end{eqnarray}

For $u\gg 1$ and $(1-m)\ll 1$, the $s$ band energy dispersions and group velocity, Eq. (\ref{equA4B}), 
behave as,
\begin{eqnarray}
\varepsilon_{s} (q) & = & - {t\over u}\,(\cos q -1) 
\nonumber \\
&& + {t\over u}\,(1-m)\sin q\,\arctan\left({1\over 2}\tan\left({q\over 2}\right)\right) 
\nonumber \\
\varepsilon_{s}^0 (q) & = & -{2t\over u} + \varepsilon_{s} (q) 
\nonumber \\
& = & - {t\over u}\,(\cos q + 1)
\nonumber \\
&& + {t\over u}\,(1-m)\sin q\,\arctan\left({1\over 2}\tan\left({q\over 2}\right)\right) 
\nonumber \\
v_{s} (q) & = & {t\over u}\sin q  + {t\over u}\,(1-m){\sin q\over 1 + 3\cos^2\left({q\over 2}\right)}
\nonumber \\
&& + {t\over u}\,(1-m)\cos q\,\arctan\left({1\over 2}\tan\left({q\over 2}\right)\right)
\nonumber \\
& & {\rm for}\hspace{0.20cm} q \in \left[-{\pi\over 2}(1+m),{\pi\over 2}(1+m)\right]
\nonumber \\
& & {\rm and}\hspace{0.20cm}(1-m)\ll 1 \, .
\label{varepsilonsulm1}
\end{eqnarray}

For $u\gg 1$ and in the $m\rightarrow 0$ limit, the $s2$ band energy dispersion 
and group velocity vanish, consistent with the momentum and energy widths of the $s2$ band vanishing.
For $u\gg 1$ and $(1-m)\ll 1$, they behave as,
\begin{eqnarray}
\varepsilon_{s2} (q) & = & {4t\over u} - {t\over 2u}\,(1 + \cos q)  
\nonumber \\
&& + {t\over 2u}\,(1-m)\sin q\{\arctan\left(2\tan\left({q\over 2}\right)\right) 
\nonumber \\
&& + \arctan\left({2\over 3}\tan\left({q\over 2}\right)\right)\}
\nonumber \\
\varepsilon_{s2}^0 (q) & = & \varepsilon_{s2} (q) - {4t\over u}
\nonumber \\
v_{s2} (q) & = & {t\over 2u}\sin q  + {t\over 2u}\,(1-m)\sin q\{{1\over 1 + 3\sin^2\left({q\over 2}\right)}
\nonumber \\
&& + {3\over 4 + 5\cos^2\left({q\over 2}\right)}\}
\nonumber \\
&& + {t\over 2u}\,(1-m)\cos q\,\{\arctan\left(2\tan\left({q\over 2}\right)\right) 
\nonumber \\
&& + \arctan\left({2\over 3}\tan\left({q\over 2}\right)\right)\}\hspace{0.20cm}{\rm for}
\nonumber \\
& & q \in [-\pi m,\pi m]\hspace{0.20cm}{\rm and}\hspace{0.20cm}(1-m)\ll 1 \, .
\label{varepsilons2ulm01}
\end{eqnarray}

For $u\gg 1$ and $(1-m)\ll 1$, the following equality holds,
\begin{equation}
v_{s} (k_{F\downarrow}) = v_{s2} (k_{F\uparrow}-k_{F\downarrow}) = {\pi t\over 2u} (1-m) \, .
\label{vvm1}
\end{equation}

The phase shifts play an important role in the spin dynamical properties. They are given by,
\begin{eqnarray}
2\pi\,\Phi_{s,\beta}(q,q') & = & 2\pi\,\bar{\Phi }_{s,\beta} \left(r,r'\right) 
\nonumber \\
{\rm where}\hspace{0.20cm}r = {\Lambda_{s}(q)\over u} & &
{\rm and}\hspace{0.20cm} r' = {\Lambda_{\beta}(q')\over u} \, .
\label{Phi-barPhi}
\end{eqnarray}
In the case of the excited energy eigenstates involved in the studies of this paper,
$\beta=s,s2$. The rapidity phase shifts 
$2\pi\bar{\Phi }_{s,\beta}\left(r,r'\right)$ on the right-hand side of the above equality are functions of the rapidity-related variables
$r=\Lambda/u$ of the $s$ and $s2$ branches. They are defined by the following integral equations,
\begin{eqnarray}
\bar{\Phi }_{s,s}\left(r,r'\right) & = & {1\over \pi}\arctan\left({r-r'\over 2}\right) 
\nonumber \\
& + & \int_{-B/u}^{B/u} dr''\,G (r,r'')\,{\bar{\Phi}}_{s,s} (r'',r') \, ,
\label{Phis1sn-m}
\end{eqnarray}
and
\begin{eqnarray}
\bar{\Phi }_{s,s2}\left(r,r'\right) & = & {1\over \pi}\arctan(r-r') + {1\over \pi}\arctan\left({r-r'\over 3}\right) 
\nonumber \\
& + & \int_{-B/u}^{B/u} dr''\,G (r,r'')\,{\bar{\Phi}}_{s,s2} (r'',r') \, .
\label{Phis1s2-m}
\end{eqnarray}
The kernel $G (r,r')$ in Eqs. (\ref{Phis1sn-m}) and (\ref{Phis1s2-m}) is 
for $u>0$ given by,
\begin{equation}
G(r,r') = - {1\over{2\pi}}\left({1\over{1+((r-r')/2)^2}}\right) \, .
\label{Gne1}
\end{equation} 

The phase shifts that appear in the expressions of the branch line exponents read,
\begin{eqnarray}
\Phi_{s,s}\left(\iota k_{F\downarrow},q\right) & = & \bar{\Phi }_{s,s} \left(\iota {B\over u},{\Lambda_s (q)\over u}\right) 
\nonumber \\
\Phi_{s,s2}\left(\iota k_{F\downarrow},q\right) & = & \bar{\Phi }_{s,s2} \left(\iota {B\over u},{\Lambda_{s2} (q)\over u}\right) 
\nonumber \\
& & {\rm where}\hspace{0.2cm}\iota = \pm 1 \, .
\label{Phis-all-qq}
\end{eqnarray}

In the $m\rightarrow 0$ limit, the phase shift
$\Phi_{s,s}(q,q')$ in units of $2\pi$ can be written as,
\begin{eqnarray}
\Phi_{s,s}(q,q') & = & \bar{\Phi }_{s,s} \left(\Lambda_ s (q),\Lambda (q')\right) \hspace{0.20cm}
{\rm where}
\nonumber \\
\bar{\Phi }_{s,s} \left(\Lambda,\Lambda'\right) & = & {1\over\pi}\int_0^{\infty}d\omega\,{\sin (\omega\,(\Lambda - \Lambda'))\over\omega
\left(1 + e^{2\omega u}\right)} \, ,
\label{Phi-barPhim0}
\end{eqnarray}
and the rapidity function $\Lambda_s (q)$ is defined in terms of its inverse function in Eq. (\ref{qLambdam0}).
The integral in Eq. (\ref{Phi-barPhim0}) can be solved for $u>0$, with the result,
\begin{eqnarray}
\bar{\Phi }_{s,s} (\Lambda,\Lambda') & = & {i\over 2\pi}\,\ln\left(
{\Gamma \left({1\over 2} + i {(\Lambda - \Lambda')\over 4u}\right)\Gamma \left(1 - i {(\Lambda - \Lambda')\over 4u}\right)
\over\Gamma \left({1\over 2} - i {(\Lambda - \Lambda')\over 4u}\right)\Gamma \left(1 + i {(\Lambda - \Lambda')\over 4u}\right)}\right)
\nonumber \\
& & {\rm for}\hspace{0.2cm}\Lambda \neq \iota\infty
\hspace{0.20cm}{\rm where}\hspace{0.2cm}\iota = \pm 1
\nonumber \\
& = & {\iota\over 2\sqrt{2}}
\hspace{0.2cm}{\rm for}\hspace{0.2cm}\Lambda= \iota\infty
\hspace{0.2cm}{\rm and}\hspace{0.2cm}\Lambda'\neq \iota\infty
\nonumber \\
& = &\iota\left({3\over 2\sqrt{2}} - 1\right)
\hspace{0.2cm}{\rm for}\hspace{0.2cm}\Lambda = \Lambda' = \iota\infty \, ,
\label{Phis1sn-m0}
\end{eqnarray}
where $\Gamma (x)$ is the usual $\gamma$ function.

The use of Eq. (\ref{Phis1sn-m0}) leads to the following expressions for the 
phase shift $\Phi_{s,s}\left(\iota k_{F},q\right)=\lim_{m\rightarrow 0}\Phi_{s,s}\left(\iota k_{F\downarrow},q\right)$
in the $m\rightarrow 0$ limit for $u>0$,
\begin{eqnarray}
\lim_{m\rightarrow 0}\Phi_{s,s}\left(\iota k_{F\downarrow},q\right) & = & \Phi_{s,s}\left(\iota\pi/2,q\right)
\nonumber \\
& = & {\iota\over 2\sqrt{2}}
\hspace{0.2cm}{\rm for}\hspace{0.2cm}q\neq \iota k_{F\downarrow}
\nonumber \\
& = &\iota\left({3\over 2\sqrt{2}} - 1\right)
\hspace{0.2cm}{\rm for}\hspace{0.2cm}q= \iota k_{F\downarrow}
\nonumber \\
{\rm for} & & u>0\hspace{0.2cm}{\rm where}\hspace{0.2cm}\iota = \pm 1 \, .
\label{Phis-all-qq-0}
\end{eqnarray}

In the $m\rightarrow 0$ limit and for $u>0$, the
phase shift $\Phi_{s,s2}\left(\iota k_{F},0\right)=\lim_{m\rightarrow 0}\Phi_{s,s2}\left(\iota k_{F\downarrow},q\right)$ 
has in units of $2\pi$ the following value,
\begin{equation}
\Phi_{s,s2}\left(\iota k_{F},0\right) = {\iota\over\sqrt{2}} \, .
\label{Phis2-all-qq-0}
\end{equation}

For $u\gg 1$ and in the $m\rightarrow 1$ limit, the phase shifts $\Phi_{s,s}\left(\iota k_{F\downarrow},q\right)$
and $\Phi_{s,s2}\left(\iota k_{F\downarrow},q\right)$ behave as,
\begin{eqnarray}
\lim_{m\rightarrow 1}\Phi_{s,s}(\iota k_{F\downarrow},q) & = & \Phi_{s,s}(0,q) 
\nonumber \\
& = & - {1\over\pi}\arctan\left({1\over 2}\tan\left({q\over 2}\right)\right) 
\nonumber \\
\lim_{m\rightarrow 1}\Phi_{s,s2}(\iota k_{F\downarrow},q) & = & \Phi_{s,s2}(0,q)
\nonumber \\
& = & - {1\over\pi}\arctan\left(2\tan\left({q\over 2}\right)\right) 
\nonumber \\
& - & {1\over\pi}\arctan\left({2\over 3}\tan\left({q\over 2}\right)\right) \, . 
\label{PhiUinfm1qF}
\end{eqnarray}

The $s$ band Fermi-points phase-shift parameters $\xi^{j}_{s\,s}$ where $j=0,1$
are given by,
\begin{equation}
\xi^{j}_{s\,s} = 1 + \sum_{\iota=\pm 1} (\iota)^{j}\,\Phi_{s,s}\left(k_{F\downarrow},\iota k_{F\downarrow}\right) \, .
\label{x-aa}
\end{equation}
They play an important role in both the spectral and static properties. For one electron per site,
the equality $\xi^{0}_{s\,s}=1/\xi^{1}_{s\,s}$ holds, so that only one of
these two parameters is needed, for instance $\xi^{1}_{s\,s}$, which is
a diagonal entry of the 1D Hubbard model dressed charge matrix \cite{Frahm,Carmelo_93}.

From manipulations of the phase-shift integral equation, Eq. (\ref{Phis1sn-m}),
one finds that the latter parameter is given by,
\begin{equation}
\xi_{s\,s}^1 = \xi_{s\,s}^1 (B/u) \, .
\label{xi1all}
\end{equation}

The function $\xi_{s\,s}^1 (r)$ on the right-hand side of this equation at $r=B/u$ is the solution
of the integral equation,
\begin{equation}
\xi_{s\,s}^1 (r) = 1 + \int_{-B/u}^{B/u} dr'\,G (r,r')\,\xi_{s\,s}^1 (r') \, .
\label{xi-ss-qq}
\end{equation}
The kernel $G (r,r')$ appearing here is given in Eq. (\ref{Gne1}).

For $u>0$, the parameter $\xi^{1}_{s\,s}$ continuously increases from $\xi^{1}_{s\,s}=1/\sqrt{2}$ as $m\rightarrow 0$ to
$\xi^{1}_{s\,s}=1$ for $m\rightarrow 1$, so that its limiting values are,
\begin{equation}
\lim_{m\rightarrow 0}\xi_{s\,s}^1 = {1\over\sqrt{2}}\hspace{0.5cm}{\rm and}
\hspace{0.5cm}
\lim_{m\rightarrow 1}\xi_{s\,s}^1 = 1 \, .
\label{Limxiss}
\end{equation}

The parameter $\xi^{1}_{s\,s}$ is also related to the phase shift $\Phi_{s,s2} (k_{F\downarrow},q)$ in Eq. (\ref{Phis-all-qq})
as follows,
\begin{eqnarray}
\xi^{1}_{s\,s} & = & - \Phi_{s,s2}(\pm k_{F\downarrow},(k_{F\uparrow}-k_{F\downarrow}))
\nonumber \\
& = & \Phi_{s,s2}(\pm k_{F\downarrow},-(k_{F\uparrow}-k_{F\downarrow})) \, .
\label{xi1Phiss2}
\end{eqnarray}

Finally the parameter $\xi_{s\,s2}^{0}$ that also appears in the momentum dependent
exponents is given by,
\begin{equation}
\xi_{s\,s2}^{0} = 2\Phi_{s,s2}(k_{F\downarrow},0) \, ,
\label{xis20}
\end{equation}
where the phase shift $\Phi_{s,s2} (k_{F\downarrow},q)$ is defined in Eq. (\ref{Phis-all-qq}).
At $q=0$ it is such that $\Phi_{s,s2}(\iota k_{F\downarrow},0)= \iota\,\Phi_{s,s2}(k_{F\downarrow},0)$.
This justifies why $\iota\,\xi_{s\,s2}^{0} = 2\Phi_{s,s2}(\iota k_{F\downarrow},0)= \iota\,2\Phi_{s,s2}(k_{F\downarrow},0)$
for $\iota=\pm 1$.

The parameter $\xi_{s\,s2}^{0}$ continuously decreases from $\xi_{s\,s2}^0=\sqrt{2}$ as $m\rightarrow 0$ to
$\xi_{s\,s2}^0=0$ for $m\rightarrow 1$. Consitent,
it follows from Eqs. (\ref{Phis2-all-qq-0}) and (\ref{PhiUinfm1qF}) that, 
\begin{equation}
\lim_{m\rightarrow 0}\xi_{s\,s2}^0 = \sqrt{2}\hspace{0.5cm}{\rm and}
\hspace{0.5cm}
\lim_{m\rightarrow 1}\xi_{s\,s2}^0 = 0 \, .
\label{Limxis20}
\end{equation}


\end{document}